\documentclass[oupdraft]{bio}
\usepackage{url}
\usepackage{xr}
\usepackage{amsmath}
\usepackage{graphicx}
\usepackage{enumerate}
\usepackage{setspace}
\usepackage{amsfonts}
\usepackage{comment}
\usepackage{amsmath}
\usepackage{graphicx}
\usepackage{algorithm}
\usepackage{tabularx}
\usepackage{amssymb}
\usepackage{amsthm}
\usepackage{array}
\usepackage{xcolor}
\usepackage{float}
\usepackage[toc,page]{appendix}
\usepackage{algpseudocode}
\usepackage{afterpage}

\newtheorem{proposition}{Proposition}
\newcommand{\PreserveBackslash}[1]{\let\temp=\\#1\let\\=\temp}
\newcolumntype{C}[1]{>{\PreserveBackslash\centering}p{#1}}
\newcommand{\indep}{\raisebox{0.05em}{\rotatebox[origin=c]{90}{$\models$}}}
\allowdisplaybreaks
\makeatletter
\newcommand{\multiline}[1]{%
	\begin{tabularx}{\dimexpr\linewidth-\ALG@thistlm}[t]{@{}X@{}}
		#1
	\end{tabularx}
}
\makeatother

\begin{document}

% Title of paper
\title{Exponential family measurement error models for single-cell CRISPR screens}

% List of authors, with corresponding author marked by asterisk
\author{TIMOTHY BARRY \\
% Author addresses
\textit{Dept.\ of Biostatistics, Harvard T.H.\ Chan School of Public Health, Boston MA}
\\
% E-mail address for correspondence
{\texttt{tbarry@hsph.harvard.edu}} \\
KATHRYN ROEDER \\
\textit{Dept.\ of Statistics and Data Science, Carnegie Mellon University, Pittsburgh, PA} \\
EUGENE KATSEVICH \\
\textit{Dept. of Statistics and Data Science, University of Pennsylvania, Philadelphia, PA}
}

% Running headers of paper:
\markboth%
% First field is the short list of authors
{T. Barry, K. Roeder, and E. Katsevich}
% Second field is the short title of the paper
{Exponential family measurement error models for single-cell CRISPR screens}

\maketitle

\begin{abstract}{CRISPR genome engineering and single-cell RNA sequencing have accelerated biological discovery. Single-cell CRISPR screens unite these two technologies, linking genetic perturbations in individual cells to changes in gene expression and illuminating regulatory networks underlying diseases. Despite their promise, single-cell CRISPR screens present considerable statistical challenges. We demonstrate through theoretical and real data analyses that a standard method for estimation and inference in single-cell CRISPR screens --``thresholded regression'' -- exhibits attenuation bias and a bias-variance tradeoff as a function of an intrinsic, challenging-to-select tuning parameter. To overcome these difficulties, we introduce GLM-EIV (``GLM-based errors-in-variables''), a new method for single-cell CRISPR screen analysis. GLM-EIV extends the classical errors-in-variables model to responses and noisy predictors that are exponential family-distributed and potentially impacted by the same set of confounding variables. We develop a computational infrastructure to deploy GLM-EIV across hundreds of processors on clouds (e.g., Microsoft Azure) and high-performance clusters. Leveraging this infrastructure, we apply GLM-EIV to analyze two recent, large-scale, single-cell CRISPR screen datasets, yielding several new insights.}{CRISPR, single-cell, GLM, mixture model, cloud computing}
\end{abstract}

\section{Introduction}

CRISPR is a genome engineering tool that has enabled scientists to precisely edit human and nonhuman genomes, opening the door to new medical therapies \citep{Musunuru2021} and accelerating biological discovery \citep{Przybyla2021}. Recently, scientists have paired CRISPR genome engineering with single-cell RNA sequencing \citep{Datlinger2017}. The resulting assays, known as ``single-cell CRISPR screens,'' link genetic perturbations in individual cells to changes in gene expression. Single-cell CRISPR screens have enabled breakthrough progress on longstanding challenges in genetics, such as causally mapping genome wide association study (GWAS) variants to target genes at genome-wide scale \citep{Morris2023}.

Despite their promise, single-cell CRISPR screens present considerable statistical challenges. One difficulty is that the ``treatment'' --- i.e., the presence or absence of a CRISPR perturbation --- is assigned randomly to cells and is not directly observable. As a consequence, one cannot know with certainty which cells were perturbed. Instead, one must leverage an indirect, quantitative proxy of perturbation presence or absence to ``guess'' which cells received a perturbation. This indirect proxy takes the form of a so-called guide RNA count, with higher counts indicating that a cell is more likely to have been perturbed. A standard approach to single-cell CRISPR screen analysis is to impute perturbation assignments onto the cells by simply thresholding the guide RNA counts; using these imputations, one can attempt to estimate the effect of the perturbation on gene expression. We call this standard approach ``thresholded regression'' or the ``thresholding method.''

We study estimation and inference in single-cell CRISPR screens from a statistical perspective, formulating the data generating mechanism using a new class of measurement error models. We assume that the response variable $y$ is a GLM of an underlying predictor variable $x^*$ and vector of confounders $z$. We do not observe $x^*$ directly; rather, we observe a noisy version $x$ of $x^*$ that itself is a GLM of $x^*$ and the same set of confounders $z$. The goal of the analysis is to estimate the effect of $x^*$ on $y$ using the observed data $(x, y, z)$ only. In the context of the biological application, $x^*$, $x$, $y$, and $z$ are CRISPR perturbations, guide RNA counts, gene expressions, and technical confounders, respectively.

Our work makes two main contributions. First, we conduct a detailed study of the thresholding method. Notably, we demonstrate on real data that the thresholding method exhibits attenuation bias and a bias-variance tradeoff as a function of the selected threshold, and we recover these phenomena in precise mathematical terms in a simplified Gaussian setting. Second, we introduce a new method, GLM-EIV (``GLM-based errors-in-variables''), for single-cell CRISPR screen analysis. GLM-EIV extends the classical errors-in-variables model \citep{Carroll2006} to responses and noisy predictors that are exponential family-distributed and potentially impacted by the same set of confounding variables. GLM-EIV thereby implicitly estimates the probability that each cell was perturbed, obviating the need to explicitly impute perturbation assignments via thresholding. We implement several statistical accelerations to bring the cost of GLM-EIV down to within about an order of magnitude of the thresholding method. We additionally develop a Docker-containerized application to deploy GLM-EIV at-scale across tens or hundreds of processors on clouds (e.g., Microsoft Azure) and high-performance clusters. 

Our analyses indicate that single-cell CRISPR screens fall into two main problem settings: the more challenging ``high background contamination'' setting and the easier ``low background contamination'' setting. GLM-EIV outperforms thresholded regression by a considerable margin in the high background contamination setting; in the low background contamination setting, by contrast, GLM-EIV and thresholded regression perform similarly, provided that accurate guide RNA-to-cell assignments are used within the thresholded regression model. We show that a simplified version of GLM-EIV can be used to obtain these guide RNA-to-cell assignments in the low background contamination setting, thereby neutralizing a tuning parameter that until this point has been challenging to select.

\section{Assay background}
There are several classes of single-cell CRISPR screen assays, each suited to answer a different set of biological questions. In this work we mostly focus on high-multiplicity of infection (MOI) single-cell CRISPR screens, which we motivate and describe here. The human genome consists of genes, enhancers (segments of DNA that regulate the expression of one or more genes), and other genomic elements. GWAS have revealed that the majority ($>90\%$) of variants associated with diseases lie outside genes and inside enhancers \citep{Gallagher2018}. These noncoding variants are thought to contribute to disease by modulating the expression of one or more disease-relevant genes. Scientists do not know the gene (or genes) through which most noncoding variants exert their effect, limiting the interpretability of GWAS results. A central open challenge in genetics, therefore, is to link enhancers that harbor GWAS variants to the genes that they target at genome-wide scale \citep{Morris2023}.

High-MOI single-cell CRISPR screens are a promising emerging technology for resolving this challenge \citep{Morris2023,Mostafavi2023}. High-MOI single-cell CRISPR screens combine CRISPR interference (CRISPRi) --- a version of CRISPR that represses a targeted region of the genome --- with single-cell sequencing. The experimental protocol is as follows. First, the scientist develops a library of several hundred to several thousand CRISPRi perturbations, each designed to target a candidate enhancer for repression. The scientist then cultures tens or hundreds of thousands of cells and delivers the CRISPRi perturbations to these cells. The perturbations assort into the cells randomly, with each cell receiving on average 10-40 distinct perturbations. Conversely, a given perturbation enters about 0.1-2\% of cells (this work).

After waiting several days for CRISPRi to take effect, the scientist profiles each cell's transcriptome (i.e., its gene expressions) and the set of perturbations that it received. Finally, the scientist conducts perturbation-to-gene association analyses. Figure \ref{analysis_challenges}a depicts this process schematically, with colored bars (blue, red, and purple) representing distinct perturbations. For a given perturbation (e.g., the perturbation represented in blue), the scientist partitions the cells into two groups: those that received the perturbation (top) and those that did not (bottom). Next, for a given gene, the scientist runs a differential expression analysis across the two groups of cells, producing an estimate for the magnitude of the gene expression change in response to the perturbation. If the estimated change in expression is large, the scientist can conclude that the enhancer \textit{targeted} by the perturbation exerts a strong regulatory effect on the gene. This procedure is repeated for a large set of preselected perturbation-gene pairs. The enhancer-by-enhancer approach is valid because the perturbations assort into cells approximately independently of one another.

The genomics literature has produced several methods for high-MOI single-cell CRISPR screen analysis \citep{Gasperini2019,Xie2019a,Barry2021,Wang2021}. For example, Gasperini et al.\ applied negative binomial GLMs (as implemented in the Monocle software; \cite{Trapnell2014}) to carry out the differential expression analysis described above. Moreover, Xie et al.\ applied chi-squared-like tests of independence for this purpose. Unfortunately, both of these approaches have limitations: the former can break down when the gene expression model is misspecified, and the latter does not adjust for the presence of technical confounders. In a prior work we introduced introduced SCEPTRE, a custom implementation of the conditional randomization test \citep{Candes2018,Liu2021} tailored to single-cell CRISPR screen data. SCEPTRE simultaneously adjusts for confounder presence and ensures robustness to expression model misspecification, thereby overcoming limitations of previous approaches and demonstrating improved sensitivity and specificity on single-cell CRISPR screen data. In this work we tackle a set of analysis challenges complimentary to those addressed by SCEPTRE. Most importantly, we seek to account for the fact that the perturbation is measured with noise. Additionally, we seek to \textit{estimate} (with confidence) the effect size of a perturbation on gene expression change, an objective that we did not consider in the original SCEPTRE study.

\section{Analysis challenges and proposed statistical model}

\begin{figure}[h!]
\centering
\includegraphics[width=1\linewidth]{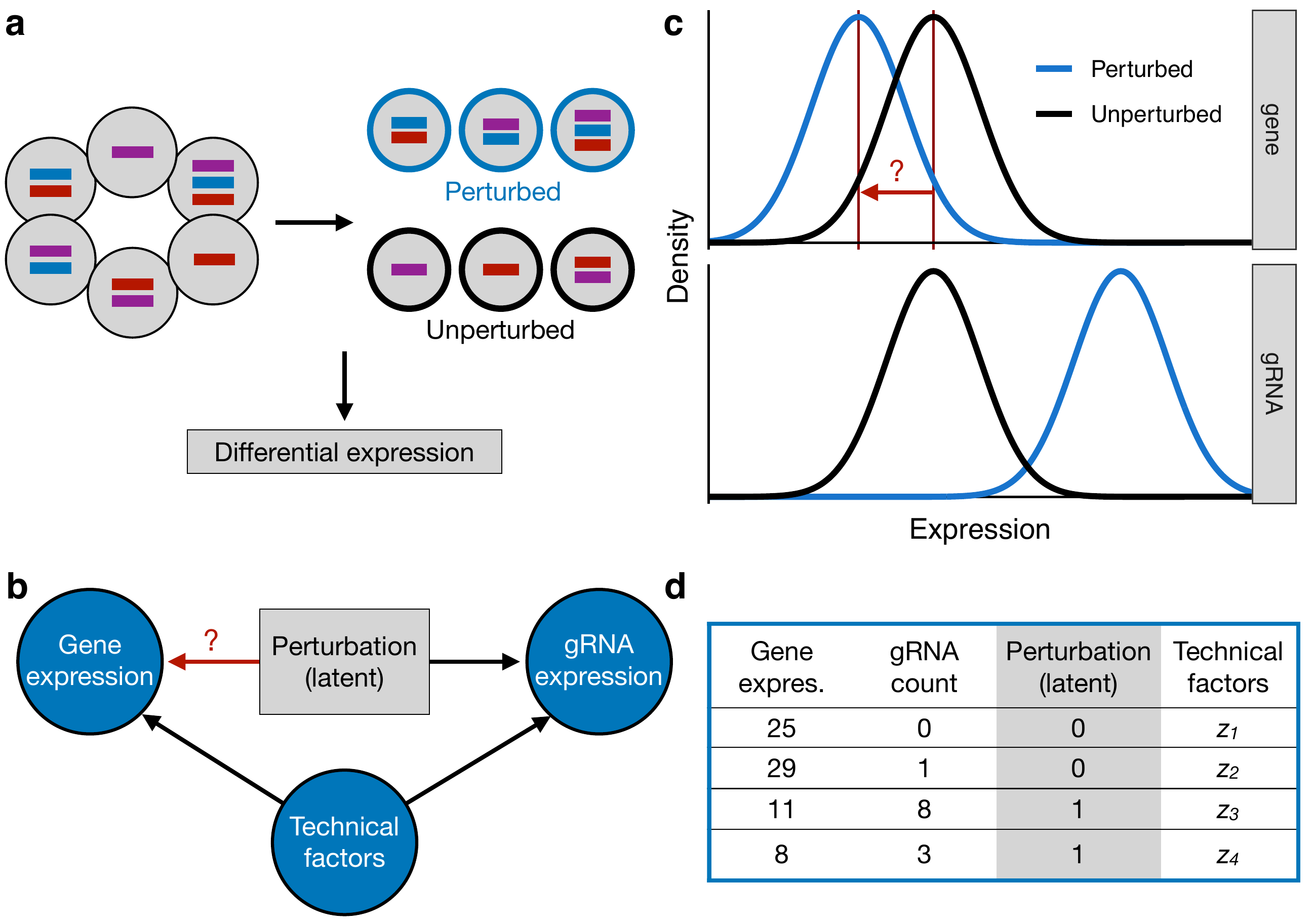}
\caption{\textbf{Experimental design and analysis challenges}: \textbf{a,} Experimental design. For a given perturbation (e.g., the perturbation indicated in blue), we partition the cells into two groups: perturbed and unperturbed. Next, for a given gene, we conduct a differential expression analysis across the two groups, yielding an estimate of the impact of the given perturbation on the given gene. \textbf{b,} DAG representing all variables in the system. The perturbation (latent) impacts both gene expression and gRNA expression; technical factors act as confounders, also impacting gene and gRNA expression. The target of estimation is the effect of the perturbation on gene expression. \textbf{c,} Schematic illustrating the ``background read'' phenomenon. Due to errors in the sequencing and alignment processes, unperturbed cells exhibit a nonzero gRNA count distribution (bottom). The target of estimation is the change in mean gene expression in response to the perturbation (top). \textbf{d}, Example data on four cells for a given perturbation-gene pair. Note that (i) the perturbation is unobserved, and (ii) the gene and gRNA data are discrete counts.}
\label{analysis_challenges}
\end{figure}

High-MOI single-cell CRISPR screens present several statistical challenges, four of which we highlight here. Throughout, we consider a single perturbation-gene pair. First, the ``treatment'' variable --- i.e., the presence or absence of a perturbation --- cannot be directly observed. Instead, perturbed cells transcribe molecules called  \textit{guide RNAs} (or \textit{gRNAs}) that serve as indirect proxies of perturbation presence. We must leverage these gRNAs to impute (explicitly or implicitly) perturbation assignments onto the cells (Figure \ref{analysis_challenges}b). Second, ``technical factors'' --- sources of variation that are experimental rather than biological in origin --- impact the measurement of both gene and gRNA expressions and therefore act as confounders (Figure \ref{analysis_challenges}b). Third, the gene and gRNA data are sparse, discrete counts. Consequently, classical statistical approaches that assume Gaussianity or homoscedasticity are not directly applicable. Finally, sequenced gRNAs sometimes map to cells that have not received a perturbation. This phenomenon, which we call ``background contamination,'' results from errors in the sequencing and alignment processes. The marginal distribution of the gRNA counts is best conceptualized as a mixture model (Figure \ref{analysis_challenges}c; Gaussian distributions used for illustration purposes only). Unperturbed and perturbed cells both exhibit nonzero gRNA count distributions, but this distribution is shifted upward for perturbed cells. Figure \ref{analysis_challenges}d shows example data on four (of possibly tens or hundreds of thousands of) cells. The analysis objective is to leverage the gene expressions and gRNA counts to estimate the effect of the (latent) perturbation on gene expression, accounting for the technical factors.

We propose to model the single-cell CRISPR screen data-generating process using a pair of GLMs. Let $n \in \mathbb{N}$ be the number of cells assayed in the experiment. Consider a single perturbation and a single gene. For cell $i \in \{1, \dots, n\}$, let $p_i \in \{0,1\}$ indicate perturbation presence or absence; let $m_i \in \mathbb{N}$ be the number of gene transcripts sequenced; let $g_i \in \mathbb{N}$ be the number of gRNA transcripts sequenced; let $d^m_i \in \mathbb{N}$ be the number of gene transcripts sequenced across \textit{all} genes (i.e., the library size or sequencing depth); let $d^g_i$ be the gRNA library size; and finally, let $z_i \in \mathbb{R}^{d-2}$ be the cell-specific covariates, including sequencing batch, percent mitochondrial reads, etc. (We note that most single-cell CRISPR screens have been carried out on cell lines consisting of a uniform cell type; however, if multiple cell types are present in the data, then cell type could be included as a covariate in the model.) The letters ``m,'' ``g'', and ``d'' stand for ``mRNA,'' ``gRNA,'' and ``depth,'' respectively.

Building on the work of several previous authors \citep{Robinson2008,Townes2019,Hafemeister2019}, \cite{Sarkar2021} proposed a simple strategy for modeling single-cell gene expression data, which, in the framework of negative binomial GLMs, is equivalent to using the log-transformed library size as an offset term. Sarkar and Stephens' framework enjoys strong theoretical and empirical support; therefore, we generalize their approach to model \textit{both} gene and gRNA modalities in single-cell CRISPR screen experiments. To this end we assume that the gene expression counts are given by
\begin{equation}\label{glmeiv_model_1}
m_i |(p_i, z_i, d^m_i) \sim \textrm{NB}_{s^m}(\mu_i^m); \quad \log(\mu^m_i) = \beta^m_0 + \beta^m_1 p_i + \gamma_m^T z_i + \log(d^m_i),
\end{equation}
where (i) $\textrm{NB}_{s^m}(\mu^m_i)$ is a negative binomial distribution with mean $\mu^m_i$ and known size parameter $s^m$; (ii) $\beta^m_0 \in \mathbb{R}, \beta^m_1 \in \mathbb{R},$ and $\gamma_m \in  \mathbb{R}^{d-2}$ are unknown parameters; and (iii) $\log(d_i^m)$ is an offset term. (We note that the ``size parameter'' is simply the inverse of the negative binomial dispersion parameter; ``size parameter'' does not refer to library size in this context.)  Similarly, we model the gRNA counts by
\begin{equation}\label{glmeiv_model_2}
g_i | (p_i, z_i, d^g_i) \sim \textrm{NB}_{s^g}\left(\mu_i^g\right); \quad \log(\mu_i^g) = \beta^g_0 + \beta^g_1p_i + \gamma^T_g z_i + \log(d^g_i),
\end{equation}
where $\mu^g_i$, $s^g$, $\beta^g_0$, $\beta^g_1$, $\gamma_g$, and $d^g_i$ are analogous. We use a negative binomial GLM to model the gRNA counts as well as the gene expressions because the gRNA transcripts are generated via the same biological mechanism as the gene transcripts \citep{Datlinger2017,Hill2018}. We model the marginal perturbation as $p_i \sim \textrm{Bern}(\pi)$, where $p_i$ is an unobserved binary variable indicating presence ($p_i = 1$) or absence ($p_i = 0$) of the perturbation. We restrict $\pi$, the probability of perturbation, to the interval $(0, 1/2]$ to ensure that the model is identifiable; this restriction is reasonable given that each perturbation infects only a small fraction of cells. The gRNA intercept term $\beta^g_0$ controls the ambient level of gRNA expression, i.e.\ the rate at which gRNA reads are generated in the absence of the perturbation. The perturbation coefficient $\beta^g_1$ controls the extent to which perturbed and unperturbed cells differentially express the gRNA; the target of inference $\beta^m_1$ is challenging to estimate when $\beta^g_1$ is close to zero, as the gRNA distributions of the perturbed and unperturbed cells are hard to differentiate in this region of the problem space. Together, (\ref{glmeiv_model_1}), (\ref{glmeiv_model_2}), and the marginal distribution of $p_i$ define the negative binomial GLM-EIV model.

The log-transformed sequencing depth $\log(d^m_i)$ is included as an offset term in (\ref{glmeiv_model_1}) so that $\beta^m_0 + \beta^m_1 p_i + \gamma^T_m z_i$ can be interpreted as a relative expression. Exponentiating both sides of (\ref{glmeiv_model_1}) reveals that the mean gene expression $\mu_i^m$ of the $i$th cell is
$\exp \left( \beta^m_0 + \beta^m_1 p_i + \gamma^T_m z_i \right) d_i^m.$ Because $d^m_i$ is the sequencing depth, $\exp \left( \beta^m_0 + \beta^m_1 p_i + \gamma^T_m z_i \right)$ is the \textit{fraction} of all transcripts sequenced in the cell produced by the gene under consideration. The target of inference $\beta^m_1$ is the log fold change in expression in response to the perturbation, controlling for the technical factors. Fold change in this context is the ratio of the mean gene expression in perturbed cells to the mean gene expression in unperturbed cells. Hence, $\exp(\beta^m_1) = 1$ (i.e., $\beta^m_1 = 0$) indicates no change in expression, whereas $\exp(\beta^m_1) > 1$ (i.e., $\beta^m_1 > 0$) and $\exp(\beta^m_1) < 1$ (i.e., $\beta^m_1 < 0$) indicate an increase and decrease in expression, respectively.

In this work we analyzed two large-scale, high-MOI, single-cell CRISPR screen datasets published by \cite{Gasperini2019} and \cite{Xie2019a}. Gasperini (resp., Xie) targeted approximately 6,000 (resp., 500) candidate enhancers in a population of approximately 200,000 (resp., 100,000) cells. Gasperini additionally designed several hundred positive control, gene-targeting perturbations and 50 non-targeting, negative control perturbations to assess method sensitivity and specificity.

\section{Analysis of the thresholding method}

We studied thresholding from empirical and theoretical perspectives, highlighting several potential limitations of the approach. In the context of the negative bionomial GLM-EIV model introduced above (\ref{glmeiv_model_1}-\ref{glmeiv_model_2}), the thresholding method leverages the gRNA counts (\ref{glmeiv_model_2})
to impute the latent perturbation indicator (\ref{glmeiv_model_2}), thereby reducing the full data generating process to a single, gene expression model (\ref{glmeiv_model_1}). We studied Gasperini et al.'s  variant of the thresholding method (i.e., thresholded negative binomial regression), as this version of the thresholding method is standard and relates most closely to GLM-EIV. The method is defined as follows:
\begin{itemize}
\item[1.] For a given threshold $c \in \mathbb{N}$, let the imputed perturbation assignment $\hat{p}_i \in \{0, 1\}$ be given by $\hat{p}_i = 0$ if  $g_i < c$ and $\hat{p}_i = 1$ otherwise.
\item[2.] Assume that $m_i$ is related to $\hat{p}_i, d^m_i,$ and $z_i$ through the following GLM:
\begin{equation}\label{thresh_glm}
m_i | (\hat{p}_i, z_i, d^m_i) \sim \textrm{NB}_{s^m}(\mu^m_i); \quad
\log(\mu^m_i) = \beta^m_0 + \beta^m_1 \hat{p}_i + \gamma^T_m z_i + \log\left(d_i^m\right).
\end{equation}
The model (\ref{thresh_glm}) is equivalent to the model (\ref{glmeiv_model_2}), but the latent perturbation indicator $p_i$ has been replaced by the imputed perturbation indicator $\hat{p}_i.$	
\item[3.] Fit a GLM to (\ref{thresh_glm}) to obtain an estimate and CI for the target of inference $\beta^m_1$.
\end{itemize}

To shed light on empirical challenges of the thresholding method, we applied thresholded negative binomial regression to analyze the set of positive control perturbation-gene pairs in the Gasperini dataset. The positive control pairs consisted of perturbations that targeted gene transcription start sites (TSSs) for inhibition. Repressing the TSS of a given gene decreases its expression; therefore, the positive control pairs \textit{a priori} are expected to exhibit a strong decrease in expression.

\begin{figure}[h!]
\centering
\includegraphics[width=1\linewidth]{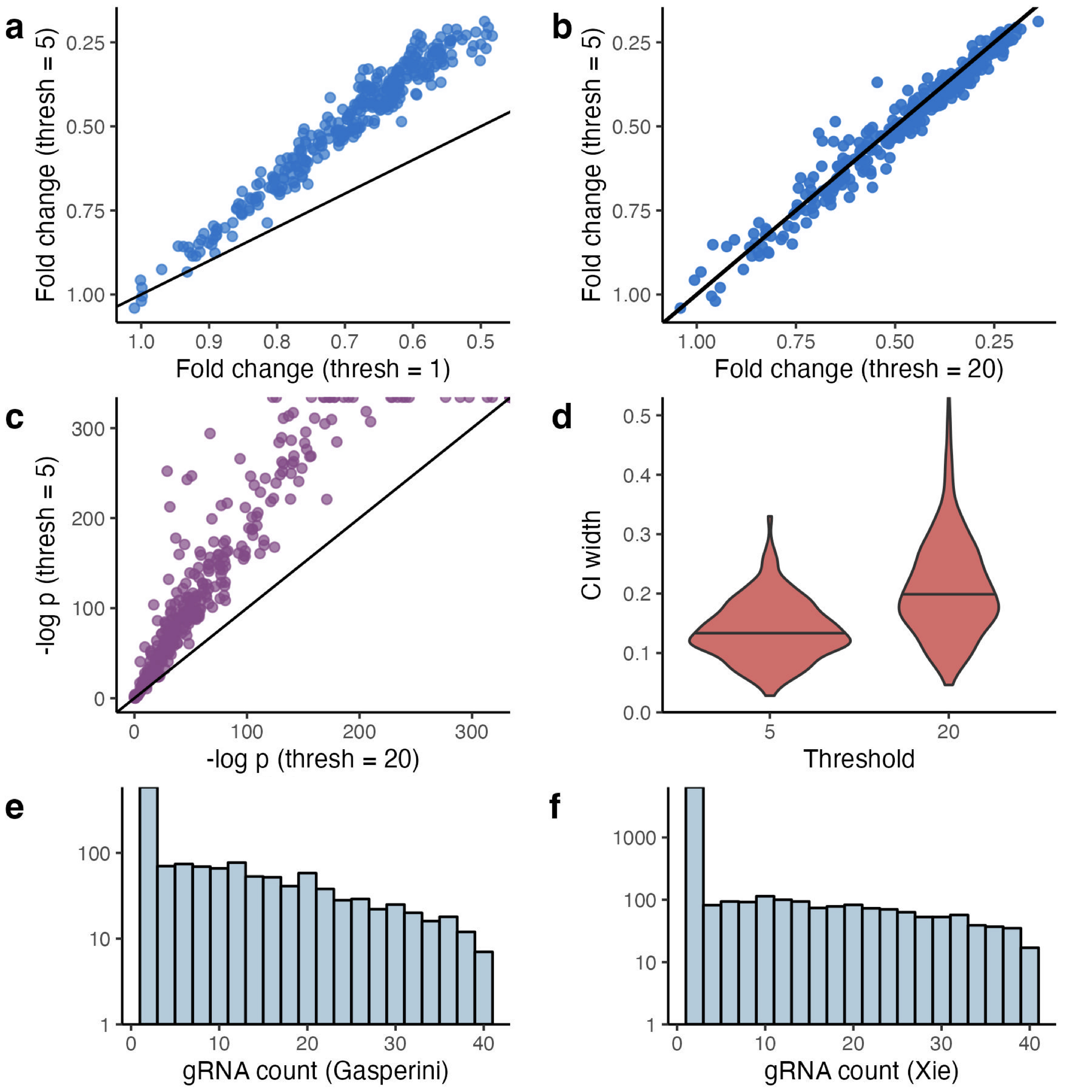}
\caption{\textbf{Empirical challenges of thresholded regression.} \textbf{a-b,} Estimates for fold change (i.e., $\exp(\beta^m_1)$ in model (\ref{thresh_glm})) produced by threshold = 5 versus threshold = 1 (a) and threshold = 5 versus threshold = 20 (b). The selected threshold substantially impacts the results. \textbf{c-d,} $p$-values (c) and CI widths (d) produced by threshold = 5 versus threshold = 20. The $p$-values correspond to a test of the null hypothesis $H_0: \beta^m_1 = 0$, i.e., a log fold change in gene expression of zero. A threshold of 5 yields more significant p-values and more confident estimates. \textbf{e-f}, Empirical distribution of a gRNA from Gasperini (e) and Xie (f) data (0 counts not shown). These gRNA count distributions do not appear to imply an obvious threshold.}
\label{thresholding_empirical}
\end{figure}

To investigate the sensitivity of the thresholding method to threshold choice, we deployed the method using three different choices for the threshold: 1, 5, and 20. We found that the chosen threshold substantially impacted the results (Figure \ref{thresholding_empirical}a-b): estimates for fold change produced by threshold = 1 were smaller in magnitude (i.e., closer to the baseline of $1$) than those produced by threshold = 5 (Figure \ref{thresholding_empirical}a). On the other hand, estimates produced by threshold = 5 and threshold = 20 were more concordant (Figure \ref{thresholding_empirical}b). 

We reasoned that thresholded regression systematically underestimated true effect sizes on the positive control pairs, especially for threshold = 1. For a given perturbation, the majority ($>98\%$) of cells are unperturbed. This imbalance leads to an asymmetry: misclassifying \textit{unperturbed} cells as \textit{perturbed} is intuitively ``worse'' than misclassifying \textit{perturbed} cells as \textit{unperturbed}. Misclassified unperturbed cells contaminate the set of truly perturbed cells, leading to attenuation bias; by contrast, misclassified perturbed cells are swamped in number and ``neutralized'' by the truly unperturbed cells. Setting the threshold to a large number reduces the unperturbed-to-perturbed misclassification rate, decreasing bias.

We hypothesized, however, that the reduction in bias obtained by selecting a large threshold causes the variance of the estimator to increase. To investigate, we compared $p$-values and confidence intervals produced by threshold = 5 and threshold = 20 for the target of inference $\beta^m_1$. We found that threshold = 5 yielded smaller (i.e., more significant) $p$-values and narrower confidence intervals than did threshold = 20 (Figures \ref{thresholding_empirical}c-d). We concluded that the threshold controls a bias-variance tradeoff: as the threshold increases, the bias of the estimator decreases and the variance increases.

Finally, to determine whether there is an ``obvious'' location at which to draw the threshold, we examined the empirical gRNA count distribution of a gRNA from the Gasperini (Figure \ref{thresholding_empirical}e) and Xie (Figure \ref{thresholding_empirical}f) dataset (counts of $0$ omitted). The distributions peaked at $1$ and then tapered off gradually; there did not exist a sharp boundary that cleanly separated the perturbed from the unperturbed cells. Overall, we concluded that the thresholding method faces several challenges: (i) the threshold is a tuning parameter that significantly impacts the results; (ii) the threshold mediates an intrinsic bias-variance tradeoff; and (iii) the gRNA count distributions may not imply a clear threshold selection strategy.

Next, we studied the thresholding method from a theoretical perspective, recovering in a simplified Gaussian setting phenomena revealed in the empirical analysis. Due to space constraints we relegate this analysis to Appendix \ref{sec:appendix_theory}, but we briefly summarize the main results here. First, we derived an exact expression for the asymptotic relative bias of the thresholding estimator $\hat{\beta}^m_1.$ Leveraging this exact expression, we showed that (i) the thresholding estimator strictly underestimates (in absolute value) the true value of $\beta^m_1$ over all choices of the threshold and over all values of the regression coefficients (an example of \textit{attenuation bias}; \cite{Stefanski2000a}); and (ii) the magnitude of the bias decreases monotonically in $\beta^g_1$, comporting with the intuition that the problem becomes easier as the gRNA mixture distribution becomes increasingly well-separated. Second, we derived an asymptotically exact bias-variance decomposition for $\hat{\beta}_m$, demonstrating that as the threshold tends to infinity, the bias decreases and the variance increases.

\section{GLM-based errors-in-variables (GLM-EIV)}

We introduce the general GLM-EIV model, which generalizes the negative binomial GLM-EIV model (\ref{glmeiv_model_1}-\ref{glmeiv_model_2}) to arbitrary exponential family response distributions and link functions, thereby providing much greater modeling flexibility. We derive efficient methods for estimation and inference in this model and develop a pipeline to deploy the model at-scale.
 
\subsection{Model and model properties}\label{sec:model}

The general GLM-EIV model uses an arbirary GLM to model the gene and gRNA modalities:

\begin{equation}\label{eqn:full_glmeiv_model_gene}
m_i |(p_i, z_i, o^m_i) \sim f_m(\mu_i^m); \quad r_m(\mu^m_i) = \beta^m_0 + \beta^m_1 p_i + \gamma_m^T z_i + o^m_i,
\end{equation}
\begin{equation}\label{eqn:full_glmeiv_model_grna}
g_i |(p_i, z_i, o^g_i) \sim f_g(\mu_i^g); \quad r_g(\mu^g_i) = \beta^g_0 + \beta^g_1 p_i + \gamma_g^T z_i + o^g_i.
\end{equation}

Here, $f_m$ (resp., $f_g$) is an exponential family distribution with mean $\mu^m_i$ (resp., $\mu^g_i$); $r_m$ and $r_g$ are the link function for the gene and gRNA models, respectively; and $o^m_i$ and $o^g_i$ are the (possibly zero) offset terms for the gene and gRNA models. In practice we typically set $o^m_i$ and $o^g_i$ to the log-transformed library sizes (i.e., $\log(d^m_i)$ and $\log(d^g_i)$). Again, we assume that the unobserved perturbation indicator $p_i$ is drawn from a $\textrm{Bern}(\pi)$ distribution. More model details are available in Appendix \ref{sec:glmeiv_details}.

The GLM-EIV model can be seen as a generalization of the simple errors-in-variables model (when the predictor is binary); the latter is defined as follows:
\begin{equation}\label{classical_eiv}
y_i = \beta_0 + \beta_1 x^*_i + \epsilon_i; \quad
x_i = x^*_i + \tau_i,
\end{equation}
where, $x^*_i \sim \textrm{Bern}(\pi), \epsilon_i, \tau_i \sim N(0,1),$ and $\epsilon_i$,$\tau_i$, and $x^*_i$ are independent. GLM-EIV extends (\ref{classical_eiv}) in at least three directions: first, GLM-EIV allows $y_i$ and $x_i$ to follow exponential family (i.e, not just Gaussian) distributions; second, GLM-EIV allows $y_i$ and $x_i$ to be related to $x^*_i$ through arbitrary (i.e., not just linear) link functions; and finally, GLM-EIV allows confounders $z_i$ to impact both $x_i$ and $y_i$. Therefore, $x_i$ and $y_i$ can be conditionally dependent given $x^*_i$, enabling GLM-EIV to capture more complex dependence relationships between $x_i$ and $y_i$ than is possible in (\ref{classical_eiv}) or other standard measurement error models.

\subsection{Estimation and inference, and computational infrastructure}\label{sec:estimation_inference}
We derived an EM algorithm (Algorithm \ref{algo:em_full}) to estimate the parameters of the GLM-EIV model. We briefly introduce some notation. Let $\beta_m = [\beta^m_0, \beta^m_1, \gamma_m]^T$ be the vector of unknown gene model parameters and $\beta_g = [\beta^g_0, \beta^g_1, \gamma_g]^T$ the vector of unknown gRNA model parameters. Let $m$, $g$, $o^m$, and $o^g$ be the vector of gene expressions, gRNA expressions, gene library sizes, and gRNA library sizes. Finally, let $X$ be the observed design matrix; let $\tilde{X}$ be the augmented design matrix that results from concatenating the column of (unobserved) $p_i$s to $X$; and let $\tilde{X}(0)$ (resp, $\tilde{X}(1)$) be the matrix that results from setting all of the $p_i$s in $\tilde{X}$ to $0$ (resp., $1$).

The E step entails computing the membership probability (i.e., the probability of perturbation) in each cell. The membership probability $T_i(1)$ of cell $i \in \{1, \dots, n\}$ given the current parameter estimates $(\beta_m^\textrm{(t)}, \beta_g^\textrm{(t)}, \pi^\textrm{(t)})$ and observed data $(m_i, g_i)$ is
$T_i(1) = \mathbb{P}(p_i = 1| M_i = m_i, G_i = g_i, \beta^\textrm{(t)}_m, \beta^\textrm{(t)}_g, \pi^\textrm{(t)}).$ We can calculate this quantity by applying (i) Bayes rule, (ii) the conditional independence property of $M_i$ and $G_i$, (iii) the density of $M_i$ and $G_i$, and (iv) a log-sum-exp-type trick to ensure numerical stability. Next, we produce updated estimates $\pi^{(\textrm{t} + 1)}$, $\beta_g^{(\textrm{t}+1)}$, and $\beta_m^{(\textrm{t}+1)}$ of the parameters by maximizing the M step objective function. It turns out that maximizing this objective function is equivalent to setting $\pi^{\textrm{(t+1)}}$ to the mean of the current membership probabilities and setting $\beta_g^{(\textrm{t}+1)}$ and $\beta_m^{(\textrm{t}+1)}$ to the fitted coefficients of a GLM weighted by the current membership probabilities (Algorithm \ref{algo:em_full}). We iterate through the E and M steps until the log likelihood (\ref{marginal_log_lik}) converges (Appendix \ref{sec:glmeiv_details}). Our EM algorithm is reminiscent of (but distinct from) that of  \cite{Ibrahim1990}, who also applied weighted GLM solvers to carry out an M step of an EM algorithm.

\begin{algorithm}
	\caption{EM algorithm for GLM-EIV model.}\label{algo:em_full}
{\fontsize{10}{10}\selectfont
	\begin{algorithmic}
		\Require Pilot estimates $\beta^\textrm{curr}_m, \beta^\textrm{curr}_g,$ and $\pi^\textrm{curr}$; data $m$, $g$, $o^m$, $o^g$, and $X$; gene expression distribution $f_m$ and link function $r_m$; gRNA expression distribution $f_g$ and link function $r_g$.  
		\While{Not converged}
		\For{$i \in \{1, \dots, n\}$} \Comment{E step}
		\State $T_i(1) \gets \mathbb{P}\left(p_i = 1 |M_i = m_i, G_i = g_i, \beta_m^\textrm{curr}, \beta_g^\textrm{curr}, \pi^\textrm{curr} \right)$
		\State $T_i(0) \gets 1 - T_i(1)$
		\EndFor
		\State $\pi^{\textrm{curr}} \gets (1/n) \sum_{i=1}^n T_i(1)$ \Comment{M step}
		\State $w \gets [T_1(0), T_2(0), \dots, T_n(0), T_1(1), T_2(1), \dots, T_n(1)]^T$
		\For{$k \in \{g,m\}$}
		\State  \multiline{ 
			Fit a GLM $GLM_k$ with responses $[k,k]^T$, offsets $[o^k, o^k]^T$, weights $w$, design matrix $[\tilde{X}(0)^T, \tilde{X}(1)^T]^T$, distribution $f_k$, and link function $r_k$.
		}
		\State Set $\beta_k^\textrm{curr}$ to the estimated coefficients of $GLM_k$.
		\EndFor
		\State Compute log likelihood using $\beta_m^\textrm{curr}$, $\beta_g^\textrm{curr}$, and $\pi^\textrm{curr}$.
		\EndWhile
		\State $\hat{\beta}_m \gets \beta_m^\textrm{curr}$; $\hat{\beta}_g \gets \beta_g^\textrm{curr}$; $\hat{\pi} \gets \pi^\textrm{curr}$.
		\State \textbf{return} $(\hat{\beta}_m, \hat{\beta}_g, \hat{\pi})$
	\end{algorithmic}
}
\end{algorithm}

After fitting the model, we perform inference on the estimated parameters. The easiest approach, given the complexity of the log likelihood, would be to run a bootstrap. This strategy, however, is prohibitively slow, as the data are large and the EM algorithm is iterative. Therefore, we derived an analytic formula for the asymptotic observed information matrix using Louis's Theorem (\cite{Louis1982}; Appendix \ref{sec:glmeiv_details}). Leveraging this analytic formula, we can calculate standard errors quickly, enabling us to perform inference in practice on real, large-scale data.

A downside of the EM algorithm (Algorithm \ref{algo:em_full}) is that it requires fitting many GLMs. Assuming that we run the algorithm 15 times using randomly-generated pilot estimates (to improve chances of convergence to the global maximum), and assuming that the algorithm iterates through E and M steps about 10 times per run, we must fit approximately 300 GLMs. (These numbers are based on exploratory applications of the method to real and simulated data.) We instead devised a strategy to produce a highly accurate pilot estimate of the true parameters, enabling us to run the algorithm once and converge upon the MLE within a few iterations. The strategy involves layering several statistical ``tricks'' on top of one another. Briefly, we first obtain pilot estimates for the nuisance parameters $\beta^m_0,\gamma_m,\beta^g_0,$ and $\gamma_g$ by regressing the gene and gRNA expression vectors onto the observed design matrix $X$; the resulting estimates are close to the full GLM-EIV model maximum likelihood estimates because the probability of perturbation is small. Next, we obtain pilot estimates for $\pi$ and the perturbation effect parameters $\beta^m_1$ and $\beta^g_1$ by estimating a simplified, ``reduced'' GLM-EIV model; this second step does not require fitting any GLMs. (See Appendix \ref{sec:statistical_accelerations} for additional details.) Overall, the statistical accelerations reduce the number of GLMs that must be fit to $<10$ in most cases.

Next, we developed a computational infrastructure to apply GLM-EIV to large-scale, single-cell CRISPR screen data. The infrastructure leverages \texttt{Nextflow}, a programming language that facilitates building data-intensive pipelines, and \texttt{ondisc}, an \texttt{R}/\texttt{C++} package that we developed (in a separate project; preprint forthcoming) to facilitate large-scale computing on single-cell data. \texttt{Nextflow} and \texttt{ondisc} together enable the construction of highly portable single-cell pipelines: one can analyze data out-of-memory on a laptop or in a distributed fashion across hundreds of processors on a cloud (e.g., Microsoft Azure, Google Cloud) or high-performance cluster. Leveraging these technologies, we built a Docker-containerized pipeline for deploying GLM-EIV at-scale. The pipeline recycles computation when possible, saving a considerable amount of compute; see Appendix \ref{sec:computing} for details. Overall, the statistical accelerations and computational infrastructure make the deployment of GLM-EIV to large-scale single-cell CRISPR screen quite feasible.

\subsection{The gRNA mixture assignment method}\label{sec:grna_mixture_method}

Thus far we have described two methods for estimating the effect of a perturbation on gene expression: the simple thresholding method and the more complex GLM-EIV method. A third approach of intermediate complexity --- which we call the ``gRNA mixture assignment'' approach --- is to (i) fit a mixture model to the gRNA count distribution, (ii) use this fitted mixture model to impute perturbation identities onto cells, and then (iii) regress the gene expressions onto the imputed perturbation indicators (as well as the remaining covariates). The gRNA mixture assignment approach enjoys at least two strengths relative to the simpler thresholding approach: the former negates the threshold tuning parameter and can account for variation across cells due to covariates.

\citet{Replogle2020} proposed a simple gRNA mixture assignment strategy that involves fitting a Poisson-Gaussian mixture model to the log-transformed gRNA counts and then assigning gRNAs to cells using the posterior perturbation probabilities of the fitted model. (We call this method the Nat.\ Biotech.\ 2020 method, representing the journal and year in which the method appeared.) Unfortunately, this method poses several conceptual and practical difficulties. First, it is unclear how the method fits the Poisson component of the mixture distribution to the log-transformed gRNA expressions, as the transformed expressions are not integer-valued. Second, due to recent changes in the Python ecosystem, we and others have had difficulty with installing the Python package upon which the Nat.\ Biotech.\ 2020 method relies. (See Appendix \ref{sec:Replogle_method} for further discussion of the Nat.\ Biotech.\ 2020 method.)

Following \cite{Replogle2020}, we devised an alternate gRNA mixture assignment strategy that is tethered more closely to the data-generating mechanism. For a given gRNA, we regress the gRNA counts onto the (latent) perturbation indicator and covariates (while ignoring the gene expressions; model \ref{eqn:full_glmeiv_model_grna}). We assign perturbation identities to cells by thresholding the posterior perturbation probabilities of the fitted model at 1/2. The latent variable gRNA model is a subset of the full GLM-EIV model (\ref{eqn:full_glmeiv_model_gene}-\ref{eqn:full_glmeiv_model_grna}). Thus, we used the GLM-EIV EM algorithm to fit the latent variable gRNA model, enabling us to exploit the various techniques that we developed in the context of GLM-EIV for obtaining fast and numerically stable estimates.

\section{Simulation study}\label{sec:simulation}

We conducted a comprehensive suite of six simulation studies to compare the empirical performance of GLM-EIV, the thresholding method, and the gRNA mixture assignment method. (We coupled the latter method to standard regression on the imputed perturbation assignments to estimate the perturbation effect size.) We describe one simulation study here and defer the remaining simulation studies to the Appendix \ref{sec:extra_sims}. We generated data on $n = 50,000$ cells from the GLM-EIV model, setting the target of inference $\beta^m_1$ to $\log(0.25)$ and the probability of perturbation $\pi$ to $0.02$. $\beta^m_1 = \log(0.25)$ represents a decrease in gene expression by a factor of $4$, which is a fairly large effect size on the order of what we might observe for a positive control pair. We included ``sequencing batch'' (modeled as a Bernoulli-distributed variable) as a covariate and sequencing depth (modeled as a Poisson-distributed variable) as an offset. We varied the log-fold change in gRNA expression, $\beta^g_1,$ over a grid on the interval $[\log(1), \log(4)];$ $\beta^g_1$ controls problem difficulty, with higher values corresponding to easier problem settings. We generated the gene expression count data from two response distributions: Poisson and negative binomial (size parameter fixed at $s = 20$ for the latter; see simulation study 3 for an exploration of different values of $s$). We generated the gRNA count data from a Poisson distribution. For each parameter setting (defined by a $\beta^g_1$-distribution pair), we synthesized $n_\textrm{sim} = 500$ i.i.d.\ datasets. Appendix \ref{sec:extra_sims} compares the parameter values used in the simulation study to those estimated from real data.

We applied four methods to the simulated data: ``vanilla'' GLM-EIV, accelerated GLM-EIV, thresholded regression, and the gRNA mixture assignment method. We used the Bayes-optimal decision boundary for classification as the threshold for the thresholding method (as derived in Section \ref{sec:non_gaussian_bayes_thresholds}). We ran all methods on the negative binomial data twice: once treating the size parameter $s$ as a known constant and once treating $s$ as unknown. In the latter case we used the \texttt{glm.nb} function from the \texttt{MASS} package to estimate $s$ before applying the methods \citep{Ripley2013}. We note that none of the methods accounts for the error in estimating $s$ when computing coefficient standard errors. We display the results of the simulation study in Figure \ref{main_text_sim}. Columns correspond to distributions (i.e., Poisson, NB with known $s$, and NB with unknown $s$), and rows correspond to performance metrics (i.e., bias, mean squared error, CI coverage rate (nominal rate $95\%$), CI width, and method run time). The $\beta^g_1$ parameter is plotted on the horizontal axis, and the methods are depicted in different colors. (GLM-EIV is masked by accelerated GLM-EIV in several panels).

We found that GLM-EIV outperformed the gRNA mixture method and that the gRNA mixture method outperformed thresholded regression across the metrics of bias, mean squaured error, and confidence interval coverage. We reasoned that GLM-EIV outperformed the gRNA mixture method because (i) GLM-EIV leveraged information from \textit{both} modalities (rather than the gRNA modality alone) to assign perturbation identities to cells, and (ii) GLM-EIV produced soft rather than hard assignments, capturing the inherent uncertainty in whether a perturbation occurred. We additionally reasoned that the gRNA mixture method outperformed thresholded regression because the gRNA mixture method better accounted for heterogeneity across cells due to the covariates. Notably, accelerated GLM-EIV performed as well as vanilla GLM-EIV on all statistical metrics (rows 1-4) despite having substantially lower computational cost (bottom row). In fact, the running time of accelerated GLM-EIV was almost within an order of magnitude of that of the thresholding method. As expected, the confidence interval coverage of the methods degraded somewhat in the negative binomial case under estimated \textit{s} as opposed to known \textit{s}, but this difference was not substantial. Appendix \ref{sec:extra_sims} presents additional simulation studies in which we generate data from a Gaussian model, vary $\beta^m_1$ and $s$, and assess the performance of the methods on data containing unmeasured covariates and outliers.
 
% Interestingly, thresholded regression exhibited better confidence interval coverage under estimated $s$ than under known $s$ (row 3). Estimating $s$ leads to slight inflation bias (i.e., overestimating the true effect size), whereas, as we showed previously, thresholding leads to attenuation bias (i.e., underestimating the true effect size). These phenomena partially canceled, yielding less biased estimates. GLM-EIV exhibited worse performance under unknown $s$ than known $s$, likely due to inaccurate $s$ estimation. We note that GLM-EIV and the thresholding method in principle are compatible with \textit{any} $s$ estimation procedure, including those based on more sophisticated techniques, such as regularization \citep{Hafemeister2019}.%We defer rigorous investigation of the impact of different $s$ estimation strategies on these methods to future work.

\section{Real data application I: estimating perturbation effects on high-MOI data}

Leveraging our computational infrastructure, we applied GLM-EIV and the thresholding method to analyze the entire Gasperini and Xie datasets. GLM-EIV ran in under two days on both datasets, using no more than 250 processors and two gigabytes of memory per process. We report only the most important aspects of the analysis and results in the main text; full details are available in Appendix \ref{sec:data_analysis_details}. We set the threshold in the thresholding method to the approximate Bayes-optimal decision boundary, as our theoretical analyses and simulation studies indicated that the Bayes-optimal decision boundary is a good choice for the threshold when the gRNA count distribution is well-separated. Operating under the assumption that the effect of the perturbation on gRNA expression is similar across pairs, we leveraged the fitted GLM-EIV models to approximate the Bayes boundary in the following way: we (i) sampled several hundred gene-perturbation pairs, (ii) extracted the fitted values $\hat{\beta}_g$ and $\hat{\pi}$ from the GLM-EIV models fitted to these pairs, (iii) computed the median $\overline{\hat{\beta}_g}$ and $\overline{\hat{\pi}}$ across the $\hat{\beta}_g$s and $\hat{\pi}$s, and (iv) used $\overline{\hat{\beta}_g}$ and $\overline{\hat{\pi}}$ to estimate a dataset-wide Bayes-optimal decision boundary (Section \ref{sec:non_gaussian_bayes_thresholds}). We repeated this procedure on both datasets, yielding a threshold of $3$ for Gasperini and $7$ for Xie.

\begin{figure}[h!]
\centering
\includegraphics[width=0.9\linewidth]{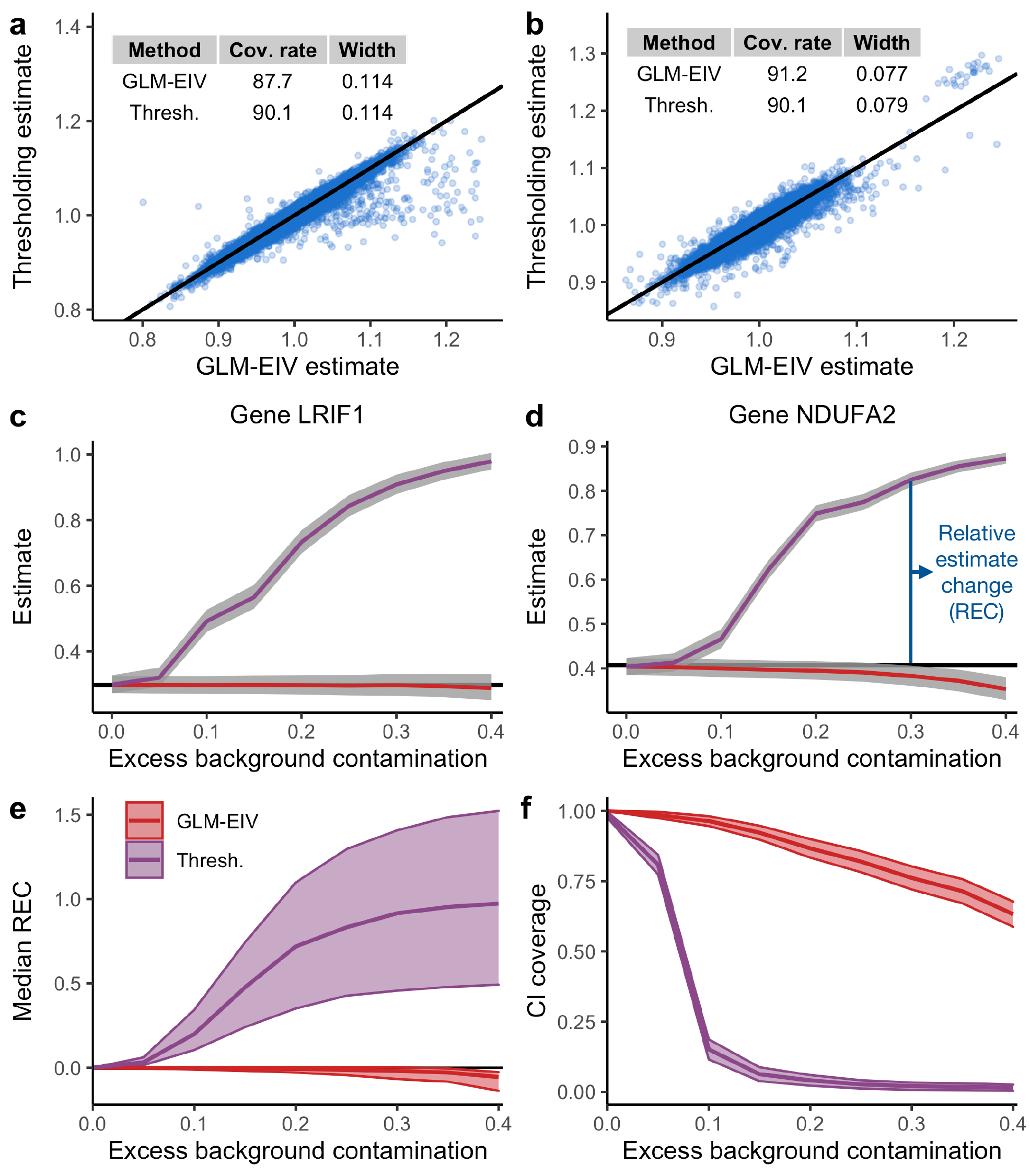}
\caption{\textbf{Applying GLM-EIV to analyze large-scale, high-MOI data}. \textbf{a-b}, Estimates for fold change produced by GLM-EIV and thresholded regression on Gasperini (\textbf{a}) and Xie (\textbf{b}) negative control pairs. \textbf{c-d}, Estimates produced by GLM-EIV and thresholded regression on two positive control pairs -- \textit{LRIF1} (\textbf{a}) and \textit{NDUFA2} (\textbf{b}) -- plotted as a function of excess background contamination. Grey bands, 95\% CIs for the target of inference outputted by the methods. \textbf{e-f}, Median relative estimate change (REC; \textbf{e}) and confidence interval coverage rate (\textbf{f}) across \textit{all} 322 positive control pairs, plotted as a function of excess background contamination. Panels (\textbf{c-f}) together illustrate that GLM-EIV demonstrated greater stability than thresholded regression as background contamination increased.}\label{fig:real_data}
\end{figure}

We compared GLM-EIV to thresholded regression on the real data, focusing specifically on the negative control pairs (i.e., gene-perturbation pairs for which the ground truth fold change is known to be $1$; Appendix \ref{sec:data_analysis_details}). We found that GLM-EIV and the thresholding method produced similar results (Figure \ref{fig:real_data}a-b): estimates, CI coverage rates, and CI widths were concordant. CI coverage rates, which ranged from 87.7\%-91.2\%, were slightly below the nominal rate of 95\%, likely due to mild model misspecification. The estimated effect of the perturbation on gRNA expression $\exp(\hat{\beta}_1^g)$ was unexpectedly large: the 95\% CI for this parameter (averaged across pairs) was $[4306, 5186]$ and $[300, 316]$ on the Gasperini and Xie data, respectively. We reasoned that the datasets lay in a region of the parameter space in which thresholding is a tenable strategy (provided the threshold is selected well). However, this was not obvious \textit{a priori} and may not be the case for other datasets. We note that GLM-EIV produced outlier estimates (defined as estimated fold change $<0.75$ or $>1.25$) on a small ($< 2.5 \%$ on Gasperini, $<0.05\%$ on Xie) number of pairs consisting of a handful of genes, likely due to non-global EM convergence. These outliers are not plotted in Figures \ref{fig:real_data}a-b but were used to compute the CI coverage reported in the inset tables.

To evaluate performance of GLM-EIV versus thresholding in more challenging settings, we increased the difficulty of the perturbation assignment problem by generating partially-synthetic datasets. First, for a given pair, we sampled gRNA counts directly from the fitted GLM-EIV model. Next, to simulate elevated background contamination, we sampled gRNA counts from a slightly modified version of the fitted model in which we increased the mean gRNA expression of \textit{unperturbed} cells while holding constant the mean gRNA expression of \textit{perturbed} cells. We defined a parameter called ``excess background contamination'' (normed to take values in $[0,1]$) to quantify the relative distance between the unperturbed and perturbed gRNA count distributions.  We held fixed the real-data gene expressions, library sizes, covariates, and fitted perturbation probabilities in all settings.

We generated partially-synthetic data in the above manner for each of the 322 positive control pairs in the Gasperini dataset, varying excess background contamination over the interval $[0,0.4].$ We then applied GLM-EIV and the thresholding method to analyze the data. We present results on two example pairs (the pair containing gene \textit{LRIF1} and the pair containing gene \textit{NDUFA2}) in Figures \ref{fig:real_data}c-d. We observed that the estimate produced by the methods on the raw data (depicted as a horizontal black line) coincided almost exactly with the estimate produced by the methods on the partially-synthetic data generated by setting excess background contamination to zero (This result replicated across nearly all pairs; average relative difference $0.003$.) We additionally observed that as excess background contamination increased, the performance of thresholded regression degraded considerably while that of GLM-EIV remained stable.

We generalized the above analysis to the entire set of positive control pairs. First, for each pair we computed the ``relative estimate change'' (REC) as a function of excess background contamination, defined as the relative difference between the estimate at a given level of excess contamination and zero excess contamination (Figure \ref{fig:real_data}d). Next, we computed the median REC across all positive control pairs (Figure \ref{fig:real_data}e; upper and lower bands indicate the pointwise interquartile range of the REC). As excess background contamination increased, thresholded regression exhibited severe attenuation bias (as reflected by large median REC values); GLM-EIV, by contrast, remained mostly stable. Finally, letting $\hat{\beta}^m_1$ denote the estimate obtained on the raw data, we computed the CI coverage of $\hat{\beta}^m_1$ as a function of excess contamination. Under the assumption that $\hat{\beta}^m_1$ is close to the true parameter $\beta^m_1$, the CI coverage of the former is similar to that of  the latter. We computed the CI coverage of $\hat{\beta}^m_1$ by calculating each individual pair's coverage of $\hat{\beta}^m_1$ (across the Monte Carlo replicates) and then averaging this quantity across all pairs. GLM-EIV exhibited significantly higher CI coverage than thresholded regression as the data became increasingly contaminated (Figure \ref{fig:real_data}f; bands indicate $95\%$ pointwise CIs). Coverage rates were slightly above the nominal level of $95\%$ in some settings because we covered an \textit{estimate} of $\beta^m_1$ rather than $\beta^m_1$ itself, leading to mild ``overfitting.'' Nonetheless, this experiment was meaningful to assess the stability of both methods to elevated background contamination.

\section{Real data application II: assigning perturbations to cells on low-MOI data}\label{sec:real_data_2}

We sought to explore whether the gRNA mixture assignment method that we proposed in Section \ref{sec:grna_mixture_method} --- which is in effect a special case of GLM-EIV --- might be an independently useful tool for assigning gRNAs to cells on real single-cell CRISPR screen data. We applied the gRNA mixture assignment method to assign gRNAs to cells on a low multiplicity-of-infection (or MOI) single-cell CRISPR screen of immune cells \citep{Papalexi2021}. (A low-MOI dataset, in contrast to a high-MOI dataset, is one in which the experimenter has aimed to insert exactly one perturbation into each cell.) We elected to assess the performance of the gRNA mixture assignment method on low-MOI data because the ``ground truth'' gRNA-to-cell mapping is easier to ascertain in low MOI than in high MOI. The majority of cells in a low-MOI screen contains a single perturbation, while a fraction of cells contains zero or two or more perturbations. Thus, if a given gRNA constitutes a large fraction (say, $>25\%$) of the gRNA reads in a given cell, we can confidently map that gRNA to that cell. Athough not foolproof, this strategy yields a reasonable approximation to the ground truth in low MOI. (There is no analogous strategy for obtaining ground truth gRNA assignments in high MOI, as each cell in high MOI contains many gRNAs, and the number of gRNAs per cell is indeterminate and variable.)

We used our proposed gRNA mixture assignment method to obtain gRNA-to-cell assignments for each gRNA in the low-MOI dataset (after restricting our attention to the $95\%$ most highly expressed gRNAs). We included the standard technical factors as covariates, including biological replicate. We compared the mixture-model-based gRNA assignments to the ground truth assignments; the latter were obtained in the manner described above. Encouragingly, we found that these two methods produced near-identical results. For example, the mixture model determined that gRNA ``CUL3g2'' was present in 141 cells (and absent in the rest), while the ground truth method indicated that ``CUL3g2'' was present in 137 cells (Figure \ref{fig:grna_mixture_model}a). Treating the ground truth assignments as a reference, we constructed a confusion matrix to assess the classification accuracy of the mixture method assignments on CUL3g2 (Figure \ref{fig:grna_mixture_model}b). The sensitivity, specificity, and balanced accuracy of the mixture method assignments were high (1.000, 0.9998, and 0.9998, respectively).

We replicated this analysis across the entire set of gRNAs, finding that the mixture method assignments exhibited consistently high concordance with the ground truth assignments as measured by sensitivity, specificity, and balanced accuracy (although there were a few outliers; Figure \ref{fig:grna_mixture_model}c). We concluded that the mixture assignment method was a statistically principled, fast, and numerically stable strategy for the recapitulating the ground truth assignments with high fidelity. We sought to compare our gRNA mixture assignment method against the Nat.\ Biotech.\ 2020 Poisson-Gaussian mixture method. Unfortunately, as discussed elsewhere (Section \ref{sec:grna_mixture_method} and Appendix \ref{sec:Replogle_method}), we were unable to get the Nat.\ Biotech.\ 2020 method (or approximations thereof written in R) working. We note that, in contrast to the Nat.\ Biotech.\ 2020 method, the proposed method allows for the inclusion of covariates (e.g., library size and batch) and models the gRNA counts directly.

\section{Discussion}

In this work we studied the problem of estimating the effect sizes of perturbations on changes in gene expression in high-MOI single-cell CRISPR screens, focusing specifically on the challenge that the perturbation is unobserved. We showed through empirical, theoretical, and simulation analyses that the commonly-used thresholding method poses several difficulties: there exist settings (i.e., high background contamination settings) in which thresholding is not a tenable strategy, and in settings in which thresholding \textit{is} a tenable strategy (i.e., low background contamination settings), selecting a good threshold is challenging and consequential. Next, we developed GLM-EIV, a method that jointly models the gene and gRNA modalities to implicitly assign perturbation identities to cells and estimate perturbation effect sizes, thereby overcoming limitations of the thresholding method. GLM-EIV demonstrated significantly improved performance relative to the thresholding method in high background contamination settings on both synthetic and realistic semi-synthetic data.

However, GLM-EIV and the thresholding method demonstrated roughly similar performance on the two real high-MOI datasets that we examined, as the real data exhibited lower background contamination than anticipated. We believe that this is an interesting finding in itself; moreover, future datasets may demonstrate higher levels of background contamination, in which case GLM-EIV could serve as an immediately applicable analytic tool. Finally, the gRNA mixture assignment method, which under the hood exploits the estimation machinery of GLM-EIV, is a statistically principled, numerically stable, fast, and accurate strategy for obtaining gRNA-to-cell assignments on real data; these assignments can used as input to downstream methods (e.g., negative binomial regression or SCEPTRE; Figure \ref{fig:grna_mixture_model}d).

We anticipate that GLM-EIV could be applied to other types of multi-modal single-cell data, such as single-cell chromatin accessibility assays. A question of interest in such experiments is whether chromatin state (i.e., closed or open) is associated with the expression of a gene or abundance of a protein \citep{Mimitou2021}. We do not directly observe the chromatin state of a cell; instead, we observe tagged DNA fragments that serve as count-based proxies for whether a given region of chromatin is open or closed. GLM-EIV might be applied in such experiments to aid in the selection of thresholds or to analyze whole datasets. The full GLM-EIV model potentially could be applied to analyze low-MOI single-cell CRISPR screen data, but we anticipate that the relative ease of assigning gRNAs to cells in low MOI (as described in section \ref{sec:real_data_2}) may obviate the need for GLM-EIV in that setting.

The closest parallels to GLM-EIV in the statistical methodology literature are \cite{Grun2008} and \cite{Ibrahim1990}. Gr\"{u}n and Leisch derived a method for estimation and inference in a $k$-component mixture of GLMs. While we prefer to view GLM-EIV as a generalized errors-in-variables method,  the GLM-EIV model is equivalent to a two-component mixture of products of GLM densities. Ibrahim proposed a procedure for fitting GLMs in the presence of missing-at-random covariates. Our method, by contrast, involves fitting two conditionally independent GLMs in the presence of a totally latent covariate. Thus, while Ibrahim and Gr\"{u}n \& Leisch are helpful references, our estimation and inference tasks are more complex than theirs. Next, \cite{Aigner1973} and \cite{Savoca2000} proposed measurement error models that consist of unobserved \textit{binary} rather than \textit{continuous} predictors; the latter are more commonly used in measurement error models. GLM-EIV likewise consists of a latent binary predictor, but unlike Aigner and Savoca, GLM-EIV handles a much broader class of exponential family-generated data. Finally, GLM-EIV accounts for a common source of measurement error between the predictor and response, a property not shared by classical measurement error models \citep{Carroll2006}. Additional related work is relayed in Appendix \ref{sec:additional_related_work}.

GLM-EIV might be applied to areas beyond genomics, such as psychology. Some psychological constructs (e.g., presence or absence of a social media addiction) are latent and can be assessed only through an imperfect proxy (e.g., the number of times one has checked social media). Researchers might use GLM-EIV to regress an outcome variable (e.g., self-reported well-being) onto the latent construct via the imperfect proxy, potentially resolving challenges related to attenuation bias and threshold selection. Applications to psychology and other areas are a topic of future investigation.

\section*{Software, code, and results}
The gRNA-only mixture assignment functionality of GLM-EIV is implemented in our \texttt{sceptre} toolkit for single-cell CRISPR screen analysis (\url{github.com/Katsevich-Lab/sceptre}). The \texttt{sceptre} user manual (\url{timothy-barry.github.io/sceptre-book/sceptre.html}) presents a detailed guide on analyzing data using the \texttt{sceptre} software, including several sections on assigning gRNAs to cells using the mixture assignment method introduced in this work.

Results are deposited at \url{upenn.box.com/v/glmeiv-files-v1}. Github repositories containing manuscript replication code, the \texttt{glmeiv} R package, and the cloud/HPC-scale GLM-EIV pipeline are available at \url{github.com/timothy-barry/glmeiv-manuscript}, \url{github.com/timothy-barry/glmeiv}, and \url{github.com/timothy-barry/glmeiv-pipeline}, respectively. Detailed replication instructions are available in the first repository.

\section*{Acknowledgements}
We thank Eric Tchetgen Tchetgen for helpful conversations, Xuran Wang for helping to process the Xie dataset, and Songcheng Dai for helping to deploy the GLM-EIV pipeline on Azure. We additionally thank three anonymous reviewers whose comments considerably improved the manuscript. This work used the Extreme Science and Engineering Discovery Environment (XSEDE; NSF grant ACI-1548562) and the Bridges-2 system (NSF grant ACI-1928147) at the Pittsburgh Supercomputing Center. This work is funded by National Institute of Mental Health (NIMH) grant R01MH123184 and NSF grant DMS-2113072.

\bibliographystyle{biorefs}
\bibliography{glmeiv}

%\printbibliography
\clearpage
\section{Figures}
% 1. Experimental design

% 2. Emprical challenges

% 3. 
\begin{figure}[h!]
\centering
\includegraphics[width=0.85\linewidth]{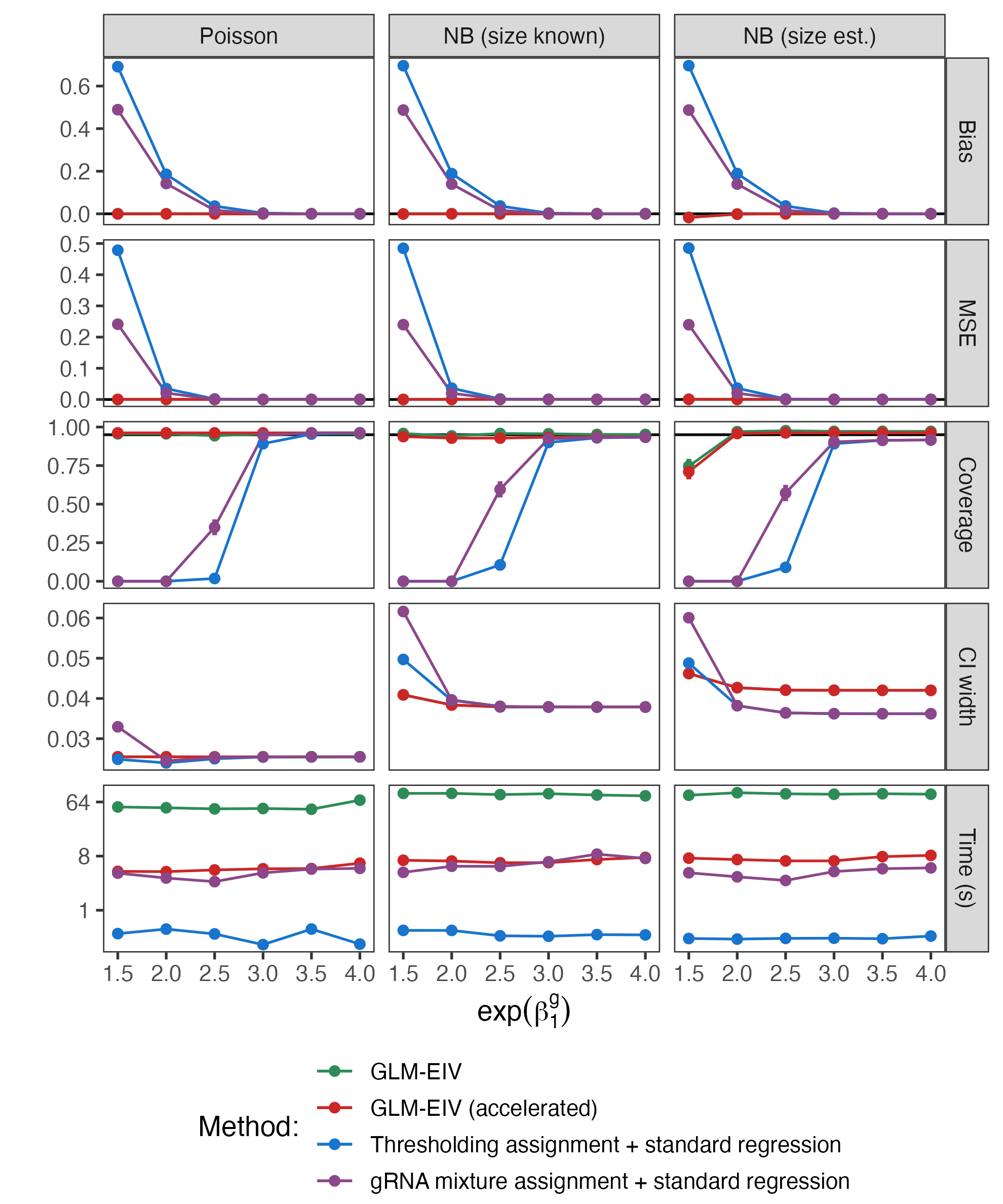}
\caption{\textbf{Simulation study}. Columns correspond to distributions (Poisson, NB with known $s$, NB with estimated $s$), and rows correspond to metrics (bias, MSE, coverage, CI width, and time). Methods are shown in different colors; GLM-EIV (green) is masked by accelerated GLM-EIV (red) in several panels. Generally, GLM-EIV (both accelerated and non-accelerated versions) outperformed the gRNA-mixture/NB-regression method, which in turn outperformed the thresholding/NB-regression method. The rejection probability (i.e., the probability of rejecting the null hypothesis $H_0: \beta_1^m = 0$ at level $\alpha = 0.05$) was strictly 1 across methods and parameter settings, likely because the effect size was fairly large.}
\label{main_text_sim}
\end{figure}

\begin{figure}[h!]
\centering
\includegraphics[width=0.9\linewidth]{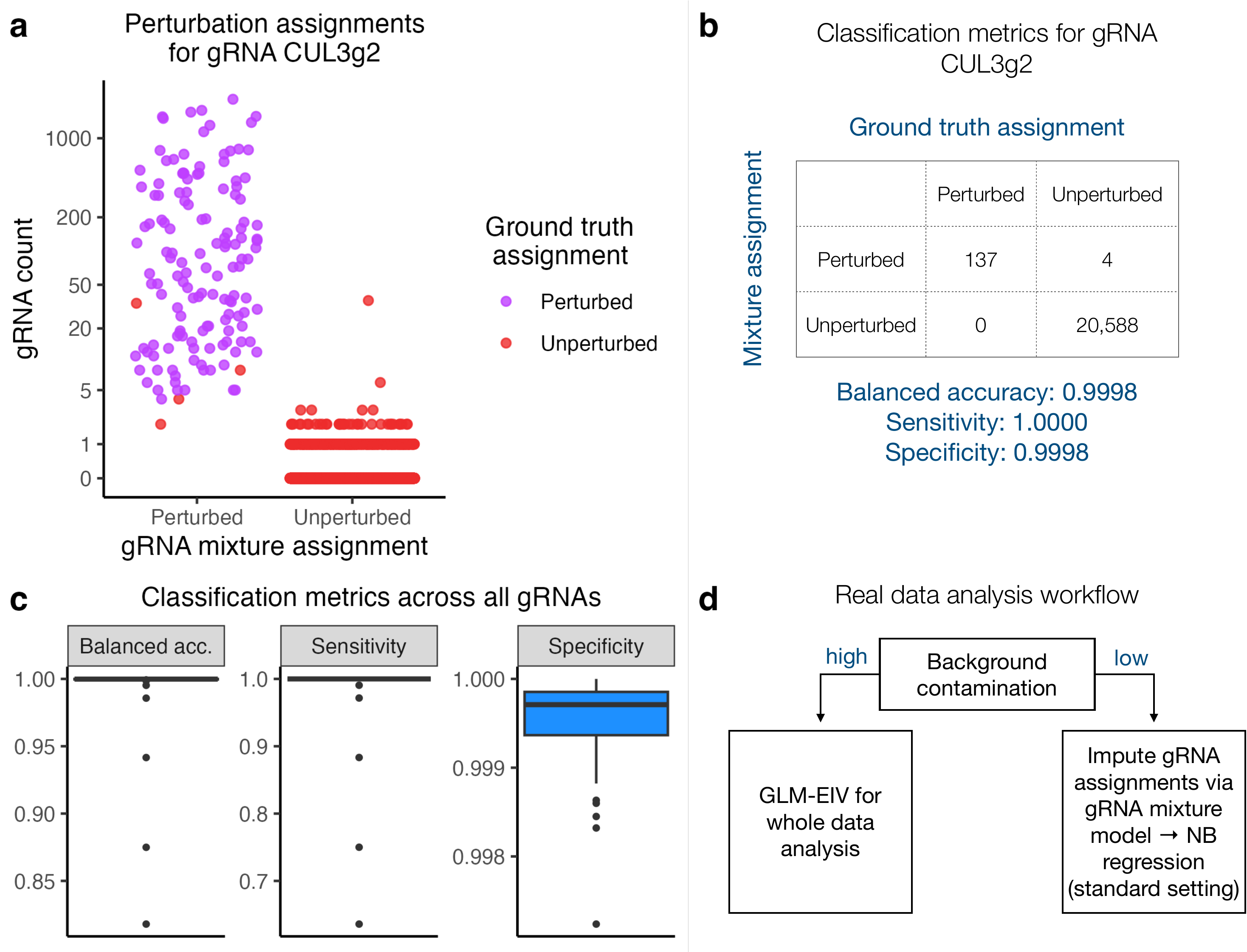}
\caption{\textbf{The gRNA-only mixture assignment functionality of GLM-EIV accurately assigns gRNAs to cells on real low-MOI data}. \textbf{a}, Each point represents a cell. The position of each cell along the vertical axis indicates the number of gRNA reads (from gRNA ``CUL3g2'') observed in that cell. Cells in the left column were classified by the gRNA mixture model as perturbed, while those in the right column were classified as unperturbed. Purple (resp., red) cells were classified by the ground truth method as perturbed (resp., unperturbed). \textbf{b}, A confusion matrix comparing the gRNA-to-cell mixture model classifications against the ground truth classifications for gRNA ``CUL3g2.'' The two sets of classifications were highly concordant, as quantified by balanced accuracy, sensitivity, and specificity metrics. \textbf{c}, The balanced accuracy (left), sensitivity (middle), and specificity (right) of the gRNA mixture assignment method across \textit{all} gRNAs. \textbf{d}, The proposed data analysis workflow. If the level of background contamination is low, then the gRNA mixture method can be used to impute perturbation identities onto cells, which can then be plugged into downstream analytic tools, such as negative binomial regression or SCEPTRE. On the other hand, if the level of background contamination is high, then the entire GLM-EIV model can be used to analyze the data.}\label{fig:grna_mixture_model}
\end{figure}

\clearpage

% \begin{appendices}
\appendix

\section{Theoretical details for thresholding estimator}\label{sec:appendix_theory}
	
We study the thresholding method from a theoretical perspective, recovering in a simplified Gaussian setting phenomena revealed in the empirical analysis. Suppose we observe gRNA expression and gene expression data $(g_1, m_1), \dots, (g_n, m_n)$ on $n$ cells from the following linear model:
\begin{equation}\label{theoretical_model}
m_i = \beta^m_0 + \beta^m_1 p_i + \epsilon_i; \quad
g_i = \beta^g_0 + \beta^g_1 p_i + \tau_i; \quad
p_i \sim \textrm{Bern}(\pi); \quad
\epsilon_i, \tau_i \sim N(0,1),
\end{equation}
where $p_i, \tau_i,$ and $\epsilon_i$ are independent. For a given threshold $c \in \mathbb{R}$, the imputed perturbation assignment $\hat{p}_i$ is $\hat{p}_i = \mathbb{I}(g_i \geq c).$ The thresholding estimator $\hat{\beta}^m_1$ is the OLS solution, i.e. $\hat{\beta}^m_1 = \left[\sum_{i=1}^n (\hat{p}_i - \overline{\hat{p}})^2\right]^{-1}\left[\sum_{i=1}^n (\hat{p}_i - \overline{\hat{p}})(m_i - \overline{m})\right].$ We derive the almost sure limit of $\hat{\beta}^m_1$: 
\begin{proposition}\label{prop:convergence}
	The almost sure limit (as $n \to \infty$) of $\hat{\beta}^m_1$ is
	\begin{equation}\label{thresh_est_intercepts}
	\hat{\beta}^m_1 \xrightarrow{a.s.} \beta^m_1 \left(\frac{ \pi( \omega - \mathbb{E}[ \hat{p}_i ])}{ \mathbb{E}[\hat{p}_i] (1 - \mathbb{E}[\hat{p}_i])}\right) \equiv \beta^m_1 \gamma(\beta^g_1, \pi, c, \beta^g_0), 
	\end{equation} where 
	$
	\mathbb{E}[\hat{p}_i] = \zeta(1-\pi) + \omega\pi$, $\omega \equiv \Phi\left(\beta_1^g + \beta_0^g -c \right)$, and $\zeta \equiv \Phi\left( \beta^g_0 - c \right).$
\end{proposition}
The function $\gamma: \mathbb{R}^4 \to \mathbb{R}$ does not depend on the gene expression parameters $\beta^m_1$ or $\beta^m_0$. The asymptotic relative bias $b: \mathbb{R}^4 \to \mathbb{R}$ of $\hat{\beta}^m_1$ is given by
$$b(\beta^g_1, \pi, c, \beta^g_0)  \equiv \frac{1}{\beta^m_1} \left(\beta^m_1 - \lim_\textrm{a.s.} \hat{\beta}^m_1 \right) = 1 - \gamma(\beta^g_1, \pi, c, \beta^g_0).$$

Having derived an exact expression for the asymptotic relative bias of $\hat{\beta}^m_1$, we can prove several results about this quantity. We fix $\pi$ to $1/2$ for simplicity. (In reality, $\pi$ is smaller, but the relevant statistical phenomena emerge for $\pi = 1/2$.) We start with informal proposition statements; we follow up with formal proposition statements below. First, the thresholding estimator strictly underestimates (in absolute value) the true value of $\beta^m_1$ over all choices of the threshold $c$ and over all values of the regression coefficients $(\beta^m_0, \beta^m_1)$ and $(\beta^g_0, \beta^g_1)$. This phenomenon, called attenuation bias, is a common attribute of estimators that ignore measurement error in errors-in-variables models \citep{Stefanski2000a}. Second, the magnitude of the bias decreases monotonically in $\beta^g_1$, comporting with the intuition that the problem becomes easier as the gRNA mixture distribution becomes increasingly well-separated. Third, the Bayes-optimal decision boundary $c_\textrm{bayes} \in \mathbb{R}$ (i.e., the most accurate decision boundary for classifying cells) is a critical value of the bias function. Finally, and most subtly, there is no universally applicable rule for selecting a threshold that yields minimal bias: when $\beta^g_1$ is small, setting the threshold to an arbitrarily large number yields smaller bias than setting the threshold to the Bayes decision boundary; when $\beta^g_1$ is large, the reverse is true. 

\begin{figure}[h!]
\centering
\includegraphics[width=1\linewidth]{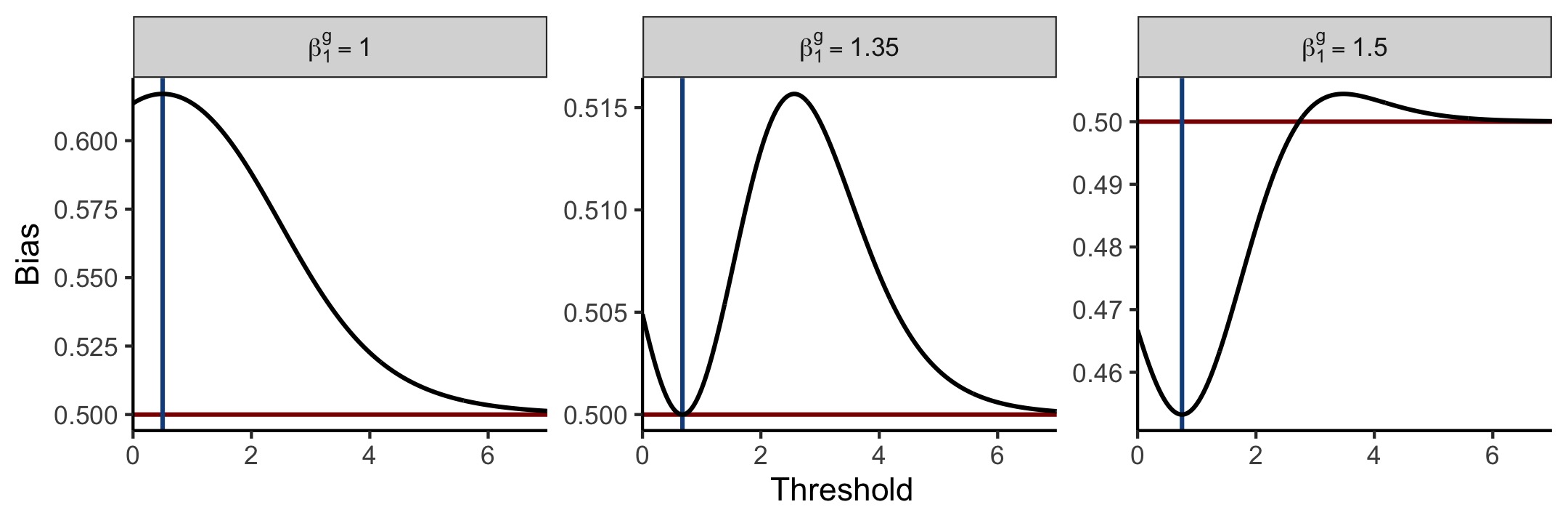}
\caption{\textbf{Bias as a function of threshold}. This figure visually depicts Propositions \ref{prop:att_bias}-\ref{prop:comparison}, which were stated informally above. Asymptotic relative bias is plotted on the vertical axis, and the threshold is plotted on the horizontal axis. Panels correspond to different values of $\beta^g_1$. Vertical blue lines indicate the Bayes-optimal decision boundary. Observe that (a) bias is strictly nonzero (proposition \ref{prop:att_bias}); (b) bias decreases monotonically in $\beta^g_1$ (Proposition \ref{prop:monotonic}); (c) the Bayes-optimal decision boundary is a critical value of the bias function (Proposition \ref{prop:bayes_opt}), in some cases a maximum and in other cases a minimum; (d) as the threshold tends to infinity, the bias converges to $1/2$ (Proposition \ref{prop:c_limit}); and (e) when $\beta^g_1 < 1.35$, an arbitrarily large number yields a smaller bias; by contrast, when $\beta^g_1 > 1.35$, the Bayes-optimal decision boundary yields a smaller bias (Proposition \ref{prop:comparison}). Together, these results illustrate that selecting a good threshold is deceptively challenging.
%Across all panels, $\beta^g_0 = 0$ and $\pi = 1/2$. When $\beta^g_1$ is small, setting the threshold to an arbitrarily large number yields a smaller bias than setting the threshold to the Bayes-optimal decision boundary; when $\beta^g_1$ is large, the opposite is true.
}
\label{fig:bias_plot}
\end{figure}

We state five propositions labeled \ref{prop:att_bias} -- \ref{prop:comparison} corresponding to the informal claims above; these propositions are depicted visually in Figure \ref{fig:bias_plot}.

\begin{proposition}\label{prop:att_bias} Fix $\pi = 1/2$. For all $(\beta^g_1, c, \beta^g_0) \in \mathbb{R}^3$, the asymptotic relative bias is positive, i.e. 
$$b(\beta^g_1, 1/2, c, \beta^g_0) > 0.$$
\end{proposition}

\begin{figure}
\centering
\includegraphics[width=0.7\linewidth]{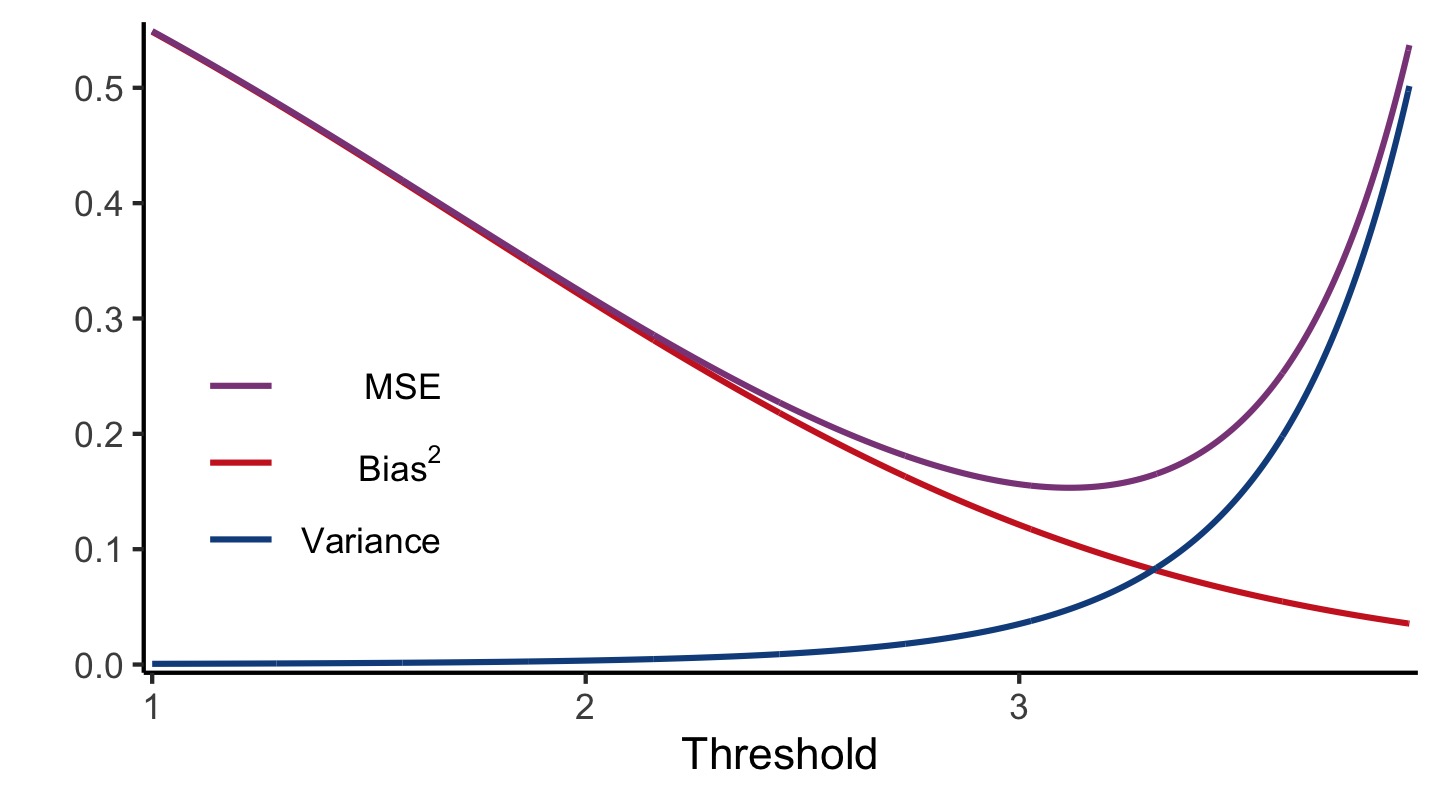}
\caption{\textbf{Thresholding method bias-variance decomposition.} Bias decreases and variance increases as the threshold tends to infinity. $\beta^g_1 = 1, \beta^m_1 = 1,$ and $\pi = 0.1$ in this plot.}
\label{thresholding_theoretical}
\end{figure}

\begin{proposition}\label{prop:monotonic} Fix $\pi = 1/2$. The asymptotic relative bias $b$ decreases monotonically in $\beta_1^g$, i.e.
$$\frac{\partial b}{\partial(\beta^g_1)}\left(\beta^g_1, 1/2, c, \beta^g_0\right) \leq 0.$$
\end{proposition}
Let $c_\textrm{bayes}$ denote the Bayes-optimal decision boundary for classifying cells as perturbed or unperturbed, i.e. $c_\textrm{bayes} = (1/2)(\beta^g_0 + \beta^g_1)$ for $\pi = 1/2$. We have that $c_\textrm{bayes}$ is a critical value of the bias function:
\begin{proposition}\label{prop:bayes_opt}
For $\pi = 1/2$ and given $(\beta^g_1, \beta^g_0) \in \mathbb{R}^2$, the Bayes-optimal decision boundary $c_\textrm{bayes}$ is a critical value of the bias function $b$, i.e.
$$ \frac{\partial b}{\partial c}\left(\beta^g_1, 1/2, c_\textrm{bayes}, \beta^g_0\right) = 0.$$
\end{proposition}
Furthermore, as the threshold tends to infinity, the asymptotic relative bias $b$ tends to $\pi$:
\begin{proposition}\label{prop:c_limit}
Assume without loss of generality that $\beta^g_1 > 0$. As the threshold $c$ tends to infinity, the asymptotic relative bias $b$ tends to $\pi,$ i.e.
$$ \lim_{ c \to \infty } b(\beta^g_1, \pi, c, \beta^g_0) = \pi.$$
\end{proposition}
As a corollary, when $\pi = 1/2$, asymptotic relative bias tends to $1/2$ as $c$ tends to infinity. Finally, we compare two threshold selection strategies head-to-head: setting the threshold to an arbitrarily large number, and setting the threshold to the Bayes-optimal decision boundary:
\begin{proposition}\label{prop:comparison} Assume without loss of generality that $\beta^g_1 > 0$. For $\beta^g_1 \in [0, 2\Phi^{-1}(3/4))$, we have that $$b(\beta^g_1, 1/2, c_\textrm{bayes}, \beta^g_0) > b(\beta^g_1, 1/2, \infty, \beta^g_0).$$ For $\beta^g_1 = 2\Phi^{-1}(3/4)$, we have that $$b(\beta^g_1, 1/2, c_\textrm{bayes}, \beta^g_0) = b(\beta^g_1, 1/2, \infty, \beta^g_0).$$ Finally, for $\beta^g_1 \in (2\Phi^{-1}(3/4), \infty)$, we have that $$b(\beta^g_1, 1/2, c_\textrm{bayes}, \beta^g_0) < b(\beta^g_1, 1/2, \infty, \beta^g_0).$$
\end{proposition}
In other words, setting the threshold to a large number yields a smaller bias when $\beta^g_1$ is small (i.e., $\beta^g_1 < 2\Phi^{-1}(3/4) \approx 1.35$; Figure \ref{thresholding_theoretical}a, left); setting the threshold to the Bayes-optimal decision boundary yields a smaller bias when $\beta^g_1$ is large (i.e., $\beta^g_1 > 2\Phi^{-1}(3/4)$; Figure \ref{thresholding_theoretical}a, right); and the two approaches coincide when $\beta^g_1$ is intermediate (i.e., $\beta^g_1 = 2\Phi^{-1}(3/4)$; Figure \ref{thresholding_theoretical}a, middle).
		
Next, we study the variance of the thresholding estimator, considering a slightly simpler model for this purpose. Suppose the intercepts in (\ref{theoretical_model}) are fixed at $0$ (i.e., $\beta^m_0$ = $\beta^g_0$ = 0). For notational simplicity we write $\beta_m = \beta^m_1$ and $\beta_g = \beta^g_1.$ The thresholding estimator $\hat{\beta}_m$ is the no-intercept OLS solution $\hat{\beta}_m = \left[\sum_{i=1}^n \hat{p}_i^2 \right]^{-1}\left[\sum_{i=1}^n \hat{p}_i m_i \right].$ The following proposition derives the scaled, asymptotic distribution of $\hat{\beta}_m:$
\begin{proposition}\label{prop:bv_decomp}
The limiting distribution of $\hat{\beta}_m$ is
$$\sqrt{n}(\hat{\beta}_m - l) \xrightarrow{d} N\left(0, \frac{ \beta_m \omega\pi(\beta_m - 2l) + \mathbb{E}[\hat{p}_i](1 + l^2) }{\left(\mathbb{E}[\hat{p}_i]\right)^2} \right),$$ where $$l \equiv \beta_m \omega \pi/[\zeta(1-\pi) + \omega \pi]; \quad
\mathbb{E}[\hat{p}_i] = \pi \omega + (1-\pi) \zeta; \quad
\omega \equiv \Phi(\beta_g - c); \quad \zeta \equiv \Phi(-c).$$
\end{proposition}
This proposition yields an asymptotically exact bias-variance decomposition for $\hat{\beta}_m$: as the threshold tends to infinity, the bias decreases and the variance increases. Figure \ref{thresholding_theoretical} plots the bias-variance decomposition as a function of the threshold.
		
\subsection{Organization} The following subsections prove all propositions. Section \ref{sec:notation} introduces some notation. Section \ref{sec:convergence} establishes almost sure convergence of the thresholding estimator in the model (\ref{theoretical_model}), proving Proposition \ref{prop:convergence}. Section \ref{sec:simplication} simplifies the expression for the attenuation function $\gamma$, and section \ref{sec:derivatives}  computes derivatives of $\gamma$ to be used throughout the proofs. Section \ref{sec:c_limit} establishes the limit in $c$ of $\gamma$, proving Proposition \ref{prop:c_limit}. Section \ref{sec:bayes_opt} establishes that the Bayes-optimal decision boundary is a critical value of $\gamma$, proving Proposition \ref{prop:bayes_opt}, and section \ref{sec:comparison} compares the competing threshold selection strategies head-to-head, proving Proposition \ref{prop:comparison}. Section \ref{sec:monotone} demonstrates that $\gamma$ is monotone in $\beta^g_1$, proving Proposition \ref{prop:monotonic}, and Section \ref{sec:att_bias} establishes attenuation bias of the thresholding estimator, proving Proposition \ref{prop:att_bias}. Finally, Section \ref{sec:bv_decomp} derives the bias-variance decomposition of the thresholding estimator in the no-intercept version of \ref{theoretical_model}, proving Proposition \ref{prop:bv_decomp}.
		
		% Labels of propositions and theorems
		% propositions:
		% prop:convergence
		% prop:att_bias
		% prop:bayes_opt
		% prop:c_limit_half
		% prop:comparison
		% prop:monotonic
		% prop:c_limit
		% prop:bv_decomp
		
		% sections
		% sec:notation
		% sec:convergence
		% sec:simplication
		% sec:derivatives
		% sec:c_limit
		% sec:bayes_opt
		% sec:comparison
		% sec:beta_lim
		% sec:monotone
		% sec:att_bias
		% sec:bv_decomp
		\subsection{Notation}\label{sec:notation}
		All notation introduced in this subsection (i.e., \ref{sec:notation}) pertains to the Gaussian model with intercepts (\ref{theoretical_model}). Recall that the attenuation function $\gamma: \mathbb{R}^4 \to \mathbb{R}$ is defined by
		$$ \gamma(\beta^g_1, c, \pi, \beta^g_0) = \frac{\pi(\omega - \mathbb{E}[\hat{p}_i])}{ \mathbb{E}[\hat{p}_i](1 -\mathbb{E}[\hat{p}_i])},$$ where $$\mathbb{E}[\hat{p}_i] = \zeta(1-\pi) + \omega\pi; \quad \omega = \Phi\left(\beta_1^g + \beta_0^g - c \right); \quad \zeta = \Phi\left( \beta^g_0 - c \right).$$ Additionally, recall that the asymptotic relative bias function $b: \mathbb{R}^4 \to \mathbb{R}$ is
		$b(\beta^g_1, c, \pi, \beta^g_0) = 1 - \gamma(\beta^g_1, c, \pi, \beta^g_0).$ Next, we define the functions $g$ and $h: \mathbb{R}^4 \to \mathbb{R}$ by
		\begin{equation}\label{def_g}
		g(\beta^g_1, c, \pi, \beta^g_0) = (1-\pi)\left( \Phi(\beta_0^g + \beta_1^g - c)\right) - (1-\pi)\left(\Phi(\beta_0^g - c)\right)\end{equation}
		and
		\begin{multline}\label{def_h}
		h(\beta^g_1, c, \pi, \beta^g_0) = \left[(1-\pi)\left( \Phi(\beta_0^g - c)\right) + \pi\left(\Phi(\beta^g_0 + \beta^g_1 - c) \right) \right] \times \\ \left[(1-\pi)\left( \Phi(c - \beta^g_0) \right) + \pi\left(\Phi(c - \beta_0^g - \beta_1^g) \right) \right].
		\end{multline}
		We use $f:\mathbb{R} \to \mathbb{R}$ to denote the $N(0,1)$ density, and we denote the right-tail probability probability of $f$ by $\bar{\Phi}$, i.e.,
		$$\bar{\Phi}(x) = \int_{x}^{\infty} f = \Phi(-x).$$
		
		The parameter $\beta^g_0$ is a given, fixed constant throughout the proofs. Therefore, to minimize notation, we typically use $\gamma(\beta^g_1, c, \pi)$ (resp., $b(\beta^g_1, c, \pi),$ $g(\beta^g_1, c, \pi),$ $h(\beta^g_1, c, \pi)$) to refer to the function $\gamma$ (resp., $b, g, h$) evaluated at $(\beta^g_1, c, \pi, \beta^g_0)$. Finally, for a given function $r: \mathbb{R}^{p} \to \mathbb{R}$, point $x \in \mathbb{R}^p$, and index $i \in \{1, \dots, p\}$, we use the symbol $D_i r(x)$ to refer to the derivative of the $i$th argument of $r$ evaluated at $x$ (\textit{sensu} \cite{fitzpatrick2009}). For example, $D_1 \gamma(\beta^g_1, c, 1/2)$ is the derivative of the first argument of $\gamma$ (the argument corresponding to $\beta^g_1$) evaluated at $(\beta^g_1, c, 1/2)$. Likewise,  $D_2g(\beta^g_1, c, \pi)$ is the derivative of the second argument of $g$ (the argument corresponding to $c$) evaluated at $(\beta^g_1, c, \pi).$
		
		\subsection{Almost sure limit of $\hat{\beta}^m_1$}\label{sec:convergence}
		
		We derive the limit in probability of $\hat{\beta}^m_1$ for the Gaussian model with intercepts (\ref{theoretical_model}). Dividing by $n$ in (\ref{thresh_est_intercepts}), we can express $\hat{\beta}^m_1$ as
		$$ \hat{\beta}^m_1 = \frac{ \frac{1}{n} \sum_{i=1}^n ( \hat{p}_i - \overline{\hat{p}_i})(m_i - \overline{m})}{ \frac{1}{n} \sum_{i=1}^n (\hat{p}_i - \overline{\hat{p}})}.$$ By weak LLN,
		$\hat{\beta}^m_1 \xrightarrow{P} \textrm{Cov}(\hat{p}_i, m_i)/\mathbb{V}\left(\hat{p}_i\right).$ To compute this quantity, we first compute several simpler quantities:
		\begin{itemize}
			\item[1.] Expectation of $m_i$: $\mathbb{E}[m_i] = \beta^m_0 + \beta^m_1\pi.$
			\item[2.] Expectation of $\hat{p}_i$: \begin{multline*}
			\mathbb{E}[\hat{p}_i] = \mathbb{P}\left[\hat{p}_i = 1\right] = \mathbb{P}\left[\beta^g_0 + \beta^g_1 p_i + \tau_i \geq c \right] = \\ \textrm{(By LOTP) } \mathbb{P}\left[ \beta^g_0 + \tau_i \geq c \right]\mathbb{P}\left[p_i = 0\right] + \mathbb{P}\left[ \beta^g_0 + \beta^g_1 + \tau_i \geq c \right] \mathbb{P}[p_i = 1] \\ = \mathbb{P}\left[ \tau_i \geq c - \beta^g_0\right](1- \pi) + \mathbb{P}\left[ \tau_i \geq c - \beta^g_1 - \beta^g_0 \right](\pi) \\ =  \left(\bar{\Phi}(c - \beta^g_0) \right) (1 - \pi) + \left( \bar{\Phi}(c - \beta^g_1 - \beta^g_0) \right)(\pi) = \\  \Phi(\beta^g_0 - c) (1-\pi) + \Phi(\beta^g_1 + \beta^g_0 - c) \pi = \zeta(1-\pi) + \omega \pi.
			\end{multline*}
			\item[3.] Expectation of $\hat{p}_i p_i$: 
			$\mathbb{E}\left[ \hat{p}_i p_i \right] = \mathbb{E}\left[\hat{p}_i | p_i = 1 \right] \mathbb{P}\left[ p_i =1 \right] = \mathbb{P}\left[ \beta^g_0 + \beta^g_1 + \tau_i \geq c \right] \pi = \omega \pi.$
			\item[4.] Expectation of $\hat{p}_i m_i$:
			\begin{multline*}
			\mathbb{E}\left[\hat{p}_i m_i\right] = \mathbb{E}[\hat{p}_i (\beta^m_0 + \beta^m_1 p_i + \epsilon_i)] = \beta^m_0 \mathbb{E}\left[\hat{p}_i\right] + \beta^m_1 \mathbb{E}\left[\hat{p}_i p_i\right] + \mathbb{E}[\hat{p}_i \epsilon_i] \\ = \beta^m_0 \mathbb{E}[\hat{p}_i] + \beta^m_1 \omega \pi + \mathbb{E}[\hat{p}_i] \mathbb{E}[\epsilon_i] = \beta^m_0 \mathbb{E}[\hat{p}_i] + \beta^m_1 \omega \pi.
			\end{multline*}
			\item[5.] Variance of $\hat{p}_i$: Because $\hat{p}_i$ is binary, we have that $\mathbb{V}[\hat{p}_i] = \mathbb{E}[\hat{p}_i]\left(1 - \mathbb{E}[\hat{p}_i]\right) .$
			\item[6.] Covariance of $\hat{p}_i, m_i$:
			\begin{multline*}
			\textrm{Cov}\left(\hat{p}_i, m_i\right) = \mathbb{E}\left[\hat{p}_i m_i\right] - \mathbb{E}[\hat{p}_i] \mathbb{E}[m_i] = \beta^m_0 \mathbb{E}[\hat{p}_i] + \beta^m_1 \omega \pi - \mathbb{E}[\hat{p}_i]( \beta^m_0 + \beta^m_1 \pi)\\ = \beta^m_1 \omega \pi - \mathbb{E}[\hat{p}_i] \beta_1^m \pi = \beta^m_1 \pi \left( \omega - \mathbb{E}[\hat{p}_i]\right).
			\end{multline*}
			Combining these expressions, we have that
			$$ \hat{\beta}^m_1 \xrightarrow{P} \frac{\beta^m_1 \pi (\omega - \mathbb{E}[\hat{p}_i])}{\mathbb{E}[\hat{p}_i](1 - \mathbb{E}[\hat{p}_i])} = \beta^m_1 \gamma(\beta^g_1, c, \pi).$$
		\end{itemize}
		
		\subsection{Re-expressing $\gamma$ in a simpler form}\label{sec:simplication}
		We rewrite the attenuation fraction $\gamma$ in a way that makes it more amenable to theoretical analysis. We leverage the fact that $f$ integrates to unity and is even. We have that
		\begin{equation}\label{thm:gamma_expression_1} \mathbb{E}\left[\hat{p}_i\right] = (1 - \pi) \bar{\Phi}(c - \beta_0^g) + \pi \bar{\Phi}(c - \beta^g_0 - \beta^g_1) = (1 - \pi) \Phi(\beta_0^g - c) + \pi\Phi(\beta^g_0 + \beta^g_1 - c), \end{equation}
		and so \begin{multline}\label{thm:gamma_expression_2} 1 - \mathbb{E}\left[\hat{p}_i\right] = (1 - \pi) + \pi - \mathbb{E}[\hat{p}_i]  = (1-\pi) \left(1 - \bar{\Phi}(c - \beta_0^g)\right)  + \pi \left(1 - \bar{\Phi}(c - \beta^g_0 - \beta^g_1) \right) \\ = (1 - \pi)\Phi(c - \beta^g_0) + \pi \Phi(c - \beta_0^g - \beta_1^g).
		\end{multline}
		Next,
		\begin{equation}\label{thm:gamma_expression_3}
		\omega = \Phi(\beta^g_1 + \beta^g_0 - c),\end{equation} and so
		\begin{multline}\label{thm:gamma_expression_4}
		\omega - \mathbb{E}[\hat{p}_i] = \Phi(\beta^g_1 + \beta^g_0 - c) - (1-\pi)\Phi(\beta^g_0 - c) - \pi \Phi(\beta^g_0 + \beta^g_1 - c)  \\ (1-\pi)\Phi(\beta^g_1 + \beta^g_0 - c)  - (1-\pi)\Phi(\beta^g_0 - c).
		\end{multline}
		Combining (\ref{thm:gamma_expression_1}, \ref{thm:gamma_expression_2}, \ref{thm:gamma_expression_3}, \ref{thm:gamma_expression_4}), we find that
		\begin{multline}\label{gamma_alternative}
		\gamma(\beta^g_1, c, \pi) = \frac{\pi(\omega - \mathbb{E}[\hat{p}_i])}{\mathbb{E}[\hat{p}_i](1 - \mathbb{E}[\hat{p}_i])} \\ = \frac{\pi \left[(1 - \pi) \Phi(\beta_0^g + \beta_1^g - c) - (1 - \pi) \Phi(\beta_0^g - c)\right]}{\left[(1-\pi)\Phi(\beta_0^g - c) + \pi \Phi(\beta^g_0 + \beta^g_1 - c) \right] \left[(1 - \pi) \Phi(c - \beta^g_0) + \pi\Phi(c - \beta_0^g - \beta_1^g) \right]}.
		\end{multline}
		As a corollary, when $\pi = 1/2$,
		\begin{equation}\label{gamma_alternative_pi_half}
		\gamma(\beta^g_1, c, 1/2)  \\ = \frac{\Phi(\beta_0^g + \beta_1^g - c) - \Phi(\beta_0^g - c) }{\left[\Phi(\beta_0^g - c) +\Phi(\beta^g_0 + \beta^g_1 - c)\right] \left[\Phi(c - \beta^g_0) + \Phi(c - \beta_0^g - \beta_1^g) \right]}.
		\end{equation}
		Recalling the definitions of $g$ (\ref{def_g}) and $h$ (\ref{def_h}), we can write $\gamma$ as
		$$ \gamma(\beta^g_1, c, \pi) = \frac{\pi g(\beta^g_1, c, \pi)}{h(\beta^g_1, c,\pi)}.$$
		The special case (\ref{gamma_alternative_pi_half}) is identical to
		\begin{equation}\label{gamma_alt2_pi_half}
		\gamma(\beta^g_1, c, 1/2) = \frac{(4)(1/2)g(\beta^g_1, c, 1/2)}{4 h(\beta^g_1, c, 1/2)} = \frac{2 g(\beta^g_1, c, 1/2)}{4h(\beta^g_1, c, 1/2)},
		\end{equation}
		i.e., the numerator and denominator of  (\ref{gamma_alt2_pi_half}) coincide with those of (\ref{gamma_alternative_pi_half}). We sometimes will use the notation $2\cdot g$ and $4\cdot h$ to refer to the numerator and denominator of (\ref{gamma_alternative_pi_half}), respectively.
		
		\subsection{Derivatives of  $g$ and $h$ in $c$}\label{sec:derivatives}
		We compute the derivatives of $g$ and $h$ in $c$, which we will need to prove subsequent results. First, by the FTC (fundamental theorem of calculus) and the evenness of $f$, we have that
		\begin{multline}\label{dg_dc}
		D_2 g(\beta^g_1, c, \pi) = -(1-\pi)f( \beta^g_0 + \beta^g_1 - c ) + (1-\pi) f(\beta^g_0 - c) \\ = (1-\pi) f(c - \beta^g_0) - (1-\pi)f(c - \beta^g_0 - \beta^g_1).
		\end{multline}
		Second, we have that
		\begin{multline}\label{dh_dc}
		D_2 h(\beta^g_1, c, \pi) = -[(1-\pi)f(\beta^g_0 - c) + \pi f( \beta^g_0 + \beta^g_1 - c )]\left[(1-\pi)\Phi(c - \beta^g_0) + \pi \Phi(c - \beta_0^g - \beta_1^g)  \right] \\ + [(1-\pi) f(c - \beta^g_0) +  \pi f(c - \beta^g_0 - \beta^g_1)] \left[(1-\pi) \Phi(\beta_0^g - c) + \pi \Phi(\beta^g_0 + \beta^g_1 - c) \right] \\ = \left[ (1-\pi) f(c - \beta^g_0) +  \pi f(c - \beta^g_0 - \beta^g_1) \right] \times \\ \bigg[ (1-\pi) \Phi(\beta_0^g - c) + \pi\Phi(\beta^g_0 + \beta^g_1 - c) - (1-\pi) \Phi(c - \beta^g_0) - \pi \Phi(c - \beta_0^g - \beta_1^g) \bigg].
		\end{multline}
		
		\subsection{Limit of $\gamma$ in $c$}\label{sec:c_limit}
		
		Assume (without loss of generality) that $\beta^g_1 > 0$. We compute $\lim_{c \to \infty} \gamma(\beta^g_1, c, \pi)$. Observe that $$\lim_{c \to \infty} g(\beta^g_1, c, \pi) = \lim_{c \to \infty} h(\beta^g_1, c, \pi)  = 0.$$ Therefore, we can apply L'H\^{o}pital's rule. We have by (\ref{dg_dc}) and (\ref{dh_dc}) that \begin{multline}\label{c_limit_product}
		\lim_{c \to \infty} \gamma(\beta^g_1, c, \pi) = \lim_{c \to \infty} \frac{\pi D_2 g(\beta^g_1, c, \pi)}{D_2h(\beta^g_1, c, \pi)} \\ = \lim_{c \to \infty} \bigg\{ \frac{(1-\pi) f(c - \beta^g_0) + \pi f(c - \beta^g_0 - \beta^g_1)}{\pi (1-\pi) f(c - \beta^g_0) - \pi (1-\pi)f(c - \beta^g_0 - \beta^g_1)} \times \\ \bigg[ (1-\pi) \Phi(\beta_0^g - c) + \pi \Phi(\beta^g_0 + \beta^g_1 - c) - (1-\pi) \Phi(c - \beta^g_0) - \pi \Phi(c - \beta_0^g - \beta_1^g) \bigg] \bigg\}^{-1}.
		\end{multline}
		We evaluate the two terms in the product (\ref{c_limit_product}) separately. Dividing by $f(c - \beta^g_0 - \beta^g_1) > 0$, we see that
		\begin{equation}\label{c_limit_product_2}
		\frac{(1-\pi) f(c - \beta^g_0) + \pi f(c - \beta^g_0 - \beta^g_1)}{\pi (1-\pi) f(c - \beta^g_0) - \pi (1-\pi)f(c - \beta^g_0 - \beta^g_1)} = \frac{\frac{(1-\pi) f(c - \beta^g_0)}{ f(c - \beta^g_0 - \beta^g_1)} + \pi}{\frac{ \pi(1-\pi) f(c - \beta^g_0)}{ f(c - \beta^g_0 - \beta^g_1)} - \pi(1-\pi)}.
		\end{equation}
		To evaluate the limit of (\ref{c_limit_product_2}), we first evaluate the limit of
		\begin{multline}\label{c_limit_product_3}
		\frac{f(c - \beta^g_0)}{f(c - \beta^g_0 - \beta^g_1)} = \frac{\exp{[-(1/2)(c - \beta_0^g)^2]}}{\exp{[-(1/2)( c - \beta^g_0 - \beta^g_1)^2]}} \\ = \frac{\exp[ -(1/2)(c^2 - 2 c \beta^g_0 + (\beta^g_0)^2)]}{\exp\left[-(1/2)( c^2 - 2c \beta^g_0 - 2 c \beta^g_1 + (\beta^g_0)^2 + 2( \beta^g_0 \beta^g_1) + (\beta^g_1)^2)\right]} \\ = \exp\big[-c^2/2 + c \beta^g_0 - (\beta^g_0)^2/2 \\ + c^2/2 - c \beta^g_0 - c \beta^g_1 + (\beta^g_0)^2/2 + \beta^g_0 \beta^g_1 + (\beta^g_1)^2/2 \big] \\ = \exp[ -c \beta^g_1 + \beta^g_0 \beta^g_1 + (\beta^g_1)^2/2] = \exp[ \beta^g_0 \beta^g_1 + (\beta^g_1)^2/2]\exp[ -c \beta^g_1]. 
		\end{multline}
		Taking the limit in (\ref{c_limit_product_3}), we obtain
		$$
		\lim_{c \to \infty} \frac{f(c - \beta^g_0)}{f(c - \beta^g_0 - \beta^g_1)} = \exp[ \beta^g_0 \beta^g_1 + (\beta^g_1)^2/2] \lim_{c \to \infty} \exp[ -c \beta^g_1] = 0
		$$ for $\beta^g_1 > 0$. We now can evaluate the limit of (\ref{c_limit_product_2}):
		$$ \lim_{c \to \infty} \frac{(1-\pi) f(c - \beta^g_0) + \pi f(c - \beta^g_0 - \beta^g_1)}{\pi (1-\pi) f(c - \beta^g_0) - \pi (1-\pi)f(c - \beta^g_0 - \beta^g_1)} = \frac{-\pi}{\pi(1-\pi)} = -\frac{1}{1 -\pi}.$$ Next, we compute the limit of the other term in the product (\ref{c_limit_product}):
		\begin{multline}\label{c_limit_product_4}
		\lim_{c \to \infty} \bigg[ (1-\pi)\Phi(\beta_0^g - c) + \pi \Phi(\beta^g_0 + \beta^g_1 - c) \\ - (1-\pi)\Phi(c - \beta^g_0) - \pi \Phi(c - \beta_0^g - \beta_1^g) \bigg] = -(1-\pi) - \pi = -1.
		\end{multline}
		Combining (\ref{c_limit_product_2}) and (\ref{c_limit_product_4}), the limit (\ref{c_limit_product}) evaluates to
		$$ \lim_{c \to \infty} \gamma(\beta^g_1, c, \pi) = \left(  \frac{ 1 }{ 1 - \pi }\right)^{-1} = 1 - \pi.$$ It follows that the limit in $c$ of the asymptotic relative bias $b$ is
		$$\lim_{c \to \infty} b(\beta^g_1, c, \pi) = 1 - \lim_{c \to \infty} \gamma(\beta^g_1, c, \pi) = \pi.$$
		A corollary is that
		$\lim_{c \to \infty} b(\beta^g_1, c, 1/2) = 1/2.$
		
		\subsection{Bayes-optimal decision boundary as a critical value of $\gamma$}\label{sec:bayes_opt}
		Let $c_\textrm{bayes} = \beta^g_0 + (1/2)\beta^g_1.$ We show that $c = c_\textrm{bayes}$ is a critical value of $\gamma$ for $\pi = 1/2$ and given $\beta^g_1$, i.e, $D_2 \gamma (\beta^g_1, c_\textrm{bayes}, 1/2) = 0.$ Differentiating (\ref{gamma_alt2_pi_half}), the quotient rule implies that
		\begin{equation}\label{quotient_rule}
		D_2\gamma(\beta^g_1, c, 1/2) \\ = \frac{D_2[2g(\beta^g_1, c, 1/2)] 4h(\beta^g_1, c, 1/2) - 2g(\beta^g_1, c, 1/2) D_2[4h(\beta^g_1, c, 1/2)]}{[4h(\beta^g_1, c, \pi)]^2}.
		\end{equation}
		We have by (\ref{dg_dc}) that
		\begin{equation}\label{dg_dc_bayes}
		D_2[2g(\beta^g_1, c_\textrm{bayes}, 1/2)] = f( \beta^g_1/2) - f( -\beta^g_1/2) = f(\beta^g_1/2) - f(\beta^g_1/2) = 0.
		\end{equation}
		Similarly, we have by (\ref{dh_dc}) that
		\begin{equation}\label{dh_dc_bayes}
		D_2[4 h(\beta^g_1, c_\textrm{bayes}, \pi)] = [f( \beta^g_1/2) + f( -\beta^g_1/2)] \left[  \Phi(-\beta^g_1/2) + \Phi(\beta^g_1/2) -  \Phi(\beta^g_1/2) - \Phi(-\beta^g_1/2) \right] = 0.
		\end{equation}
		Plugging in (\ref{dh_dc_bayes}) and (\ref{dg_dc_bayes}) to (\ref{quotient_rule}), we find that 
		$D_2[\gamma(\beta^g_1, c_\textrm{bayes}, 1/2)] = 0.$ Finally, because
		$$b(\beta^g_1, c, 1/2) = 1 - \gamma(\beta^g_1, c, 1/2),$$ it follows that
		$$D_2[b(\beta^g_1, c_\textrm{bayes}, 1/2)] = -D_2[\gamma(\beta^g_1, c_\textrm{bayes}, 1/2)] = 0.$$
		
		\subsection{Comparing Bayes-optimal decision boundary and large threshold}\label{sec:comparison}
		
		We compare the bias produced by setting the threshold to a large number to the bias produced by setting the threshold to the Bayes-optimal decision boundary. Let $r: \mathbb{R}^{\geq 0} \to \mathbb{R}$ be the value of attenuation function evaluated at the Bayes-optimal decision boundary $c_\textrm{bayes} = \beta^g_0 + (1/2) \beta^g_1$, i.e.
		\begin{multline*}
		r(\beta^g_1) = \gamma(\beta^g_1, \beta^g_0 + (1/2)\beta^g_1, 1/2) = \frac{\Phi(\beta^g_1/2) - \Phi(-\beta^g_1/2)}{\left[\Phi(-\beta^g_1/2) + \Phi( \beta^g_1/2) \right] \left[\Phi(\beta^g_1/2) + \Phi( -\beta^g_1/2)\right]} \\ = \frac{\int_{-\beta^g_1/2}^{\beta^g_1/2} f}{\left[ 1 - \Phi(\beta^g_1/2) + \Phi(\beta^g_1/2) \right]\left[ \Phi(\beta^g_1/2) + 1 - \Phi(\beta^g_1/2) \right]} = 2 \int_{0}^{\beta^g_1/2} f = 2 \Phi(\beta^g_1/2) - 1.
		\end{multline*}
		We set $r$ to $1/2$ and solve for $\beta^g_1$:
		\begin{multline*}
		r(\beta^g_1) = 1/2 \iff 2\Phi(\beta^g_1/2) -1 = 1/2 \iff \Phi(\beta^g_1/2) = 3/4 \iff \beta^g_1 = 2 \Phi^{-1}(3/4) \approx 1.35.
		\end{multline*}
		Because $r$ is a strictly increasing function, it follows that $r(\beta^g_1) < 1/2$ for $\beta^g_1 < 2\Phi^{-1}(3/4)$ and $r(\beta^g_1) > 1/2$ for $\beta^g_1 > 2\Phi^{-1}(3/4).$ Next, because $$b(\beta^g_1, c_\textrm{bayes}, 1/2) = 1 - \gamma(\beta^g_1, c_\textrm{bayes}, 1/2) = 1 - r(\beta^g_1),$$ we have that $b(\beta^g_1, c_\textrm{bayes}, 1/2) > 1/2$ for $\beta^g_1 < 2 \Phi^{-1}(3/4)$ and $b(\beta^g_1, c_\textrm{bayes}, 1/2) < 1/2$ for $\beta^g_1 > 2 \Phi^{-1}(3/4)$. Recall that the bias induced by sending the threshold to infinity (as stated in Proposition \ref{prop:c_limit} and proven in Section \ref{sec:c_limit}) is $1/2$, i.e. $$b(\beta^g_1, \infty, 1/2) = 1/2.$$ We conclude that $b(\beta^g_1, c_\textrm{bayes},1/2) > b(\beta^g_1, \infty, 1/2)$ on $\beta^g_1 \in [0, 2\Phi^{-1}(3/4))$; $b(\beta^g_1, c_\textrm{bayes},1/2) = b(\beta^g_1, \infty, 1/2)$ for $\beta^g_1 = 2\Phi^{-1}(3/4)$; and $b(\beta^g_1, c_\textrm{bayes},1/2) < b(\beta^g_1, \infty, 1/2)$ on $\beta^g_1 \in (2\Phi^{-1}(3/4), \infty)$.
		
		\subsection{Monotonicity in $\beta^g_1$}\label{sec:monotone}
		We show that $\gamma$ is monotonically increasing in $\beta^g_1$ for $\pi = 1/2$ and given threshold $c$. We begin by stating and proving two lemmas. The first lemma establishes an inequality that will serve as the basis for the proof.
		
		\begin{lemma}
			The following inequality holds: 
			\begin{multline}\label{basic_ineq_cp}
			\left[\Phi(\beta^g_0 - c) + \Phi(\beta^g_0 + \beta^g_1 - c) \right] \cdot \left[\Phi(\beta_0^g + \beta_1^g - c) - \Phi(\beta_0^g - c) + \Phi(c - \beta^g_0) + \Phi(c - \beta_0^g - \beta_1^g) \right] \\ \geq \left[\Phi(\beta_0^g + \beta_1^g - c) - \Phi(\beta_0^g - c)\right]\left[\Phi(c - \beta^g_0) + \Phi(c - \beta_0^g - \beta_1^g)\right].
			\end{multline}
		\end{lemma}
		
		\textbf{Proof}: We take cases on the sign on $\beta^g_1$.
		
		\underline{Case 1}: $\beta^1_g < 0$. Then $ \beta^g_1 + (\beta^g - c) < (\beta^g_0 - c),$ implying $\Phi(\beta^g_0 + \beta^g_1 - c) < \Phi(\beta^g_0 - c),$ or $[\Phi(\beta^g_0 + \beta^g_1 - c) - \Phi(\beta^g_0 - c)] < 0.$ Moreover, $[\Phi(c - \beta^g_0) + \Phi(c - \beta_0^g - \beta_1^g)]$ is positive. Therefore, the right-hand side of (\ref{basic_ineq_cp}) is negative.
		
		Turning our attention of the left-hand side of (\ref{basic_ineq_cp}), we see that
		\begin{equation}\label{basic_ineq_cp_2}
		\Phi(\beta^g_0 + \beta^g_1 - c) + \Phi( c - \beta^g_0 - \beta^g_1) = 1 -\Phi(\beta^g_0 + \beta^g_1 - c) + \Phi( c - \beta^g_0 - \beta^g_1) = 1.
		\end{equation}
		Additionally, $\Phi(\beta^g_0 - c) < 1$ and $ \Phi(c - \beta^g_0) > 0$. Combining these facts with (\ref{basic_ineq_cp_2}), we find that
		$$ \left[\Phi(\beta_0^g + \beta_1^g - c) - \Phi(\beta_0^g - c) + \Phi(c - \beta^g_0) + \Phi(c - \beta_0^g - \beta_1^g) \right] > 0. $$ Finally, because $\left[\Phi(\beta^g_0 - c) + \Phi(\beta^g_0 + \beta^g_1 - c) \right] > 0,$ the entire left-hand side of (\ref{basic_ineq_cp}) is positive. The inequality holds for $\beta^g_1 < 0$.
		
		\noindent
		\underline{Case 2}: $\beta^1_g \geq 0.$  We will show that the first term on the LHS of (\ref{basic_ineq_cp}) is greater than the first term on the RHS of (\ref{basic_ineq_cp}), and likewise that the second term on the LHS is greater than the second term on the RHS, implying the truth of the inequality. Focusing on the first term, the positivity of $\Phi(\beta^g_0 -c)$ implies that
		$\Phi(\beta^g_0 - c) \geq - \Phi(\beta^g_0 - c),$ and so
		$$\Phi(\beta^g_0 - c) + \Phi(\beta^g_0 + \beta^g_1 - c) \geq \Phi(\beta^g_0 - \beta^g_1 - c) - \Phi(\beta^g_0 - c).$$
		Next, focusing on the second term, $\beta^g_1 \geq 0$ implies that 
		\begin{equation}\label{basic_ineq_cp_3}
		\beta^g_1 + \beta^g_0 - c \geq \beta^g_0 - c \implies \Phi(\beta^g_1 + \beta^g_0 - c) - \Phi(\beta^g_0 - c) \geq 0.
		\end{equation}
		Adding $\Phi(c - \beta^g_0) + \Phi(c - \beta^g_0 - \beta^g_1)$ to both sides of (\ref{basic_ineq_cp_3}) yields
		\begin{equation*}
		\Phi(\beta^g_1 + \beta^g_0 - c) - \Phi(\beta^g_0 - c) + \Phi(c - \beta^g_0) + \Phi(c - \beta^g_0 - \beta^g_1) \geq \Phi(c - \beta^g_0) + \Phi(c - \beta^g_0 - \beta^g_1). \textrm{ }
		\end{equation*}
		The inequality holds for $\beta^g_1 \geq 0$. Combining the cases, the inequality holds for all $\beta^g_1 \in \mathbb{R}$. $\square$
		
		The second lemma establishes the derivatives of the functions $2\cdot g$ and $4 \cdot h$ in $\beta^g_1$.
		\begin{lemma}
			The derivatives in $\beta^g_1$ of $2\cdot g$ and $4\cdot h$ are
			\begin{equation}\label{dg_dbeta}
			\textcolor{violet}{D_1[2g(\beta^g_1, c, 1/2)] = f(\beta^g_0 + \beta^g_1 - c)},
			\end{equation}
			\begin{multline}\label{dh_dbeta}
			\textcolor{teal}{D_1[4h(\beta^g_1, c, 1/2)] = f(\beta^g_0 + \beta^g_1 - c) \left[\Phi(c - \beta^g_0) + \Phi(c - \beta_0^g - \beta_1^g) \right]} \\ \textcolor{teal}{- f(\beta^g_0 + \beta^g_1 - c) \left[\Phi(\beta_0^g - c) + \Phi(\beta^g_0 + \beta^g_1 - c) \right]}.\end{multline}
		\end{lemma}
		\textbf{Proof}: Apply FTC and product rule. $\square$
		
		We are ready to prove the monotonicity of $\gamma$ in $\beta^g_1$. Subtracting $$\left[\Phi(\beta_0^g - c) + \Phi(\beta^g_0 + \beta^g_1 - c) \right]\left[\Phi(\beta_0^g + \beta_1^g - c) - \Phi(\beta_0^g - c)\right]$$ from both sides of (\ref{basic_ineq_cp}) and multiplying by $f(\beta^g_0 + \beta^g_1 - c) > 0$ yields
		\begin{multline}\label{basic_ineq_cp_4}
		\textcolor{violet}{f(\beta^g_0 + \beta^g_1 - c)} \textcolor{red}{ \left[\Phi(\beta^g_0 - c) + \Phi\left(\beta^g_0 + \beta^g_1 - c \right) \right] \left[ \Phi(c - \beta^g_0) + \Phi(c - \beta^g_0 - \beta^g_1) \right]}  \\ \geq \textcolor{teal}{f(\beta^g_0 + \beta^g_1 - c) \left[\Phi(c - \beta^g_0) + \Phi(c - \beta_0^g - \beta_1^g)\right]}\textcolor{blue}{\left[\Phi(\beta_0^g + \beta_1^g - c) - \Phi(\beta_0^g - c)\right]} \\ -\textcolor{teal}{f(\beta^g_0 + \beta^g_1 - c)   \left[\Phi(\beta_0^g - c) + \Phi(\beta^g_0 + \beta^g_1 - c) \right]} \textcolor{blue}{\left[\Phi(\beta_0^g + \beta_1^g - c) - \Phi(\beta_0^g - c)\right]}.
		\end{multline}
		Next, recall that
		\begin{equation}\label{def_2g}
		\textcolor{blue}{2g(\beta^g_1,c,1/2) = \Phi(\beta^g_0 + \beta^g_1 - c) - \Phi(\beta^g_0 - c)}.
		\end{equation}
		and
		\begin{equation}\label{def_4h}
		\textcolor{red}{4h(\beta^g_1, c, 1/2) = \left[ \Phi(\beta^g_0 - c) + \Phi(\beta^g_0 + \beta^g_1 - c) \right] \left[\Phi(c - \beta^g_0) + \Phi( c - \beta^g_0 - \beta^g_1) \right]}.
		\end{equation}
		Substituting (\ref{dg_dbeta}, \ref{dh_dbeta}, \ref{def_2g}, \ref{def_4h}) into (\ref{basic_ineq_cp_4}) produces
		\begin{equation*}
		\textcolor{violet}{D_1[2g(\beta^g_1, c, 1/2)]}\textcolor{red}{4h(\beta^g_1, c, 1/2)} \geq \textcolor{blue}{2g(\beta^g_1, c, 1/2)}\textcolor{teal}{D_1[4h(\beta^g_1, c, 1/2)]},
		\end{equation*}
		or 
		\begin{equation}\label{basic_ineq_cp_5}
		\textcolor{violet}{D_1[2g(\beta^g_1, c, 1/2)]}\textcolor{red}{4h(\beta^g_1, c, 1/2)} - \textcolor{blue}{2g(\beta^g_1, c, 1/2)}\textcolor{teal}{D_1[4h(\beta^g_1, c, 1/2)]} \geq 0.
		\end{equation}
		The quotient rule implies that
		\begin{equation}\label{d_gamma_d_beta}
		D_1 \gamma(\beta^g_1, c, 1/2) = \frac{ \textcolor{violet}{D_1[2g(\beta^g_1, c, 1/2)]}\textcolor{red}{4h(\beta^g_1, c, 1/2)} - \textcolor{blue}{2g(\beta^g_1, c, 1/2)}\textcolor{teal}{D_1[4h(\beta^g_1, c, 1/2)]} }{[4h(\beta^g_1, c, 1/2)]^2}.
		\end{equation}
		We conclude by (\ref{basic_ineq_cp_5}) and (\ref{d_gamma_d_beta}) that $\gamma$ is monotonically increasing in $\beta^g_1$. Finally, $b(\beta^g_1, c, \pi) = 1 - \gamma(\beta^g_1, c, \pi)$ is monotonically decreasing in $\beta^g_1$.
		
		\subsection{Strict attenuation bias}\label{sec:att_bias}
		
		We begin by computing the limit of $\gamma$ in $\beta^g_1$ given $\pi = 1/2$.  First,
		\begin{multline*}
		\lim_{\beta^g_1 \to \infty} \gamma(\beta^g_1, c, 1/2) = \frac{1 - \Phi(\beta^g_0 - c)}{\left[1 + \Phi(\beta^g_0 - c) \right] \left[\Phi(c - \beta^g_0) \right]} \\ = \frac{\Phi(c - \beta^g_0)}{ \left[1 + \Phi(\beta^g_0 - c) \right] \left[\Phi(c - \beta^g_0) \right]} = \frac{1}{1 + \Phi(\beta^g_0 - c)} < 1.
		\end{multline*}
		Similarly,
		\begin{equation*}
		\lim_{\beta^g_1 \to -\infty} \gamma(\beta^g_1, c, 1/2) = \frac{ - \Phi(\beta^g_0 - c)}{\left[\Phi(\beta^g_0 - c)\right] \left[\Phi(c - \beta^g_0) + 1 \right]} = \frac{-1}{1 + \Phi(c - \beta^g_0)} > -1.
		\end{equation*}
		The function $\gamma(\beta^g_1, c, 1/2, \beta^g_0)$ is monotonically increasing in $\beta^g_1$ (as stated in Proposition \ref{prop:monotonic} and proven in section \ref{sec:monotone}). It follows that 
		$$-1 < -\frac{1}{1 + \Phi(c - \beta^g_0)} \leq \gamma(\beta^g_1, c, 1/2, \beta^g_0) \leq \frac{1}{1 - \Phi(\beta^g_0 - c)} < 1$$ for all $\beta^g_1 \in \mathbb{R}$. But $\beta^g_0$ and $c$ were chosen arbitrarily, and so
		$$-1 < \gamma(\beta^g_1, c, 1/2, \beta^g_0) < 1$$ for all $(\beta^g_1, c, \beta^g_0) \in \mathbb{R}^3$. Finally, because $b(\beta^g_1, c, 1/2, \beta^g_0) = 1 - \gamma(\beta^g_1, c, 1/2, \beta^g_0)$, it follows that
		$$ 0 < b(\beta^g_1, c, 1/2, \beta^g_0) < 2$$ for all $(\beta^g_1, c, \beta^g_0) \in \mathbb{R}^3$
		
		\subsection{Bias-variance decomposition in no-intercept model}\label{sec:bv_decomp}
		
		We prove the bias-variance decomposition for the no-intercept version of (\ref{theoretical_model}). Define $l$ (for ``limit'') by
		$$l = \beta_m \left(\frac{\omega \pi}{\zeta(1-\pi) + \omega \pi}\right),$$ where
		$$
		\omega = \bar{\Phi}(c - \beta_g) = \Phi(\beta_g - c); \quad
		\zeta = \bar{\Phi}(c) = \Phi(-c).
		$$
		We have that
		\begin{equation*}
		\hat{\beta}_m - l = \frac{\sum_{i=1}^n \hat{p}_i m_i}{ \sum_{i=1}^n \hat{p}^2_i} - l = \frac{\sum_{i=1}^n \hat{p}_i m_i}{ \sum_{i=1}^n \hat{p}^2_i} - \frac{l \sum_{i=1}^n \hat{p}_i^2 }{ \sum_{i=1}^n \hat{p}_i^2} \\ = \frac{\sum_{i=1}^n \hat{p}_i(m_i - l \hat{p}_i)}{ \sum_{i=1}^n \hat{p}_i^2}.
		\end{equation*}
		Therefore,
		\begin{equation}\label{bc_decomp_1}
		\sqrt{n}(\hat{\beta}_m - l) = \frac{(1/\sqrt{n})\sum_{i=1}^n \hat{p}_i(m_i - l \hat{p}_i)}{(1/n)\sum_{i=1}^n \hat{p}_i^2}.
		\end{equation}
		Next, we compute the expectation and variance of $\hat{p}_i(m_i - l\hat{p}_i)$. To do so, we first compute several simpler quantities:
		\begin{enumerate}
			\item Expectation of $\hat{p}_i$: $
			\mathbb{E}[\hat{p}_i] = \mathbb{P}(p_i\beta_g + \tau_i \geq c) =  \mathbb{P}(\beta_g + \tau_i \geq c)\pi + \mathbb{P}(\tau_i \geq c)(1-\pi) = \pi \omega + (1-\pi)\zeta.$
			
			\item Expectation of $\hat{p}_i p_i$: $\mathbb{E}\left[\hat{p}_i p_i\right] = \mathbb{E}\left[\hat{p}_i | p_i = 1 \right]\mathbb{P}\left[p_i = 1\right] = \omega \pi.$
			\item Expectation of $\hat{p}_i m_i$:
			\begin{multline*}
			\mathbb{E}[\hat{p}_i m_i] = \mathbb{E}\left[\hat{p}_i(\beta_m p_i + \epsilon_i)\right] = \mathbb{E}\left[\beta_m \hat{p}_i p_i + \hat{p}_i \epsilon_i \right] \\ = \beta_m \mathbb{E}\left[ \hat{p}_i p_i \right] + \mathbb{E}[\hat{p}_i]\mathbb{E}[\epsilon_i] = \beta_m \omega \pi + 0 = \beta_m \omega \pi.
			\end{multline*}
			\item Expectation of $\hat{p}_i m_i^2$: \begin{multline*}
			\mathbb{E}\left[\hat{p}_i m_i^2\right] = \mathbb{E} \left[ \hat{p}_i( \beta_m p_i + \epsilon_i )^2 \right] = \mathbb{E}\left[ \hat{p}_i \left( \beta_m^2 p_i^2 + 2 \beta_m p_i \epsilon_i + \epsilon_i^2 \right)  \right] \\ = \mathbb{E}\left[ \hat{p}_i p_i \beta^2_m + 2 \beta_m p_i \hat{p}_i \epsilon_i + \hat{p}_i \epsilon_i^2 \right] = \beta^2_m \mathbb{E}[ \hat{p}_i p_i] + 2 \beta_m \mathbb{E}[p_i\hat{p}_i] \mathbb{E}[\epsilon_i] + \mathbb{E}[\hat{p}_i] \mathbb{E}[ \epsilon^2_i ] \\ = \beta^2_m \mathbb{E}[ \hat{p}_i p_i] + \mathbb{E}[\hat{p}_i] = \beta^2_m \omega \pi + \mathbb{E}[ \hat{p}_i]. 
			\end{multline*}
		\end{enumerate}
		
		Now, we can compute the expectation and variance of $\hat{p}_i(m_i - l\hat{p}_i)$. First,
		\begin{equation}\label{bv_decomp_2}
		\mathbb{E}\left[\hat{p}_i(m_i - l\hat{p}_i) \right] = \mathbb{E}[\hat{p}_i m_i] - l \mathbb{E}[\hat{p}_i] = \beta_m \omega \pi - \left(\frac{\beta_m \omega \pi}{\zeta (1-\pi) + \omega \pi}\right)[\zeta (1-\pi) + \omega \pi] = 0.
		\end{equation}
		Additionally,
		\begin{multline}\label{bv_decomp_3}
		\mathbb{V}\left[\hat{p}_i(m_i - l\hat{p}_i)\right] = \mathbb{E}\left[\hat{p}_i^2(m_i - l\hat{p}_i)^2\right] - (\mathbb{E}\left[ \hat{p}_i(m_i - l\hat{p}_i)\right])^2 \\ = \mathbb{E}\left[ \hat{p}_i m_i^2\right] - 2l \mathbb{E}[m_i\hat{p}_i] +l^2 \mathbb{E}[\hat{p}_i]= \beta^2_m \omega \pi + \mathbb{E}[ \hat{p}_i] -2l \beta_m \omega \pi + l^2 \mathbb{E}[\hat{p}_i] \\ = \beta_m\omega\pi(\beta_m - 2l) + \mathbb{E}[\hat{p}_i](1 + l^2).
		\end{multline}
		Therefore, by CLT, (\ref{bv_decomp_2}), and (\ref{bv_decomp_3}),
		\begin{equation}\label{bv_decomp_4}
		(1/\sqrt{n})\sum_{i=1}^n \hat{p}_i(m_i - l \hat{p}_i) \xrightarrow{d} N\left(0, \beta_m\omega\pi(\beta_m - 2l) + \mathbb{E}[\hat{p}_i](1 + l^2) \right).
		\end{equation}
		Next, by weak LLN,
		\begin{equation}\label{bv_decomp_5}
		(1/n) \sum_{i=1}^n \hat{p}_i^2 = (1/n) \sum_{i=1}^n \hat{p}_i \xrightarrow{P} \mathbb{E}[\hat{p}_i].
		\end{equation}
		
		Finally, by (\ref{bc_decomp_1}), (\ref{bv_decomp_4}), (\ref{bv_decomp_5}), and Slutsky's Theorem,
		$$ \sqrt{n}(\hat{\beta}_m - l) \xrightarrow{d} N\left(0, \frac{ \beta_m\omega\pi(\beta_m - 2l) + \mathbb{E}[\hat{p}_i](1 + l^2) }{\left(\mathbb{E}[\hat{p}_i]\right)^2} \right).$$ Thus, for large $n \in \mathbb{N}$, we have that 
		$$
		\mathbb{E} [\hat{\beta}_m] \approx l; \quad
		\mathbb{V}[\hat{\beta}_m] \approx \left[\beta_m\omega\pi(\beta_m - 2l) + \mathbb{E}[\hat{p}_i](1 + l^2)\right]/[n\mathbb{E}^2[\hat{p}_i]],$$
		completing the bias-variance decomposition.

\subsection{Bayes-optimal decision boundary for non-Gaussian mixture distributions and GLMs}\label{sec:non_gaussian_bayes_thresholds}

We report the Bayes-optimal decision boundary for gRNA count distributions that are non-Gaussian. First, consider a simple two-component Poisson mixture model with means $\mu_0$ and $\mu_1$ and mixing probability $\pi$:

$$p(k; \mu_0, \mu_1, \pi) = (1 - \pi) f(k; \mu_0) + \pi f(k; \mu_1),$$ where $f(k;\mu) = (\mu^k e^{-k})/\mu!$ is a Poisson density. Suppose we draw an observation from this distribution. The Bayes-optimal threshold for classifying the observation as having been drawn from the first or second component is
\begin{equation}\label{eqn:poisson_bayes}
\frac{\mu_0 - \mu_1 + \log(\pi) - \log(1 - \pi)}{\log(\mu_0) - \log(\mu_1)}.
\end{equation}
Next, consider the slightly more complex Poisson mixture GLM:
$$ g_i |(p_i, z_i, o_i) \sim \textrm{Pois}(\mu_i); \quad r(\mu_i) = \beta_0 + \beta_1 p_i + \gamma^T z_i + o_i,$$ where $p_i \sim \textrm{Bern}(\pi)$ is unobserved. Conditional on the covariates and offset, the mean of the unperturbed component is $\mu_i(1) = r^{-1}(\beta_0 + \gamma^Tz_i + o_i),$ and that of the perturbed component is $\mu_i(1) = r^{-1}(\beta_0 + \beta_1 + \gamma^T z_i + o_i.)$ The Bayes-optimal threshold is obtained by plugging in $\mu_i(1)$ for $\mu_1$ and $\mu_i(0)$ for $\mu_0$ in (\ref{eqn:poisson_bayes}). To obtain a fixed gRNA assignment threshold across cells, we compute the Bayes-optimal decision boundary for each cell and then take the average across cells. The situation is similar for the negative binomial (with known size $s$) distribution; the Bayes-optimal decision boundary in this case is
$$
\frac{s \left[ \log(\mu_0 + s) - \log(\mu_1 + s) \right] + \log(\pi) - \log(1 - \pi)}{\log(\mu_0 (\mu_1 + s)) - \log(\mu_1 (\mu_0 + s))}.
$$

\section{Estimation and inference in the GLM-EIV model}\label{sec:glmeiv_details}

\subsection{Detailed specification of the model}

We provide a more precise and technical specification of the GLM-EIV model than provided in the main text. Let $\tilde{x_i} = [1, p_i, z_i]^T \in \mathbb{R}^d$ be the vector of covariates (including an intercept term) for the $i$th cell. (We use the tilde as a reminder that the vector is partially unobserved.) Let $\beta_m = [\beta^m_0, \beta^m_1, \gamma_m]^T \in \mathbb{R}^d$ and $\beta_g = [\beta^g_0, \beta^g_1, \gamma_g]^T \in \mathbb{R}^d$ be the unknown coefficient vectors corresponding to the gene and gRNA expression models, respectively. Finally, let $o^m_i$ and $o^g_i$ be the (possibly zero) offset terms for the gene and gRNA models; in practice, we typically set $o^m_i$ and $o^g_i$ to the log-transformed library sizes (i.e., $\log(d^m_i)$ and $\log(d^g_i)$, respectively).

We use a pair of GLMs to model the gene and gRNA expressions. Considering first the gene expression model, let the $i$th linear component $l^m_i$ of the model be $l^m_i \equiv \langle \tilde{x}_i, \beta_m \rangle + o^m_i.$ Next, let the mean $\mu^m_i$ of the $i$th observation be $r_m(\mu^m_i) \equiv l^m_i,$ where $r_m:\mathbb{R} \to \mathbb{R}$ is a strictly increasing, differentiable link function. Let $\psi_m: \mathbb{R} \to \mathbb{R}$ be the differentiable, cumulant-generating function of the selected exponential family distribution. We can express the canonical parameter $\eta^m_i$ in terms of $\psi_m$ and $r_m$ by
$\eta^m_i = \left([\psi'_m]^{-1} \circ r^{-1}_m\right)(l_i^m) \equiv h_m(l_i^m).$ Finally, let $c_m: \mathbb{R} \to \mathbb{R}$ be the carrying density of the selected exponential family distribution. The density $f_m$ of $m_i$ conditional on the the canonical parameter $\eta_i$ is
$f_m(m_i; \eta^m_i) = \exp\left\{m_i \eta^m_i - \psi_m(\eta^m_i) + c_m(m_i) \right\}.$  We implicitly set the ``scale'' term (i.e., the $a(\phi)$ term in \cite{McCullagh1990}, Eqn.  2.4, p.\ 28) to unity; this slightly simplified model encompasses the most important distributions for our purposes, including the Poisson, negative binomial, and Gaussian (with unit variance) distributions. %The function $c_m$ appears as a constant in the log likelihood of $m_i$; therefore, the only functions relevant to inference are $\psi_m$ and $r_m$.

Let the terms $l^g_i, o^g_i, \mu^g_i, \eta^g_i, \psi_g, r_g, h_g$ and $c_g$ be defined in an analogous way for the gRNA model, i.e. $l^g_i \equiv \langle \tilde{x}_i, \beta_g \rangle + o^g_i$, $r_g(\mu^g_i) \equiv l^g_i$, and $\eta^g_i = \left([\psi'_g]^{-1} \circ r^{-1}_g\right)(l_i^g) \equiv h_g(l_i^g).$ The density $f_g$ of $g_i$ given the canonical parameter is $f_g(m_i; \eta^g_i) = \exp\left\{g_i \eta^g_i - \psi_g(\eta^g_i) + c_g(g_i)\right\}.$
Finally, the unobserved variable $p_i$ is assumed to follow a Bernoulli distribution with mean $\pi \in (0, 1/2]$. Its marginal density $f_p$ is given by $f_p(p_i) = \pi^{p_i}(1-\pi)^{1 - p_i}.$
The unknown parameters in the model are
$\theta = [\beta_m, \beta_g, \pi]^{T}  \in \mathbb{R}^{2d + 1}.$

\subsection{Notation} We briefly introduce notation that we will use throughout. For $j \in \{0,1\}$, let $\tilde{x}_i(j) \equiv [1, j, z_i]^T$ denote the value of $\tilde{x}_i$ that results from setting $p_i$ to $j$. Next, let  $l^m_i(j)$, $\eta^m_i(j),$ and $\mu^m_i(j)$ be the values of $l^m_i$, $\eta^m_i$, and $\mu^m_i$, respectively, that result from setting $p_i$ to $j$, i.e.,
$l^m_i(j) \equiv \langle \tilde{x}_i(j), \beta_m \rangle + o^m_i$, $\eta^m_i(j) \equiv h_m(l^m_i(j))$, and
$\mu_i^m(j) \equiv r_m^{-1}(l^m_i(j)).$ Let the corresponding gRNA quantities $l^g_i(j)$, $\eta_i^g(j)$, and $\mu^g_i(j)$ be defined analogously. Next, let $X \in \mathbb{R}^{n \times (d-1)}$ be the observed design matrix, and let $\tilde{X} \in \mathbb{R}^{n \times d}$ be the augmented design matrix that results from concatenating the column of (unobserved) $p_i$s to $X$, i.e.
$$ X \equiv \begin{bmatrix} 
1 & z_1 \\
\vdots & \vdots \\
1 & z_n
\end{bmatrix}; \quad 
\tilde{X} \equiv 
\begin{bmatrix}
1 & p_1 & z_1 \\
\vdots & \vdots & \vdots \\
1 & p_n & z_n
\end{bmatrix} = \begin{bmatrix}
\tilde{x}_1^T \\ \vdots \\ \tilde{x}_n^T
\end{bmatrix}.
$$ 
Furthermore, for $j \in \{0,1\}$, let $\tilde{X}(j) \in \mathbb{R}^{n \times d}$ be the matrix that results from setting $p_i$ to $j$ for all $i \in \{1, \dots, n\}$ in $\tilde{X}$, and let  $[\tilde{X}(0)^T, \tilde{X}(1)^T]^T$ denote the $\mathbb{R}^{2n \times d}$ matrix that results from vertically concatenating $\tilde{X}(0)$ and $\tilde{X}(1)$. Furthermore, define $m := [m_1, \dots, m_n]$, and let $g$, $p$, $o^m$, and $o^g$ be defined analogously. Finally, let $[m,m]^T \in \mathbb{R}^{2n}$ be the vector that results from concatenating $m$ to itself, i.e.
$[m,m]^T \equiv [m_1, \dots, m_n, m_1, \dots, m_n],$ and let $[g,g]^T$, $[o^g,o^g]^T$, and $[o^m,o^m]^T$ be defined similarly. 

\subsection{Log likelihood and estimation}

 We conduct estimation and inference conditional on the library sizes and technical factors $l^m_i, l^g_i,$ and $z_i$; therefore, we treat these quantities as fixed constants. We assume that the gene expression $m_i$ and gRNA expression $g_i$ are conditionally independent given the perturbation $p_i$. The model log-likelihood is
\begin{equation}\label{marginal_log_lik}
\mathcal{L}(\theta; m, g) = \sum_{i=1}^n \log\left[(1-\pi) f_m(m_i; \eta^m_i(0)) f_g(g_i; \eta^g_i(0)) + \pi f_m(m_i; \eta^m_i(1)) f_g(g_i; \eta^g_i(1)) \right].
\end{equation}
We see from (\ref{marginal_log_lik}) that the GLM-EIV model is equivalent to a two-component mixture of \textit{products} of GLM densities. We estimate the parameters of the GLM-EIV model using an EM algorithm.

\subsubsection*{E step}
The E step entails computing the membership probability of each cell. Let $\theta^{(t)} = (\beta_m^{(t)}, \beta_g^{(t)}, \pi^{(t)})$ be the parameter estimate at the $t$-th iteration of the algorithm. For $k \in \{0,1\}$, let $[\eta^m_i(k)]^{(t)}$ be the $i$th canonical parameter at the $t$-th iteration of the algorithm of the gene expression distribution that results from setting $p_i$ to $k$, i.e.
$[\eta^m_i(k)]^{(t)} \equiv h_m\left( \langle \tilde{x}_i(k) , \beta_m^{(t)} \rangle + o^m_i \right).
$ Similarly, let $\left[\eta^g_i(k)\right]^{(t)}$ be defined by
$\left[\eta^g_i(k)\right]^{(t)} \equiv  h_g\left( \langle \tilde{x}_i(k) , \beta_g^{(t)} \rangle + o^g_i \right).$
Next, for $k \in \{0,1\},$ define $\alpha^{(t)}_i(k)$ by
\begin{multline*}
\alpha^{(t)}_i(k) \equiv \mathbb{P}\left( M_i = m_i, G_i = g_i | P_i = k, \theta^{(t)} \right) \\ = \mathbb{P}\left( M_i = m_i | P_i = k, \theta^{(t)} \right) \mathbb{P}\left(G_i = g_i | P_i = k, \theta^{(t)} \right) \textrm{ (because $G_i \indep M_i | P_i$)} \\ = f_m\left(m_i; \left[ \eta^m_i(k) \right]^{(t)}\right) f_g\left(g_i; \left[ \eta^g_i(k) \right]^{(t)} \right).
\end{multline*}
Finally, let $\pi^{(t)}(1) \equiv \pi^{(t)} = \mathbb{P}\left(P_i = 1 | \theta^{(t)} \right)$ and $\pi^{(t)}(0) \equiv 1 - \pi^{(t)} = \mathbb{P}\left(P_i = 0 | \theta^{(t)} \right)$.
The $i$th membership probability $T^{(t)}_i(1)$ is
\begin{multline}\label{e_step_1}
T^{(t)}_i(1) = \mathbb{P}(P_i = 1 | M_i = m_i, G_i = g_i, \theta^{(t)})  = \frac{\pi^{(t)}(1) \alpha^{(t)}_i(1)}{ \sum_{k=0}^1 \pi^{(t)}(k) \alpha^{(t)}_i(k)} \textrm{ (by Bayes rule)} \\ = \frac{1}{\frac{ \pi^{(t)}(0) \alpha_i(0)}{\pi^{(t)}(1) \alpha_i(1)} + 1} = \frac{1}{ \exp\left(\log\left(\frac{\pi^{(t)}(0) \alpha_i(0)}{\pi^{(t)}(1) \alpha_i(1)}\right)\right) + 1} = \frac{ 1 }{ \exp\left(q^{(t)}_i\right) + 1},
\end{multline}
where we set 
\begin{equation}\label{e_step_2}
q_i^{(t)} := \log\left(\frac{\pi^{(t)}(0) \alpha_i^{(t)}(0)}{\pi^{(t)}(1) \alpha_i^{(t)}(1)}\right).
\end{equation}
Next, we have that
\begin{multline*}
q^{(t)}_i = \log\left[ \pi^{(t)}(0) \right] + \log\left[ f_m\left(m_i; \left[ \eta^m_i(0) \right]^{(t)}\right) \right] + \log\left[ f_g\left(g_i; \left[ \eta^g_i(0) \right]^{(t)}\right) \right] \\ - \log\left[ \pi^{(t)}(1) \right] - \log\left[ f_m\left(m_i; \left[ \eta^m_i(1) \right]^{(t)}\right) \right] - \log\left[ f_g\left(g_i; \left[ \eta^g_i(1) \right]^{(t)}\right) \right],
\end{multline*}
We therefore conclude that
$T_i^{(t)} = 1/\left(\exp\left(q^{(t)}_i\right) + 1\right),$ which is easily computable.
\subsection*{M step}
The complete-data log-likelihood of the GLM-EIV model  is
\begin{equation}\label{full_log_lik}
\mathcal{L}(\theta; m, g, p) = \sum_{i=1}^n \left[ p_i \log(\pi) + (1-p_i) \log(1-\pi) \right] + \sum_{i=1}^n \log\left( f_m(m_i; \eta^m_i)\right) + \sum_{i=1}^n \log\left( f_g(g_i; \eta_i^g) \right).
\end{equation}
Define $Q(\theta | \theta^{(t)}) = \mathbb{E}_{\left(P |M = m, G = g, \theta^{(t)}\right)}\left[ \mathcal{L}(\theta; m, g, p) \right].$ We have that
		\begin{multline}\label{Q_funct}
		Q(\theta |\theta^{(t)}) = \sum_{i=1}^n \left[T^{(t)}_i(1)\log(\pi) + T_i^{(t)}(0) \log(1 - \pi)\right] \\ + \sum_{k=0}^1 \sum_{i=1}^n T^{(t)}_i(k) \log \left[ f_m(m_i; \eta_i^m(k)) \right] + \sum_{k=0}^1 \sum_{i=1}^n T^{(t)}_i(k) \log \left[ f_g( g_i; \eta^{g,b}_i(k)) \right].
		\end{multline}
		The three terms of (\ref{Q_funct}) are functions of different parameters: the first is a function of $\pi,$ the second is a function of $\beta_m,$ and the third is a function of $\beta_g$. Therefore, to find the maximizer $\theta^{(t+1)}$ of (\ref{Q_funct}), we maximize the three terms separately. Differentiating the first term with respect to $\pi$, we find that
		\begin{equation*}
		\frac{ \partial }{\partial \pi } \sum_{i=1}^n \left[ T^{(t)}_i(1)\log(\pi) + T_i^{(t)}(0) \log(1 - \pi)\right]  = \frac{\sum_{i=1}^n T_i^{(t)}(1)}{\pi} - \frac{ \sum_{i=1}^n T_i^{(t)}(0) }{ 1 - \pi}.
		\end{equation*} Setting the derivative equal to $0$ and solving for $\pi$,
		\begin{multline*}
		\frac{\sum_{i=1}^n T_i^{(t)}(1)}{\pi} - \frac{ \sum_{i=1}^n T_i^{(t)}(0) }{ 1 - \pi} = 0 \iff \sum_{i=1}^n T_i^{(t)}(1) - \pi \sum_{i=1}^n T^{(t)}_i(1) = \pi \sum_{i=1}^n T_i(0) \\ \iff \sum_{i=1}^n T^{(t)}_i(1) - \pi\sum_{i=1}^n T_i^{(t)}(1) = \pi n - \pi\sum_{i=1}^n T_i(1) \iff \pi = \frac{ \sum_{i=1}^n T_i^{(t)} (1) }{n}.
	\end{multline*}

	Thus, the maximizer $\pi^{(t+1)}$ of (\ref{Q_funct}) in $\pi$ is $\pi^{(t+1)} = (1/n)\sum_{i=1}^n T^{(t)}_i(1)$. Next, define $w^{(t)} = [T^{(t)}_1(0), \dots, T^{(t)}_n(0), T^{(t)}_1(1), \dots, T^{(t)}_n(1)]^T \in \mathbb{R}^{2n}$. We can view the second term of (\ref{Q_funct}) as the log-likelihood of a GLM -- call it $\textrm{GLM}^{(t)}_m$ -- that has exponential family density $f_m$, link function $r_m$, responses $[m,m]^T$, offsets $[o^m, o^m]^T$, weights $w^{(t)}$, and design matrix $[ \tilde{X}(0)^T, \tilde{X}(1)^T ]^T.$ Therefore, the maximizer $\beta^{(t+1)}_m$ of the second term of (\ref{Q_funct}) is the maximizer of $\textrm{GLM}^{(t)}_m$, which we can compute using the iteratively reweighted least squares (IRLS) procedure, as implemented in R's GLM function. Similarly, the maximizer $\beta^{(t+1)}_g$ of the third term of (\ref{Q_funct}) is the maximizer of the GLM with exponential family density $f_g$, link function $r_g$, responses $[g,g]^T$, offsets $[o^g, o^g]^T$, weights $w^{(t)}$, and design matrix $[ \tilde{X}(0)^T, \tilde{X}(1)^T ]^T.$
		
		\subsection{Inference}
		We derive the asymptotic observed information matrix of the GLM-EIV log likelihood, enabling us to perform inference on the parameters. First, we define some notation. For $i \in \{1, \dots, n\}$, $j \in \{0, 1\}$, and $\theta = (\pi, \beta_m, \beta_g),$ let $T^\theta_i(j)$ be defined by
		$$T^\theta_i(j) = \mathbb{P}_\theta\left(P_i = j | M_i = m_i, G_i = g_i\right).$$ Let the $n \times n$ matrix $T^\theta(j)$ be given by $T^\theta(j) = \textrm{diag}\left\{T^\theta_1(j), \dots, T^\theta_n(j)\right\}.$
		Next, define the diagonal $n \times n$ matrices $\Delta^m$, $[\Delta']^m$, $V^m$, and $H^m$ by
		$$
		\begin{cases}
		\Delta^m = \textrm{diag} \{h_m'(l_1^m), \dots, h_m'(l_n^m)\} \\
		[\Delta']^m = \textrm{diag} \{h_m''(l_1^m), \dots, h_m''(l_n^m) \} \\
		V^m = \textrm{diag} \{ \psi''_m( \eta^m_1), \dots, \psi''_m( \eta^m_n) \} \\
		H^m = \textrm{diag} \{ m_1 - \mu_1^m, \dots, m_n - \mu_n^m\}.
		\end{cases} 
		$$ Define the $n \times n$ matrices $\Delta^g, [\Delta']^g, V^g,$ and $H^g$ analogously. These matrices are \textit{unobserved}, as they depend on $\{p_1, \dots, p_n\}$. Next, for $j \in \{0,1\}$, let the diagonal $n \times n$ matrices $\Delta^m(j), [\Delta']^m(j), V^m(j),$ and $H^m(j)$ be given by
		$$\begin{cases}
		\Delta^m(j) = \textrm{diag} \{ h_m'(l_1^m(j)), \dots, h_m'( l_n^m(j) ) \} \\
		[\Delta']^m(j) = \textrm{diag} \{ h_m''(l_1^m(j)), \dots, h_m''( l_n^m(j)) \} \\
		V^m(j) = \textrm{diag} \{\psi''_m( \eta^m_1(j)), \dots, \psi''_m( \eta^m_n(j))\} \\
		H^m(j) = \textrm{diag} \{m_1 - \mu_1^m(j), \dots, m_n - \mu_n^m(j)\} .
		\end{cases}
		$$
		Define the matrices $\Delta^g(j)$, $[\Delta']^{g}(j)$, $V^g(j),$ and $H^g(j)$ analogously. Finally, define the vectors $s^m(j), w^m(j) \in \mathbb{R}^n$ by 
		$$ \begin{cases}
		s^m(j) = [m_1 - \mu_1^m(j), \dots, m_n - \mu_n^m(j) ]^T \\ w^m(j) = [ T_1(0)T_1(1)\Delta^m_1(j) H^m_1(j), \dots, T_n(0)T_n(1)\Delta_n^m(j) H_n^m(j)]^T,
		\end{cases} $$
		and let the vectors $s^g(j)$ and $w^g(j)$ be defined analogously. The quantities $\Delta^m(j), [\Delta']^m(j), V^m(j),$ $H^m(j),$ $s^m(j),$ $w^m(j),$ $\Delta^g(j), [\Delta']^g(j), V^g(j),$ $H^g(j),$ $s^g(j),$ and $w^g(j)$ are all \textit{observed}. 
		
\begin{figure}
\centering
\includegraphics[width=0.45\linewidth]{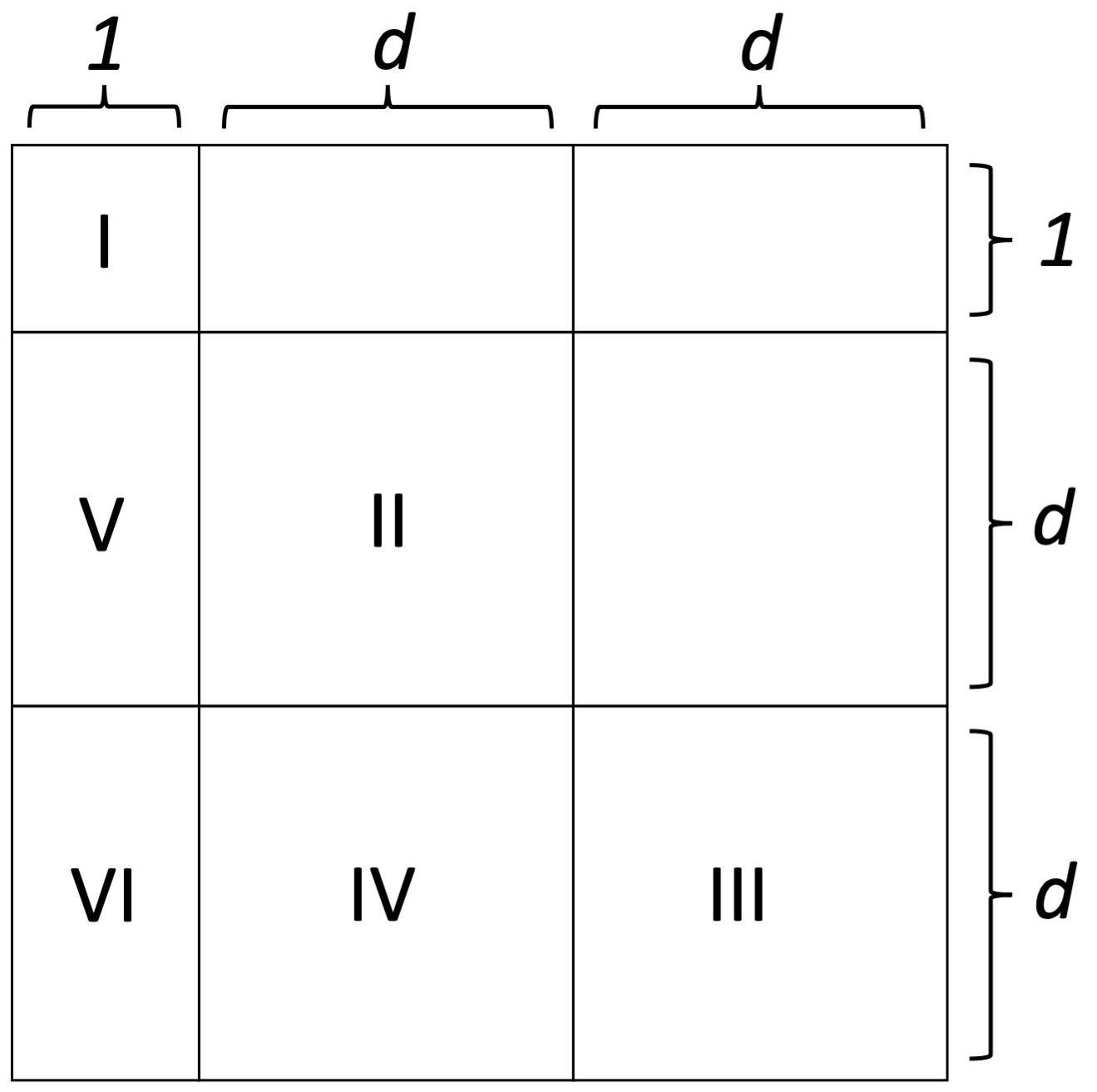}
\caption{Block structure of the observed information matrix $J(\theta; m, g) = -\nabla^2 \mathcal{L}(\theta; m, g)$. The matrix is symmetric, and so we only need to compute submatrices I-VI to compute the entire matrix.}
\label{infomatrixbackground}
\end{figure}

		The observed information matrix $J(\theta; m, g)$ evaluated at $\theta = (\pi, \beta_m, \beta_g)$ is the negative Hessian of the log likelihood (\ref{marginal_log_lik}) evaluated at $\theta$, i.e.\
		$J(\theta; m, g) = - \nabla^2\mathcal{L}(\theta; m, g) .$ This quantity, unfortunately, is hard to compute, as the log likelihood (\ref{marginal_log_lik}) is a complicated mixture. \cite{Louis1982} showed that $J(\theta; m, g)$ is equivalent to the following quantity:
		\begin{multline}\label{zero_inf_info_mat}
		J(\theta; m, g) = -\mathbb{E} \left[\nabla^2 \mathcal{L}(\theta; m, g, p) | G = g, M = m \right] \\ + \mathbb{E}\left[\nabla \mathcal{L}(\theta; m, g, p) | G = g, M = m \right] \mathbb{E}\left[\nabla \mathcal{L}(\theta; m, g, p) | G = g, M = m \right]^T \\ - \mathbb{E}\left[ \nabla\mathcal{L}(\theta; m, g, p) \nabla \mathcal{L}(\theta; m, g, p)^T | G = g, M = m \right].
		\end{multline}
		The observed information matrix $J(\theta; m, g)$ has dimension $(2d+1) \times (2d + 1).$ Recall that the complete-data log-likelihood (\ref{full_log_lik}) is the sum of three terms. The first term depends only on $\pi$, the second on $\beta_m$, and the third on $\beta_g$. Therefore, the observed information matrix can be viewed as block matrix consisting of nine submatrices (Figure \ref{infomatrixbackground}; only six submatrices labelled). Submatrix I depends on $\pi$, submatrix II on $\beta_m$, submatrix III on $\beta_g$, submatrix IV on $\beta_m$ and $\beta_g$, submatrix V on $\pi$ and $\beta_m$, and submatrix VI on $\pi$ and $\beta_g$. We only need to compute these six submatrices to compute the entire matrix, as the matrix is symmetric. The following sections derive formulas for submatrices I-VI. All expectations are understood to be \textit{conditional} on $m$ and $g$. The notation $\nabla_v$  and $\nabla^2_v$  represent the gradient and Hessian, respectively, with respect to the vector $v$.
		
		\subsubsection*{Submatrix I}
		Denote submatrix I by $J_{\pi}(\theta; m, g).$ The formula for $J_{\pi}(\theta; m, g)$ is 
		\begin{equation}\label{sub_mat_pi}
		J_{\pi}(\theta; m, g) = -\mathbb{E}\left[\nabla^2_\pi \mathcal{L}(\theta; m, g, p) \right] + \left(\mathbb{E}\left[ \nabla_\pi \mathcal{L}(\theta; m, g, p) \right] \right)^2 - \mathbb{E}\left[(\nabla_\pi \mathcal{L}(\theta; m, g, p))^2 \right].
		\end{equation}
		
		We begin by calculating the first and second derivatives of the log-likelihood $\mathcal{L}$ with respect to $\pi$. The first derivative is
		\begin{multline}\label{d_L_d_pi}
		\nabla_\pi \mathcal{L}(\theta; m, g, p) = \frac{\partial }{\partial \pi } \left( \sum_{i=1}^n p_i \log(\pi) + \sum_{i=1}^n (1 - p_i) \log(1 - \pi) \right) \\ = \frac{ \sum_{i=1}^n p_i }{\pi} - \frac{ \sum_{i=1}^n (1 - p_i) }{ 1 - \pi } = \frac{\sum_{i=1}^n p_i}{\pi} - \frac{n - \sum_{i=1}^n p_i}{1 - \pi} = \left( \frac{1}{\pi} + \frac{1}{1 - \pi} \right) \sum_{i=1}^n p_i - \frac{n}{1-\pi}.
		\end{multline}
		The second derivative is
		\begin{multline*}
		\nabla^2_\pi \mathcal{L}(\theta; m, g, p)  = \frac{\partial^2}{\partial^2\pi} \left( \frac{ \sum_{i=1}^n p_i }{ \pi } - \frac{ n - \sum_{i=1}^n p_i }{1 - \pi}  \right) = \frac{\left( \sum_{i=1}^n p_i \right) - n}{(1 - \pi)^2} - \frac{\sum_{i=1}^n p_i }{ \pi^2 }.
		\end{multline*}
		We compute the expectation of the first term of (\ref{sub_mat_pi}):
		\begin{multline}\label{submat_pi_1}
		\mathbb{E} \left[ -\nabla^2_{\pi} \mathcal{L}(\theta; m, g, p)\right] = - \mathbb{E}\left[\frac{ ( \sum_{i=1}^n p_i) - n}{(1 - \pi)^2} - \frac{\sum_{i=1}^n p_i}{\pi^2} \right] \\ = - \mathbb{E}\left\{\left[\frac{1}{(1-\pi)^2} - \frac{1}{\pi^2} \right] \sum_{i=1}^n p_i - \frac{n}{ (1 - \pi)^2 } \right\} = - \left\{\left[ \frac{1}{(1-\pi)^2} - \frac{1}{\pi^2} \right] \sum_{i=1}^n T^\theta_i(1) - \frac{n}{ (1 - \pi)^2}  \right\} \\ = \left[ \frac{1}{\pi^2} - \frac{1}{(1 - \pi)^2} \right] \sum_{i=1}^n T^\theta_i(1) + \frac{n}{(1-\pi)^2}.
		\end{multline}
		Next, we compute the difference of the second two pieces of (\ref{sub_mat_pi}). To this end, define $a \equiv 1/(1-\pi) + 1/\pi$ and $b \equiv n/(1-\pi).$ We have that
		\begin{multline*}
		\mathbb{E} \left[\nabla_\pi \mathcal{L}(\theta; m, g, p)^2 \right] = \mathbb{E} \left[ \left( a \sum_{i=1}^n p_i - b\right)^2 \right]  =  \mathbb{E} \left[ a^2 \left( \sum_{i=1}^n p_i \right)^2 - 2ab \sum_{i=1}^n p_i + b^2 \right] \\ = a^2 \sum_{i=1}^n \sum_{j=1}^n \mathbb{E}[p_i p_j] -2ab \sum_{i=1}^n \mathbb{E} [p_i] + b^2.
		\end{multline*}
		Next,
		\begin{equation*}
		\left( \mathbb{E} \left[\nabla_\pi \mathcal{L}(\theta; m, g, x) \right] \right)^2 = \left( a \sum_{i=1}^n \mathbb{E} [p_i] - b \right)^2 = a^2 \sum_{i=1}^n \sum_{j=1}^n \mathbb{E}[p_i]  E[p_j] - 2ab \sum_{i=1}^n \mathbb{E}[p_i] + b^2.
		\end{equation*}
		Therefore,
		\begin{multline}\label{submat_pi_2}
		(\mathbb{E} [\nabla_\pi \mathcal{L}(\theta; m, g, p)])^2 - \mathbb{E} \left[\nabla_\pi \mathcal{L}(\theta; m, g, p)^2 \right] \\ = a^2 \sum_{i=1}^n \sum_{j=1}^n \mathbb{E}[p_i] \mathbb{E}[p_j] - a^2 \sum_{i=1}^n \sum_{j=1}^n \mathbb{E}[p_i p_j] = a^2 \left( \sum_{i=1}^n \mathbb{E}[p_i]^2 - \mathbb{E}[p_i^2]\right) \\ = a^2 \left( \sum_{i=1}^n [T^\theta_i(1)]^2 - T^\theta_i(1) \right) = \left( \frac{1}{(1 - \pi)} + \frac{1}{\pi} \right)^2 \left(\sum_{i=1}^n [T^\theta_i(1)]^2 - T^\theta_i(1) \right).
		\end{multline}
		Stringing (\ref{sub_mat_pi}), (\ref{submat_pi_1}) and (\ref{submat_pi_2}) together, we obtain
		\begin{multline}\label{sub_mat_1_formula}
		J_\pi(\theta; m, g) = 
		\left[ \frac{1}{\pi^2} - \frac{1}{(1 - \pi )^2} \right] \sum_{i=1}^n T^\theta_i(1) + \frac{n}{(1-\pi )^2} \\ + \left( \frac{1}{(1 - \pi )} + \frac{1}{\pi} \right)^2 \left( \sum_{i=1}^n [T^\theta_i(1)]^2 - T^\theta_i(1) \right).
		\end{multline}
		
		\subsubsection*{Submatrix II}
		Denote submatrix II by $J_{\beta^m}(\theta; m, g).$ The formula for $J_{\beta^m}(\theta; m, g)$ is
		\begin{multline}\label{sub_mat_2}
		J_{\beta^m}(\theta; m, g) = -\mathbb{E} \left[\nabla_{\beta^m}^2 \mathcal{L}(\theta; m, g, p) \right] \\ + \mathbb{E}\left[\nabla_{\beta^m} \mathcal{L}(\theta; m, g, p) \right] \mathbb{E}\left[ \nabla_{\beta^m} \mathcal{L}(\theta; m, g, p) \right]^T - \mathbb{E}\left[ \nabla_{\beta^m} \mathcal{L}(\theta; m, g, p) \nabla_{\beta^m} \mathcal{L}(\theta; m, g, p)^T  \right].
		\end{multline}
		Standard GLM results imply that
		$ -\nabla_{\beta^m}^2 \mathcal{L}(\theta; m, g, p) = \tilde{X}^T ( \Delta^m V^m \Delta^m - [\Delta']^m H^m ) \tilde{X}$ and $\nabla_{\beta^m}\mathcal{L}(\theta; m, g, p) = \tilde{X}^T \Delta^m s^m.$ We compute the first term of (\ref{sub_mat_2}). The $(k,l)$th entry of this matrix is
		\begin{multline*}
		\left( \mathbb{E}\left[-\nabla_{\beta^m}^2 \mathcal{L}(\theta; m, g, p)\right]\right)[k,l] = \mathbb{E} \left\{\tilde{X}[,k]^T (\Delta^m V^m \Delta^m - [\Delta']^mH^m) \tilde{X}[,l] \right\} \\ = \sum_{i=1}^n \mathbb{E} \left\{ \tilde{x}_{i,k} (\Delta^m_{i} V^m_{i} \Delta^m_{i} - [\Delta']^m_{i} H^m_{i}) \tilde{x}_{i,l} \right\} \\ = \sum_{i=1}^n \tilde{x}_{i,k}(0) T_i^{\theta}(0) [{\Delta}^m_i(0)  {V}^m_i(0) {\Delta}^m_i(0) - [\Delta']^m_i(0) {H}^m_i(0)] \tilde{x}_{i,l}(0) \\ + \sum_{i=1}^n \tilde{x}_{i,k}(1) T_i^{\theta}(1) [ {\Delta}^m_i(1)  {V}^m_i(1) {\Delta}^m_i(1) - [{\Delta}']^m_i(1) {H}^m_i(1)] \tilde{x}_{i,l}(1) \\ = \sum_{s = 0}^1 \tilde{X}(s)[,k]^T {T}^{\theta}(s) \left[ {\Delta}^m(s) {V}^m(s) {\Delta}^m(s) - [{\Delta}']^m(s) {H}^m(s) \right] \tilde{X}(s)[,l].
		\end{multline*}
		We therefore have that
		\begin{multline}\label{sub_mat_2_1}
		\mathbb{E}\left[-\nabla_{\beta^m}^2 \mathcal{L}(\theta; m, g, p)\right] = \sum_{s=0}^1 \tilde{X}(s)^T T^{\theta}(s) \left[ {\Delta}^m(s) {V}^m(s) {\Delta}^m(s) - [{\Delta}']^m(s) {H}^m(s) \right] \tilde{X}(s).
		\end{multline}
		Next, we compute the difference of the last two terms of (\ref{sub_mat_2}). The $(k,l)$th entry is
		\begin{multline*}
		\bigg[ \mathbb{E} \left[ \nabla_{\beta^m} \mathcal{L}(\theta; m, g, p) \right] \mathbb{E} \left[ \nabla_{\beta^m} \mathcal{L}(\theta; m, g, p) \right]^T \\ - \mathbb{E} \left[\nabla_{\beta^m} \mathcal{L}(\theta; m, g, p) \nabla_{\beta^m} \mathcal{L}(\theta; m, g, p)^T \right] \bigg] [k,l] \\ 
		= \left[ \mathbb{E} \left[\tilde{X}^T \Delta^m s^m \right] \mathbb{E} \left[\tilde{X}^T \Delta^m s^m \right]^T\right][k,l] - \mathbb{E} \left[\tilde{X}^T \Delta^m s^m (s^m)^T \Delta^m \tilde{X} \right][k,l] \\
		= \mathbb{E}\left[ \tilde{X}[,k]^T \Delta^m s^m \right] \mathbb{E} \left[ \tilde{X}[,l]^T \Delta^m s^m \right] - \mathbb{E} \left[ \tilde{X}[,k]^T \Delta^m s^m (s^m)^T \Delta^m \tilde{X}[,l ] \right] \\ 
		=\mathbb{E}\left(\sum_{i=1}^n \tilde{x}_{ik} \Delta^m_i s^m_{i} \right) \mathbb{E} \left( \sum_{j=1}^n \tilde{x}_{jl} \Delta^m_j s^m_j \right) - \mathbb{E} \left( \sum_{i=1}^n \sum_{j=1}^n \tilde{x}_{ik} \Delta^m_i s^m_i s^m_j \Delta^m_j \tilde{x}_{jl} \right) \\
		= \sum_{i=1}^n \sum_{j=1}^n \mathbb{E}[ \tilde{x}_{ik} \Delta^m_is^m_i] \mathbb{E} [\tilde{x}_{jl} \Delta^m_j s^m_j]  -  \sum_{i=1}^n \sum_{j=1}^n \mathbb{E} [ \tilde{x}_{ik} \Delta^m_i s^m_i s^m_j \Delta^m_j \tilde{x}_{jl}] \\
		= \sum_{i=1}^n \sum_{j=1}^n \mathbb{E}[ \tilde{x}_{ik} \Delta^m_i s^m_i] \mathbb{E} \left[\tilde{x}_{jl} \Delta^m_j s^m_j \right]  - \sum_{i \neq j} \mathbb{E} [ \tilde{x}_{ik} \Delta^m_i s^m_i] \mathbb{E}[s^m_j \Delta^m_j \tilde{x}_{jl}] \\ - \sum_{i=1}^n \mathbb{E}[\tilde{x}_{ik} \Delta^m_i s^m_i s^m_i \Delta^m_i \tilde{x}_{il}] \\ 
		= \sum_{i=1}^n \mathbb{E}[\tilde{x}_{ik} \Delta^m_i s^m_i ] \mathbb{E} [ \tilde{x}_{il} \Delta^m_i s^m_i] - \sum_{i=1}^n \mathbb{E}[\tilde{x}_{ik} (\Delta_i^m)^2 (H_i^m)^2 \tilde{x}_{il}] \\ = \sum_{i=1}^n \left[\tilde{x}_{ik}(0) {\Delta}^m_i(0) T_i^{\theta}(0) {H}^m_i(0) + \tilde{x}_{ik}(1) {\Delta}^m_i(1) T_i^{\theta}(1) {H}^m_i(1) \right] \\ \cdot \left[ \tilde{x}_{il}(0) {\Delta}^m_i(0) T_i^{\theta}(0) {H}^m_i(0) + \tilde{x}_{il}(1) {\Delta}^m_i(1) T_i^{\theta}(1) {H}^m_i(1) \right] \\ - \sum_{i=1}^n \left[ \tilde{x}_{ik}(0) T_i^{\theta}(0) (\Delta_i^m(0))^2 ({H}_i^m(0))^2 \tilde{x}_{il}(0)  + \tilde{x}_{ik}(1) T_i^{\theta}(1) ({\Delta}^m_i(1))^2 ({H}_i^m(1))^m \tilde{x}_{il}(1) \right] \\ = \sum_{s=0}^1 \sum_{t=0}^1 \left[ \sum_{i=1}^n \tilde{x}_{ik}(s) T^{\theta}_i(s) {\Delta}^m_i(s) {H}^m_i(t) T_i^{\theta}(t){\Delta}^m_i(t) {H}^m_i(t) \tilde{x}_{il}(t) \right] \\ - \sum_{s=0}^1 \left[\sum_{i=1}^n \tilde{x}_{ik}(s) T_i^\theta(s) ({\Delta}^m_i(s))^2 ({H}_i^m(s))^2 \tilde{x}_{il}(s) \right] \\ = \sum_{s=0}^1 \sum_{t=0}^1 \tilde{X}(s)[,k]^T {T}^\theta(s) {\Delta}^m(s) {H}^m(s) {T}^\theta(t) {\Delta}^m(t) {H}^m(t) \tilde{X}(k)[,l] \\ - \sum_{s=0}^1{X}(s)[,k]^T {T}^\theta(s) ({\Delta}^m(s))^2 ({H}^m(s))^2 \tilde{X}(s)[,l].
		\end{multline*}
		The sum of the last two terms on the right-hand side of (\ref{sub_mat_2}) is therefore
		\begin{multline}\label{sub_mat_2_2}
		\mathbb{E} \left[ \nabla_{\beta^m} \mathcal{L}(\theta; m, g, p) \right] \mathbb{E} \left[ \nabla_{\beta^m} \mathcal{L}(\theta; m, g, p) \right]^T - \mathbb{E} \left[\nabla_{\beta^m} \mathcal{L}(\theta; m, g, p) \nabla_{\beta^m} \mathcal{L}(\theta; m, g, p)^T \right] \\ =
		\sum_{s=0}^1 \sum_{t=0}^1 \tilde{X}(s)^T {T}^\theta(s) {\Delta}^m(s) {H}^m(s) T^{\theta}(t) {\Delta}^m(t) {H}^m(t) \tilde{X}(t) \\ - \sum_{s=0}^1 \tilde{X}(s)^T {T}^\theta(s) ({\Delta}^m(s))^2 ({H}^m(s))^2 \tilde{X}(s). \end{multline}
		Combining (\ref{sub_mat_2}), (\ref{sub_mat_2_1}), (\ref{sub_mat_2_2}), we find that
		\begin{multline}\label{sub_mat_2_formula}
		J_{\beta^m}(\theta; m, g) = \sum_{s=0}^1 \tilde{X}(s)^T T^{\theta}(s) \left[ {\Delta}^m(s) {V}^m(s) {\Delta}^m(s) - [{\Delta}']^m(s) {H}^m(s) \right] \tilde{X}(s) \\ + \sum_{s=0}^1 \sum_{t=0}^1 \tilde{X}(s)^T {T}^\theta(s) {\Delta}^m(s) {H}^m(s) {T}^\theta(t) {\Delta}^m(t) {H}^m(t) \tilde{X}(t) \\ - \sum_{s=0}^1 \tilde{X}(s)^T T^{\theta}(s) ({\Delta}^m(s))^2 ({H}^m(s))^2 \tilde{X}(s).
		\end{multline}
		
		\subsubsection*{Submatrix III}
		Denote submatrix III by $J_{\beta^g}(\theta; m, g).$ The formula for sub-matrix III is similar to that of sub-matrix II (\ref{sub_mat_2_formula}). Substituting $g$ for $m$ in this equation yields
		\begin{multline}\label{sub_mat_3_formula}
		J_{\beta^g}(\theta; m, g) = \sum_{s=0}^1 \tilde{X}(s)^T T^{\theta}(s) \left[{\Delta}^g(s) {V}^g(s) {\Delta}^g(s) - [{\Delta}']^g(s) {H}^g(s) \right] \tilde{X}(s) \\ + \sum_{s=0}^1 \sum_{t=0}^1 \tilde{X}(s)^T {T}^\theta(s) {\Delta}^g(s) {H}^g(s) {T}^\theta(t) {\Delta}^g(t) {H}^g(t) \tilde{X}(t) \\ - \sum_{s=0}^1 \tilde{X}(s)^T T^{\theta}(s) ({\Delta}^g(s))^2 ({H}^g(s))^2 \tilde{X}(s).
		\end{multline}
		
		\subsubsection*{Submatrix IV}
		Denote sub-matrix IV by $J_{(\beta^g, \beta^m)}(\theta; m, g)$. The formula for $J_{(\beta^g, \beta^m)}(\theta; m, g)$ is 
		\begin{multline}\label{sub_mat_4}
		J_{(\beta^g, \beta^m)}(\theta; m,g) = \mathbb{E} \left[-\nabla_{\beta^g} \nabla_{\beta^m} \mathcal{L}(\theta; m, g, p) \right] \\ + \mathbb{E}\left[ \nabla_{\beta^{g}}\mathcal{L}(\theta ; m,g,p) \right] \mathbb{E} \left[\nabla_{\beta^m}\mathcal{L} (\theta ; m,g,p)  \right]^T - \mathbb{E} \left[ \nabla_{\beta^{g}}\mathcal{L} (\theta; m,g,p) \nabla_{\beta^m}\mathcal{L}(\theta; m,g,p)^T  \right].
		\end{multline}
		First, we have that
		\begin{equation}\label{sub_mat_4_1}
		\mathbb{E}\left[-\nabla_{\beta^g} \nabla_{\beta^m} \mathcal{L}(\theta; m, g, p) \right] = 0,
		\end{equation}
		as differentiating $\mathcal{L}$ with respect to $\beta^g$ yields a vector that is a function of $\beta^g$, and differentiating this vector with respect to $\beta^m$ yields $0$. Next, recall from GLM theory that $\nabla_{\beta^g} \mathcal{L}(\theta; m, g, p) =   \tilde{X}^T\Delta^g s^g$ and $\nabla_{\beta^m} \mathcal{L}(\theta; m, g, p) = \tilde{X}^T \Delta^m s^m.$ The $(k,l)$th entry of the last two terms of (\ref{sub_mat_4}) is
		\begin{multline}\label{sub_mat_4_2}
		\bigg[ \mathbb{E} \left[\nabla_{\beta^g} \mathcal{L}(\theta; m, g, p) \right] \mathbb{E} \left[\nabla_{\beta^m} \mathcal{L}(\theta; m, g, p) \right]^T \\ - \mathbb{E} \left[ \nabla_{\beta^g} \mathcal{L}(\theta; m, g, p) \nabla_{\beta^m} \mathcal{L}(\theta; m, g, p)^T \right] \bigg][k,l] \\ 
		= \left[ \mathbb{E} \left[ \tilde{X}^T \Delta^g s^g \right]\mathbb{E} \left[ \tilde{X}^T \Delta^m s^m \right]^T\right][k,l] - \mathbb{E} \left[ \tilde{X}^T \Delta^g s^g (s^m)^T \Delta^m \tilde{X} \right][k,l] \\ 
		= \mathbb{E}\left[\tilde{X}[,k]^T \Delta^g s^g \right] \mathbb{E} \left[\tilde{X}[,l]^T \Delta^m s^m \right] - \mathbb{E} \left[\tilde{X}[,k]^T \Delta^g s^g (s^m)^T \Delta^m \tilde{X}[,l ] \right] \\
		=\mathbb{E}\left( \sum_{i=1}^n \tilde{x}_{ik} \Delta^g_i s^g_i \right) \mathbb{E} \left( \sum_{j=1}^n \tilde{x}_{jl} \Delta^m_j s^m_j \right) - \mathbb{E} \left( \sum_{i=1}^n \sum_{j=1}^n \tilde{x}_{ik} \Delta^g_i s^g_i s^m_j \Delta^m_j \tilde{x}_{jl} \right) \\ 
		= \sum_{i=1}^n \sum_{j=1}^n \mathbb{E}[\tilde{x}_{ik} \Delta^g_is^g_i] \mathbb{E}[ \tilde{x}_{jl} \Delta^m_j s^m_j ] - \sum_{i=1}^n \sum_{j=1}^n \mathbb{E}[ \tilde{x}_{ik} \Delta^g_i s^g_i s^m_j \Delta^m_j \tilde{x}_{jl}]  \\
		= \sum_{i=1}^n \sum_{j=1}^n \mathbb{E}[ \tilde{x}_{ik} \Delta^g_i s^g_i] \mathbb{E} \left[\tilde{x}_{jl} \Delta^m_j s^m_j \right]  - \sum_{i \neq j} \mathbb{E}[\tilde{x}_{ik} \Delta^g_i s^g_i] \mathbb{E}[\tilde{x}_{jl}\Delta^m_j  s^m_j ] \\ - \sum_{i=1}^n \mathbb{E}[\tilde{x}_{ik} \Delta^g_i s^g_i s^m_i \Delta^m_i \tilde{x}_{il}] \\
		= \sum_{i=1}^n \mathbb{E}[\tilde{x}_{ik} \Delta^g_i H^g_i] \mathbb{E}[\tilde{x}_{il} \Delta_i^m H^m_i] - \sum_{i=1}^n \mathbb{E}[\tilde{x}_{ik} H_i^g \Delta_i^g \Delta_i^m H_i^m \tilde{x}_{il}] \\ 
		= \sum_{i=1}^n \left[\tilde{x}_{ik}(0) {\Delta}^g_i(0) T^\theta_i(0) {H}^g_i(0) + \tilde{x}_{ik}(1) {\Delta}^g_i(1) T^\theta_i(1) {H}^g_i(1)\right] \\ 
		\cdot \left[\tilde{x}_{il}(0) {\Delta}^m_i(0) T^\theta_i(0) {H}^m_i(0) + \tilde{x}_{il}(1) {\Delta}^m_i(1) T^\theta_i(1) {H}^m_i(1)\right] 
		\\ - \sum_{i=1}^n [\tilde{x}_{ik}(0) T^\theta_i(0) {\Delta}^g_i(0) {H}^g_i(0) {\Delta}^m_i(0) {H}^m_i(0) \tilde{x}_{il}(0) \\ + \tilde{x}_{ik}(1) T^\theta_i(1) {\Delta}^g_i(1) {H}^g_i(1) {\Delta}^m_i(1) {H}^m_i(1) \tilde{x}_{il}(1) ] 
		\\ = \sum_{s=0}^1 \sum_{t=0}^1 \left[\sum_{i=1}^n \tilde{x}_{ik}(s) T^\theta_i(s) {\Delta}^g_i(s) {H}^g_i(s) T^\theta_i(t){\Delta}^m_i(t) {H}^m_i(t) \tilde{x}_{il}(t) \right]
		\\ - \sum_{s=0}^1 \left[\sum_{i=1}^n \tilde{x}_{ik}(s) T^\theta_i(s) {\Delta}^g_i(s) {H}^g_i(s) {\Delta}^m_i(s) {H}^m_i(s) \tilde{x}_{il}(s)\right] 
		\\ = \sum_{s=0}^1 \sum_{t=0}^1 \left[ \tilde{X}(s)[,k]^T T^\theta(s) {\Delta}^g(s) {H}^g(s) T^\theta(t){\Delta}^m(t) {H}^m(t) \tilde{X}(t)[,l] \right]\\ - \sum_{s=0}^1 \left[ \tilde{X}[,k]^T T^\theta(s) {\Delta}^g(s) {H}^g(s) {\Delta}^m(s) {H}^m(s) \tilde{X}[,l](s)\right].
		\end{multline}
		Combining (\ref{sub_mat_4}), (\ref{sub_mat_4_1}), and (\ref{sub_mat_4_2}) produces
		\begin{multline}\label{sub_mat_4_formula}
		J_{(\beta^g, \beta^m)}(\theta; m, g) = \sum_{s=0}^1 \sum_{t=0}^1 \tilde{X}(s)^T  T^\theta(s) {\Delta}^g(s) {H}^g(s) T^\theta(t){\Delta}^m(t) {H}^m(t) \tilde{X}(t) \\ - \sum_{s=0}^1 \tilde{X}(s)^T T^\theta(s) {\Delta}^g(s) {H}^g(s) {\Delta}^m(s) {H}^m(s) \tilde{X}(s).
		\end{multline}
		\subsubsection*{Submatrix V}
		Denote submatrix V by $J_{(\beta^m,\pi)}(\theta; m, g).$ The formula for $J_{(\beta^m,\pi)}(\theta; m, g)$ is
		\begin{multline}\label{sub_mat_5}
		J_{(\beta^m,\pi)}(\theta; m, g) = \mathbb{E} \left[ - \nabla_{\beta^m} \nabla_{ \pi } \mathcal{L}(\theta; m, g, p) \right] \\ + \mathbb{E}\left[ \nabla_{\beta^m}\mathcal{L}(\theta ; m,g,p) \right] \mathbb{E} \left[ \nabla_{\pi}\mathcal{L}(\theta ; m,g,p) \right]^T  - \mathbb{E} \left[ \nabla_{\beta^m}\mathcal{L}(\theta; m,g,p) \nabla_{\pi}\mathcal{L}(\theta; m,g,p)^T \right].
		\end{multline}
		We have that
		\begin{equation}\label{sub_mat_5_1}
		\mathbb{E} \left[ - \nabla_{\beta^m} \nabla_{ \pi } \mathcal{L}(\theta; m, g, p) \right] = 0,
		\end{equation}
		as $\beta^m$ and $\pi$ separate in the log likelihood. Next, set $a \equiv 1/\pi + 1/(1 - \pi)$ and $b \equiv n/(1 - \pi).$ Recall from GLM theory that
		$\nabla_{\beta^m} \mathcal{L}(\theta; m, g, p) = \tilde{X}^T \Delta^m s^m$ and from (\ref{d_L_d_pi}) that
		$a \sum_{i=1}^n p_i - b.$
		The $k$th entry of the last two terms of (\ref{sub_mat_5}) is
		\begin{multline}\label{sub_mat_5_2}
		\mathbb{E} \left[\nabla_\pi \mathcal{L}(\theta; m, g, p) \right] \mathbb{E}\left[\nabla_{\beta^m} \mathcal{L}(\theta; m, g, p)[k] \right] - \mathbb{E} \left[\nabla_{\pi}\mathcal{L}(\theta; m,g,p) \nabla_{\beta^m}\mathcal{L}(\theta; m,g,p)[k] \right] \\= \left(\mathbb{E} \left[ a \sum_{i=1}^n p_i - b \right] \right) \left(\mathbb{E}\left[ \tilde{X}[,k]^T \Delta^m s^m \right] \right) - \mathbb{E} \left[ \left( a \sum_{i=1}^n p_i - b \right) \tilde{X}[,k]^T \Delta^m s^m \right] \\ = \left( a \sum_{i=1}^n \mathbb{E}[p_i] - b \right) \left( \sum_{j=1}^n \mathbb{E} [ \tilde{x}_{jk}\Delta^m_js^m_j] \right) - \mathbb{E} \left[ \left( a \sum_{i=1}^n p_i - b \right) \left( \sum_{j=1}^n \tilde{x}_{jk} \Delta^m_j s^m_j \right) \right] \\ = a \sum_{i=1}^n \sum_{j=1}^n \mathbb{E} [p_i] \mathbb{E}[ \tilde{x}_{jk} \Delta^m_j s^m_j] - b \sum_{j=1}^n \mathbb{E}[\tilde{x}_{jk} \Delta^m_j s^m_j] \\ - \left[ a \sum_{i=1}^n \sum_{j=1}^n \mathbb{E} [ p_i \tilde{x}_{jk} \Delta^m_j s^m_j] - b \sum_{j=1}^n \mathbb{E}[\tilde{x}_{jk} \Delta^m_j s^m_j] \right] \\ =  a \sum_{i=1}^n \sum_{j=1}^n \mathbb{E}[p_i] \mathbb{E}[\tilde{x}_{jk} \Delta^m_j s^m_j] - a\sum_{i \neq j} \mathbb{E}[p_i] \mathbb{E}[\tilde{x}_{jk} \Delta^m_j s^m_j] - a\sum_{i=1}^n \mathbb{E}[ p_i \tilde{x}_{ik} \Delta^m_i s^m_i ] \\ = a \sum_{i=1}^n \mathbb{E}[p_i] \mathbb{E}[ \tilde{x}_{ik} \Delta^m_i s^m_i] - a \sum_{i=1}^n \mathbb{E}[p_i \tilde{x}_{ik} \Delta^m_i s^m_i] \\ = a \sum_{i=1}^n T^\theta_i(1) [T^\theta_i(0) \Delta^m_i(0) s^m_i(0) \tilde{x}_{ik}(0) + T^\theta_i(1) \Delta^m_i(1) s^m_i(1) \tilde{x}_{ik}(1)] - a \sum_{i=1}^n T^\theta_i(1)\Delta^m_i(1)s^m_i(1)\tilde{x}_{ik}(1) \\ = a \sum_{i=1}^n T^\theta_i(0)T^\theta_i(1) \Delta_i^m(0)H^m_i(0)\tilde{x}_{ik}(0) \\ + a \sum_{i=1}^n \left( [T^\theta_i(1)]^2 \Delta^m_i(1)H^m_i(1) - T^\theta_i(1)\Delta^m_i(1)H^m_i(1) \right) \tilde{x}_{ik}(1)  \\ =a \left[ \sum_{i=1}^n T^\theta_i(0) T^\theta_i(1) \Delta^m_i(0) H^m_i(0) \tilde{x}_{ik}(0) + \sum_{i=1}^n T^\theta_i(1)\Delta^m_i(1)H^m_i(1)[T^\theta_i(1) - 1] \tilde{x}_{ik}(1) \right] \\ = a \left[ \sum_{i=1}^n T^\theta_i(0) T^\theta_i(1) \Delta^m_i(0) H^m_i(0) \tilde{x}_{ik}(0) - \sum_{i=1}^n T^\theta_i(0) T^\theta_i(1) \Delta^m_i(1) H^m_i(1) \tilde{x}_{ik}(1) \right] \\ = a\left(\tilde{X}(0)[,k]^T w^m(0) - \tilde{X}(1)[,k]^T w^m(1)  \right).
		\end{multline}
		Combining (\ref{sub_mat_5}), (\ref{sub_mat_5_1}), and (\ref{sub_mat_5_2}), we conclude that
		\begin{equation}\label{sub_mat_5_formula} J_{(\beta^m, \pi)}(\theta; m, g, p) = \left( \frac{1}{\pi} + \frac{1}{1 - \pi} \right) \left( \tilde{X}(0)^T w^m(0) - \tilde{X}(1)^T w^m(1)\right). \end{equation}
		
		\subsubsection*{Submatrix VI}
		Denote submatrix VI by $J_{(\beta^g,\pi)}(\theta; m, g).$ Calculations similar to those for submatrix V show that
		\begin{equation}\label{sub_mat_6_formula} J_{(\beta^g, \pi)}(\theta; m, g, p) = \left(\frac{1}{\pi} + \frac{1}{1 - \pi} \right) \left( \tilde{X}(0)^T w^g(0) - \tilde{X}(1)^T w^g(1)\right). \end{equation}
		
		\subsubsection*{Combining submatrices}
		To summarize, the formulas for submatrices I-VI are as follows:
		\begin{itemize}
			\item[I]\begin{multline*}
			J_\pi(\theta; m, g) = 
			\left[ \frac{1}{\pi^2} - \frac{1}{(1 - \pi )^2} \right] \sum_{i=1}^n T^\theta_i(1) + \frac{n}{(1-\pi )^2} \\ + \left( \frac{1}{(1 - \pi )} + \frac{1}{\pi} \right)^2 \left( \sum_{i=1}^n [T^\theta_i(1)]^2 - T^\theta_i(1) \right).
			\end{multline*}
			\item[II] \begin{multline*}
			J_{\beta^m}(\theta; m, g) = \sum_{s=0}^1 \tilde{X}(s)^T T^{\theta}(s) \left[ {\Delta}^m(s) {V}^m(s) {\Delta}^m(s) - [{\Delta}']^m(s) {H}^m(s) \right] \tilde{X}(s) \\ + \sum_{s=0}^1 \sum_{t=0}^1 \tilde{X}(s)^T {T}^\theta(s) {\Delta}^m(s) {H}^m(s) {T}^\theta(t) {\Delta}^m(t) {H}^m(t) \tilde{X}(t) \\ - \sum_{s=0}^1 \tilde{X}(s)^T T^{\theta}(s) ({\Delta}^m(s))^2 ({H}^m(s))^2 \tilde{X}(s).
			\end{multline*}
			\item[III] \begin{multline*}
			J_{\beta^g}(\theta; m, g) = \sum_{s=0}^1 \tilde{X}(s)^T T^{\theta}(s) \left[{\Delta}^g(s) {V}^g(s) {\Delta}^g(s) - [{\Delta}']^g(s) {H}^g(s) \right] \tilde{X}(s) \\ + \sum_{s=0}^1 \sum_{t=0}^1 \tilde{X}(s)^T {T}^\theta(s) {\Delta}^g(s) {H}^g(s) {T}^\theta(t) {\Delta}^g(t) {H}^g(t) \tilde{X}(t) \\ - \sum_{s=0}^1 \tilde{X}(s)^T T^{\theta}(s) ({\Delta}^g(s))^2 ({H}^g(s))^2 \tilde{X}(s).
			\end{multline*}
			\item[IV] \begin{multline*}
			J_{(\beta^g, \beta^m)}(\theta; m, g) = \sum_{s=0}^1 \sum_{t=0}^1 \tilde{X}(s)^T  T^\theta(s) {\Delta}^g(s) {H}^g(s) T^\theta(t){\Delta}^m(t) {H}^m(t) \tilde{X}(t) \\ - \sum_{s=0}^1 \tilde{X}(s)^T T^\theta(s) {\Delta}^g(s) {H}^g(s) {\Delta}^m(s) {H}^m(s) \tilde{X}(s).
			\end{multline*}
			\item[V] $$ J_{(\beta^m, \pi)}(\theta; m, g, p) = \left( \frac{1}{\pi} + \frac{1}{1 - \pi} \right) \left( \tilde{X}(0)^T w^m(0) - \tilde{X}(1)^T w^m(1)\right).  $$
			\item[VI] $$ J_{(\beta^g, \pi)}(\theta; m, g, p) = \left(\frac{1}{\pi} + \frac{1}{1 - \pi} \right) \left( \tilde{X}(0)^T w^g(0) - \tilde{X}(1)^T w^g(1)\right).$$
		\end{itemize}
		We stitch these pieces together and transpose submatrices IV, V, and VI to produce the whole information matrix $J(\theta; m, g)$. Evaluating this matrix at the EM estimate $\theta^\textrm{EM}$ and inverting yields the asymptotic covariance matrix, which we can use to compute standard errors.
		
		\subsection{Implementation}
		To evaluate the observed information matrix, we need to compute the matrices $\Delta^m(j),$ $[\Delta']^m(j),$ $V^m(j),$ and $H^m(j)$ and the vectors $s^m(j)$ and $w^m(j)$ for $j \in \{0,1\}$. We likewise need to compute the analogous gRNA quantities. The procedure that we propose for this purpose is general, but for concreteness, we describe how to implement this procedure using the \texttt{glm} function in R by extending base family objects. We implicitly condition on $p_i$, $z^m_i$, and $o^m_i$.
		
		An \texttt{R} family object contains several functions, including \texttt{linkinv}, \texttt{variance}, and \texttt{mu.eta}. \texttt{linkinv} is the inverse link function $r_m^{-1}$. \texttt{variance} takes as an argument the mean $\mu^m_i$ of the $i$th example and returns its variance $[\sigma_i^m]^2$. \texttt{mu.eta} is the derivative of the inverse link function $[r^{-1}_m]^{'}$. We extend the \texttt{R} family object by adding two additional functions: \texttt{skewness} and \texttt{mu.eta.prime}. \texttt{skewness} returns the skewness $\gamma^m_i$ of the distribution as a function of the mean $\mu_i$, i.e. $$\texttt{skewness}\left(\mu_i\right) = \mathbb{E} \left[\left(\frac{m_i - \mu_i^m}{ \sigma_i^m}\right)^3\right] := \gamma_i^m.$$ Finally, \texttt{mu.eta.prime} is the second derivative of the inverse link function $[r^{-1}_m]''.$ Algorithm \ref{algo:computing_info_matrices} computes the matrices $\Delta^m(j)$, $[\Delta']^m(j)$, $V^m(j)$, and $H^m(j)$ and vector $s^m(j)$ for given $\beta_m$ and given family object. (The vector $w^m(j)$ can be computed in terms of $\Delta^m(j)$ and $H^m(j)$.) We use $\sigma^m_i(j)$ (resp. $\gamma^m_i(j)$) to refer to the standard deviation (resp. skewness) of the gene expression distribution the $i$th cell when the perturbation $p_i$ is set to $j$.

		All steps of the algorithm are obvious except the calculation of $h'_m(l^m_i(j))$ (line 6), $h''(l^m_i(j))$ (line 9), and $V^m_i(j)$ (line 12). We omit the $(j)$ notation for compactness. First, we prove the correctness of the expression for $h'_m(l^m_i)$. Recall the basic GLM identities
		\begin{equation}\label{computing_info_matrix_1}
		 \psi_m''(\eta_i^m) =  [\sigma^m_i]^2
		\end{equation}
		and, for all $t \in \mathbb{R}$, 
		\begin{equation}\label{computing_info_matrix_2}
		r_m^{-1}(t) = \psi_m'(h_m(t)).
		\end{equation}
		Differentiating (\ref{computing_info_matrix_2}) in $t$, we find that
		\begin{equation}\label{computing_info_matrix_3}
		(r_m^{-1})'(t) = \psi_m''(h_m(t))h_m'(t) \iff h_m'(t) = \frac{(r_m^{-1})'(t) }{\psi_m''(h_m(t))}.
		\end{equation}
	Finally, plugging in $l^m_i$ for $t$,
		$$ h_m'(l_i) = \frac{(r_m^{-1})'(l^m_i)}{\psi''(h_m(l^m_i))} = \frac{(r_m^{-1})'(l^m_i)}{\psi_m''(\eta^m_i)} = \textrm{ by (\ref{computing_info_matrix_1}) } \frac{(r_m^{-1})'(l_i^m)}{[\sigma_i^m]^2}.$$
		
		Next, we prove the correctness for the expression for $h_m''(l_i^m)$. Recall the exponential family identity 
		\begin{equation}\label{computing_info_matrix_4}
		\psi'''_m(\eta^m_i) = \gamma^m_i  ([\sigma^m_i]^2)^{3/2}.
		\end{equation}
		Differentiating (\ref{computing_info_matrix_3}) in $t$, we obtain
		$$(r_m^{-1})''(t) = \psi_m'''(h_m(t)) [h_m'(t)]^2 + \psi_m''(h_m(t)) h_m''(t) \iff h_m''(t) =\frac{(r_m^{-1})''(t) - \psi'''(h_m(t))[h_m'(t)]^2}{\psi_m''(h_m(t))}.$$ Plugging in $l^m_i$ for $t$, we find that
		\begin{equation*}
		h_m''(l^m_i) = \frac{(r_m^{-1})''(l^m_i) - \psi_m'''(\eta^m_i) [h_m'(l_i^m)]^2}{[\sigma_i^m]^2} = \textrm{ (by \ref{computing_info_matrix_4}) }  \frac{(r_m^{-1})''(l^m_i) - ([\sigma_i^m]^2)^{3/2} (\gamma_i^m) [h_m'(l^m_i)]^2 }{[\sigma_i^m]^2}.
		\end{equation*}
		Finally, the expression for $V^m_i$ follows from (\ref{computing_info_matrix_1}). We can apply a similar algorithm to compute the analogous matrices for the gRNA modality. Table \ref{family_object_functions} shows the \texttt{linkinv}, \texttt{variance}, \texttt{mu.eta}, \texttt{skewness}, and \texttt{mu.eta.prime} functions for several common family objects (which are defined by a distribution and link function). 
				
			\begin{algorithm}
			\caption{Computing the matrices $\Delta^m(j)$, $[\Delta']^m(j)$, $V^m(j)$, $H^m(j)$, and $s^m(j)$ given given $\beta_m$.}\label{algo:computing_info_matrices}
			\begin{algorithmic}[3]
				\Require A coefficient vector $\beta_m$; data $[m_1, \dots, m_n]$, $[o^m_1, \dots, o^m_n]$, and $[z_1, \dots, z_n]$; and a family object containing functions \texttt{linkinv}, \texttt{variance}, \texttt{mu.eta}, \texttt{mu.eta.prime}, and \texttt{skewness}.
				\For{$j \in \{0, 1\}$}
				\For{$i \in \{1, \dots, n\}$}
				\State $l^m_i(j) \gets \langle \beta_m, \tilde{x}_i(j) \rangle + o^m_i$
				\State $\mu^m_i(j) \gets \texttt{linkinv}(l^m_i(j))$
				\State $[\sigma_i^m(j)]^2 \gets \texttt{variance}(\mu_i^m(j))$
				\State $h_m'(l_i^m(j)) \gets \texttt{mu.eta}(l_i^m(j))/[\sigma_i^m(j)]^2$
				\State $\gamma^m_i(j) \gets \texttt{skewness}(\mu^m_i(j))$
				\State $[r_m^{-1}]''(l_i^m(j)) \gets \texttt{mu.eta.prime}(l^m_i(j))$
				\State $$h_m''(l_i^m(j)) \gets \frac{[r^{-1}]''(l_i^m(j)) - [([\sigma_i^m(j)]^2)^{3/2}][\gamma^m_i(j)] [h_m'(l_i^m(j))]^2}{[\sigma_i^m(j)]^2}$$
				\Comment Assign quantities to matrices
				\State $\Delta_{i}^m(j) \gets h_m'( l_i^m(j))$
				\State $[\Delta']^m_{i}(j) \gets h''(l^m_i(j))$
				\State $V^m_{i}(j) \gets [\sigma^m_i(j)]^2$
				\State $H^m_{i}(j) \gets s^m_i(j) \gets m_i - \mu^m_i(j)$
				\EndFor
				\EndFor
			\end{algorithmic}
		\end{algorithm}
		
		\begin{table}
			\centering
			\caption{\texttt{linkinv}, \texttt{variance}, \texttt{mu.eta}, \texttt{skewness}, \texttt{mu.eta.prime} for common family objects (i.e., pairs of distributions and link functions).}\label{family_object_functions}
			\begin{tabular}{|C{2.5cm}|C{3.5cm}|C{3.4cm}|C{2.5cm}|}
				\hline 
				& Gaussian response, identity link & Poisson response, log link & NB response ($s > 0$ fixed), log link \\ 
				\hline 
				\texttt{linkinv} & $x$ & $\exp(x)$ & $\exp(x)$  \\ 
				\hline 
				\texttt{variance} & $x$ & $x$ & $x + x^2/s$ \\ 
				\hline 
				\texttt{mu.eta} & $1$ & $x$  & $\exp(x)$ \\ 
				\hline 
				\texttt{skewness} & $0$ & $x^{-1/2}$ & $\frac{2 x + s}{\sqrt{s x} \sqrt{x + s}}$ \\ 
				\hline 
				\texttt{mu.eta.prime} & $0$ & $\exp(x)$ & $\exp(x)$ \\ 
				\hline 
			\end{tabular}
		\end{table}
		
\section{Statistical accelerations and computing}\label{sec:statistical_accelerations}
		
\subsection{Statistical accelerations}

We describe in detail the procedure for obtaining the pilot parameter estimates $(\pi^\textrm{pilot}, \beta_m^\textrm{pilot}, \beta_g^\textrm{pilot})$. This procedure consists of two subroutines, which we label Algorithm \ref{algo:pilot_estimates_1} and Algorithm \ref{algo:pilot_estimates_2}. The first step (Algorithm \ref{algo:pilot_estimates_1}) is to obtain good parameter estimates for $[\beta^m_0, \gamma_m]^T$ and $[\beta^g_0, \gamma_g]^T$ via regression. Recall that the underlying gene expression parameter vector $\beta_m$ is $\beta_m = [\beta^m_0, \beta^m_1, \gamma_m]^T \in \mathbb{R}^d$, where $\beta^m_0$ is the intercept, $\beta^m_1$ is the effect of the perturbation, and $\gamma_m^T$ is the effect of the technical factors. To produce estimates $[\beta^m_0]^\textrm{pilot}$ and $[\gamma_m^T]^\textrm{pilot}$, we regress the gene expressions $m$ onto the technical factors $X$. The intuition for this procedure is as follows: the probability of perturbation $\pi$ is very small. Therefore, the true log likelihood is approximately equal to the log likelihood that results from omitting $p_i$ from the model:
\begin{multline*}
\sum_{i=1}^n f_m(m_i; \eta^m_i) = \underbrace{\sum_{i : p_i =1} f_m(m_i; h_m(\beta_0 + \beta_1 + \gamma^T z_i + o^m_i))}_\textrm{few terms}  + \underbrace{\sum_{i : p_i = 0} f_m(m_i; h_m(\beta_0 + \gamma^T z_i + o^m_i))}_\textrm{many terms} \\ \approx \sum_{i = 1}^n f_m(m_i; h_m(\beta_0 + \gamma^T z_i + o^m_i)).
\end{multline*}
We similarly can obtain pilot estimates $[\beta^g_0]^\textrm{pilot}$ and $[\gamma^T_g]^\textrm{pilot}$ by regressing the gRNA counts $g$ onto the technical factors $X$. We extract the fitted values (on the scale of the linear component) for use in a subsequent step: $\hat{f}^k_i = [\beta^k_0]^\textrm{pilot} + \langle [\gamma^T_k]^\textrm{pilot}, z_i \rangle + o^k_i,$ for $k \in \{m,g\}$.

\begin{algorithm}
	\caption{Computing $[\beta_0^m]^\textrm{pilot}$, $[\gamma^T_m]^\textrm{pilot}$, $[\beta_0^g]^\textrm{pilot}$, and $[\gamma^T_g]^\textrm{pilot}$ .}\label{algo:pilot_estimates_1}
	\begin{algorithmic}[2]
		\Require Data $m$, $g$, $o^m$, $o^g$, and $X$; gene expression distribution $f_m$ and link function $r_m$; gRNA expression distribution $f_g$ and link function $r_g$; number of EM starts $B$.
		\For {$k \in \{m,g\}$}
		\State \multiline{Fit a GLM $GLM_k$ with responses $k$, offsets $o^k$, design matrix $X$, distribution $f_k$, and link function $r_k$.}
		\State Set $[\beta_0^k]^\textrm{pilot}$ and $[\gamma_k^T]^\textrm{pilot}$ to the fitted coefficients of $GLM_k$.
		\For{$i \in \{1, \dots, n\}$}
		\State $\hat{f}^k_i \gets [\beta_0^k]^\textrm{pilot} + \langle [\gamma_k^T]^\textrm{pilot} , z_i \rangle + o_i^k$ \Comment{untransformed fitted values}
		\EndFor
		\EndFor
		\State \textbf{return} $([\beta_0^m]^\textrm{pilot}$, $\hat{f}^m$, $[\gamma^T_m]^\textrm{pilot}$, $[\beta_0^g]^\textrm{pilot}$, $[\gamma^T_g]^\textrm{pilot}$, $\hat{f}^g)$
	\end{algorithmic}
\end{algorithm}

Next, we obtain estimates $[\beta_1^m]^\textrm{pilot},$ $[\beta_1^g]^\textrm{pilot},$ and $\pi^\textrm{pilot}$ for $\beta^m_1$, $\beta^g_1$, and $\pi$ by fitting a ``reduced'' GLM-EIV (Algorithm \ref{algo:pilot_estimates_2}). The log likelihood of the no-intercept, univariate GLM with predictor $p_i$ and offset $\hat{f}^m_i$ is approximately equal to the true log likelihood:
$$ \sum_{i=1}^n f_m(m_i; \eta^m_i) =
\sum_{i=1}^n f_m(m_i; h_m(\beta_0 + \beta_1 p_i + \gamma^T z_i + o^m_i)) \approx \sum_{i=1}^n f_m(m_i; h_m( \beta_1 p_i + \hat{f}^m_i)).$$

\begin{algorithm}
			\caption{Computing $\pi^\textrm{pilot}$, $[\beta^m_1]^\textrm{pilot}$, $[\beta^m_1]^\textrm{pilot}$.}\label{algo:pilot_estimates_2}
			\begin{algorithmic}[2]
				\Require Data $m$, $g$; fitted offsets $\hat{f}^m,$ $\hat{f}^g$.
				\State \texttt{bestLik} $\gets -\infty$ \Comment{Reduced GLM-EIV}
				\For{$i \in \{1,\dots,B\}$}
				\State Randomly generate starting parameters $\pi^\textrm{curr}, [\beta_1^m]^\textrm{curr}, [\beta^g_1]^\textrm{curr}.$
				\While{Not converged}
				\For{$i \in \{1, \dots, n\}$} \Comment{E step}
				\State $T_i(1) \gets \mathbb{P}(P_i = 1 | M_i = m_i, G_i = g_i, \pi^\textrm{curr}, [\beta^g_1]^\textrm{curr}, [\beta_1^m]^\textrm{curr})$
				\State $T_i(0) \gets 1 - T_i(1)$
				\EndFor
				\State $\pi^{\textrm{curr}} \gets (1/n) \sum_{i=1}^n T_i(1)$ \Comment{M step}
				\State $w \gets [T_1(0), T_2(0), \dots, T_n(0), T_1(1), T_2(1), \dots, T_n(1)]^T$
				\For{$k \in \{g,m\}$}
				\State \multiline{Fit no-intercept, univariate GLM $GLM_k$ with predictors $[\underbrace{0, \dots, 0}_\textrm{n}, \underbrace{1, \dots, 1}_\textrm{n}]$, responses $[k,k]^T$, offsets $[\hat{f}^k, \hat{f}^k]^T,$ and weights $w$.}
				\State Set $[\beta^k_1]^\textrm{curr}$ to fitted coefficient of $GLM_k$.
				\EndFor
				\State \multiline{
					Compute log likelihood \texttt{currLik} using $\pi^\textrm{curr}$,$ [\beta^m_1]^\textrm{curr}$, and $[\beta^g_1]^\textrm{curr}.$}
				\EndWhile
				\If{\texttt{currLik} $>$ \texttt{bestLik}}
				\State bestLik $\gets$ currLik
				\State $\pi^\textrm{pilot} \gets \pi^\textrm{curr};$ $[\beta^m_1]^\textrm{pilot} \gets [\beta^m_1]^\textrm{curr}; [\beta^g_1]^\textrm{pilot} \gets [\beta^g_1]^\textrm{curr}$
				\EndIf
				\EndFor
				\State \textbf{return} $(\pi^\textrm{pilot}, [\beta^m_1]^\textrm{pilot}, [\beta^g_1]^\textrm{pilot})$
			\end{algorithmic}
\end{algorithm}

Therefore, to estimate $\beta^m_1$, $\beta^g_1$, and $\pi$, we fit a GLM-EIV model with gene expressions $m$, gRNA counts $g$, gene offsets $\hat{f}^m := [\hat{f}^m_1, \dots, \hat{f}^m_n]^T$, gRNA offsets $\hat{f}^g := [\hat{f}^g_1, \dots, \hat{f}^g_n]^T$, and \textit{no} intercept or covariate terms. Intuitively, we ``encode'' all information about technical factors, library sizes, and baseline expression levels into $\hat{f}^m$ and $\hat{f}^g$. We run the algorithm $B \approx 15$ times over randomly-selected starting values for $\beta^m$, $\beta^g$, and $\pi$ and select the solution with greatest the log likelihood.

The M step of the reduced GLM-EIV algorithm requires fitting two no-intercept, univariate GLMs with offsets. We derive analytic formulas for the MLEs of these GLMs in the three most important cases: Gaussian response with identity link, Poisson response with log link, and negative binomial response with log link (see section \ref{sec:int_plus_offset}; the latter formula is asymptotically exact). Consequently, we do not need to run the relatively slow IRLS procedure to carry out the M step of the reduced GLM-EIV algorithm. Overall, the proposed method for obtaining the full set of pilot parameter estimates requires fitting only two GLMs (via IRLS).

\subsection{Intercept-plus-offset models}\label{sec:int_plus_offset}
		
		A key step in the algorithm for computing the pilot parameter estimates (Algorithm \ref{algo:pilot_estimates_2}) is to fit a weighted, no-intercept, univariate GLM with nonzero offset terms and a binary predictor variable. We derive an analytic formula for the MLE of this GLM for three important pairs of response distributions and link functions: Gaussian response with identity link, Poisson response with log link, and negative binomial response with log link. The GLM that we seek to estimate has responses $[m,m]^T$, predictors $[\underbrace{0, \dots, 0}_\textrm{n}, \underbrace{1, \dots, 1}_\textrm{n}]$, offsets $[\hat{f}^m, \hat{f}^m],$ and weights $w = [T_1(0), \dots, T_n(0), T_1(1), \dots, T_n(1)]^T.$ Throughout, $C$ denotes a universal constant. The log likelihood of this GLM is
		\begin{multline}\label{stat_acc_1}
		\mathcal{L}(\beta_1; m) = \sum_{i=1}^n T_i(0) f_m(m_i; h_m(\beta_1 + \hat{f}^m_i )) + \sum_{i=1}^n T_i(1) f_m(m_i; h_m(\hat{f}^m_i)) \\ = \sum_{i=1}^n T_i(1) f_m(m_i; h_m(\beta_1 + \hat{f}^m_i )) + C.
		\end{multline}
		Thus, finding the MLE $\hat{\beta_1}$ is equivalent to estimating a GLM with intercept $\beta_1$, offsets $\hat{f}^m$, weights $T_i(1)$, and \textit{no} covariate terms. We term such a GLM a \textit{intercept-plus-offset} model. Below, we study intercept-plus-offset models in generality.
		
\paragraph{General formulation}	Let $\beta \in \mathbb{R}$ be an unknown constant. Let $o_1, \dots, o_n \sim \mathcal{P}_1$, where $\mathcal{P}_1$ is a distribution. Let $Y_i|o_i, \dots, Y_n|o_i$ be exponential family-distributed random variables with identity sufficient statistic. Suppose the mean $\mu_i$ of $Y_i|o_i$ is given by $r(\mu_i) = \beta + o_i,$ where $r: \mathbb{R} \to \mathbb{R}$ is a strictly increasing, differentiable link function. We call this model the \textit{intercept-plus-offset} model.
		
		We derive the (weighted) log likelihood of this model. Let $w_1, \dots, w_n \sim \mathcal{P}_2$ be weights, where $\mathcal{P}_2$ is a distribution bounded above by $1$ and below by $0$. (A special case, which corresponds to no weights, is $w_i = 1$ for all $i \in \{1, \dots, n\}$.) Throughout, we assume that $y_iw_i$ and $\exp(o_i)w_i$ have finite first moment.  Suppose the cumulant-generating function and carrying density of the exponential family distribution are $\psi:\mathbb{R} \to \mathbb{R}$ and $c: \mathbb{R} \to \mathbb{R}$, respectively. The canonical parameter $\eta_i$ of the $i$th observation is 
		\begin{equation}\label{can_param}
		\eta_i = ([\psi']^{-1} \circ r^{-1})(\beta + o_i) := h(\beta + o_i),
		\end{equation}
		and the density $f$ of $Y_i | \eta_i$ is
		$f(y_i; \eta_i) = \exp\{y_i \eta_i - \psi(\eta_i) + c(y_i)\}.$ The weighted log likelihood is
		\begin{equation}\label{m_plus_o_mle}
		\mathcal{L}(\beta;y_i) = \sum_{i=1}^n w_i\log\left[f(y_i;\eta_i)\right] = C + \sum_{i=1}^n w_i(y_i \eta _i - \psi(\eta_i)).
		\end{equation}
		Our goal is to find the weighted MLE $\hat{\beta}$ of $\beta$. We consider three important choices for the exponential family distribution and link function. % This amounts to maximizing (\ref{m_plus_o_mle}) for three specific choices of $r$ and $\psi$.
		In the first two cases -- Gaussian distribution with identity link and Poisson distribution with log link -- we find the \textit{finite-sample} maximizer of (\ref{m_plus_o_mle}); by contrast, in the third case -- negative binomial distribution with log link -- we find an \textit{asymptotically exact} maximizer.
		
		\paragraph{Gaussian}
		First, consider a Gaussian response distribution and identity link function $r(\mu) = \mu$. The cumulant-generating function $\psi$ is $\psi(\eta) = \eta^2/2$, and so, by (\ref{can_param}),
		$$h(t) = [\psi']^{-1}( r^{-1}(t)) = [\psi']^{-1}(t) = t.$$
		Plugging $\eta_i = h(\beta + o_i) = \beta + o_i$ and $\psi(\eta_i) = (1/2)(\beta + o_i)^2$ into (\ref{m_plus_o_mle}), we obtain
		$$\mathcal{L}(\beta; y) = \sum_{i=1}^n w_i (y_i(\beta + o_i) - (\beta + o_i)^2/2).$$ The derivative of this expression in $\beta$ is
		$$\frac{\partial \mathcal{L}(\beta;y)}{\partial\beta} = \sum_{i=1}^n w_i (y_i - \beta - o_i) = \sum_{i=1}^n w_i(y_i - o_i) - \beta \sum_{i=1}^n w_i.$$ Setting this quantity to 0 and solving for $\beta$, we find that the MLE $\hat{\beta}^\textrm{gauss}$ is
		$$\hat{\beta}^\textrm{gauss} = \frac{\sum_{i=1}^n w_i (y_i - o_i)}{\sum_{i=1}^n w_i}.$$
		
		\paragraph{Poisson}
		Next, consider a Poisson response distribution and log link function $r(\mu) = \log(\mu).$ The cumulant-generating function $\psi$ is $\psi(\eta) = e^\eta.$ Therefore, by (\ref{can_param}),
		$$h(t) = [\psi']^{-1}(r^{-1}(t)) = [\psi']^{-1} \left(\exp(t) \right) = \log(\exp(t)) = t.$$ Plugging $\eta_i = h(\beta + o_i) = \beta + o_i$ and $\psi(\eta_i) = \exp(\beta + o_i)$ into (\ref{m_plus_o_mle}), we obtain
		$$ \mathcal{L}(\beta; y) = \sum_{i=1}^n w_i \left( y_i(\beta + o_i) - \exp(\beta + o_i) \right).$$ The derivative of this function in $\beta$ is 
		$$\frac{\partial \mathcal{L}(\beta; y)}{\partial \beta} = \sum_{i=1}^n w_iy_i - w_i \exp(\beta + o_i) = \sum_{i=1}^n w_i y_i - \exp(\beta) \sum_{i=1}^n w_i \exp(o_i).$$
		Setting to zero and solving for $\beta$, we find that the MLE $\hat{\beta}^\textrm{pois}$ is
		\begin{equation}\label{pois_mle}
		\hat{\beta}^\textrm{pois} = \log\left(\frac{\sum_{i=1}^n w_i y_i}{\sum_{i=1}^n w_i e^{o_i}}\right).
		\end{equation}
		
		\paragraph{Negative binomial}
		Finally, we consider a negative binomial response distribution (with fixed size parameter $s > 0$) and log link function $r(\mu) = \log(\mu)$. The cumulant-generating function $\psi$ is
		$\psi(\eta) = -s \log(1 - e^\eta).$ The derivative $\psi'$ of $\psi$ is
		$$ \psi'(t) = s \left(\frac{e^t}{1 - e^t}\right) = \frac{s}{e^{-t} - 1}.$$ Define the function $\delta: \mathbb{R} \to \mathbb{R}$ by $\delta(t) = -\log\left(s/t + 1 \right).$ We see that
		$$\psi'(\delta(t)) = \frac{s}{\exp\left(\log(s/t + 1 )\right) - 1} = t,$$ implying $\delta = [\psi']^{-1}.$ By (\ref{can_param}), we have that
		$$
		h(t) = [\psi']^{-1}(r^{-1}(t)) = -\log\left(\frac{s}{\exp(t)} + 1 \right) = \log\left(\frac{\exp(t)}{s + \exp(t)}\right).
		$$
		Therefore,
		\begin{equation}\label{nb_mo_1}
		\eta_i = h(\beta + o_i) = \log\left(\frac{\exp(\beta+o_i)}{s + \exp(\beta + o_i)} \right) = \beta + o_i - \log\left(s + e^{\beta}e^{o_i}\right) = \beta - \log\left(s + e^{\beta}e^{o_i} \right) + C,
		\end{equation}
		and
		\begin{multline}\label{nb_mo_2}
		\psi(\eta_i) = -s\log\left(1 - \frac{\exp(\beta+o_i)}{s + \exp(\beta + o_i)} \right) = -s \log \left(\frac{s}{s + \exp(\beta + o_i)} \right) \\ = -s \log (s) + s \log[s + \exp(\beta + o_i)] = s \log(s + e^{s}e^{o_i}) + C.
		\end{multline}
		Plugging (\ref{nb_mo_1}) and (\ref{nb_mo_2}) into (\ref{m_plus_o_mle}), the log-likelihood (up to a constant) is
		\begin{multline*}
		\mathcal{L}(\beta; y) = \beta \sum_{i=1}^n w_i y_i - \sum_{i=1}^n w_i y_i \log(s + e^\beta e^{o_i}) - s \sum_{i=1}^n w_i \log(s + e^\beta  e^{o_i}) \\ = \beta \sum_{i=1}^n w_i y_i - \sum_{i=1}^n (y_i + s)w_i\log(s + e^\beta e^{o_i}).
		\end{multline*}
		The derivative of $\mathcal{L}$ in $\beta$ is
		$$\frac{\partial \mathcal{L}(\beta;y)}{\partial \beta} = \sum_{i=1}^n w_i y_i  - \sum_{i=1}^n \frac{w_i(y_i + s) e^{\beta} e^{o_i}}{s + e^{\beta} e^{o_i}}.$$
		Setting the derivative to zero, the equation defining the MLE is
		\begin{equation}\label{nb_mle}
		e^\beta \sum_{i=1}^n \frac{w_i e^{o_i} (y_i + s)}{e^\beta e^{o_i} + s} = \sum_{i=1}^n w_i y_i.
		\end{equation}
		We cannot solve for $\beta$ in (\ref{nb_mle}) analytically. However, we can derive an asymptotically exact solution. By the law of total expectation,
		$$ \mathbb{E} \left[\frac{w_i  e^{o_i} (y_i + s)}{e^{\beta + o_i} + s} \right] = \mathbb{E} \left[\mathbb{E}\left[\frac{w_i e^{o_i} (y_i + s) }{e^{\beta + o_i} + s} \bigg| (o_i, w_i) \right] \right] = \mathbb{E} \left[\frac{w_i e^{o_i} (e^{\beta + o_i} + s)}{ e^{\beta + o_i} + s} \right] = \mathbb{E} [w_i e^{o_i}];
		$$
		the second equality holds because $\mathbb{E}[y_i | o_i ] = \mu_i = e^{\beta + o_i}.$ Dividing by $n$ on both sides of (\ref{nb_mle}) and rearranging,
		\begin{equation}\label{nb_mo_3}
		\beta = \log\left( \frac{ (1/n) \sum_{i=1}^n w_i e^{o_i} (y_i + s)/(e^\beta e^{o_i} + s)}{ (1/n) \sum_{i=1}^n w_i y_i. } \right).
		\end{equation}
		By weak LLN, the limit (in probability) of the MLE $\hat{\beta}^\textrm{NB}$ is
		\begin{equation}
		\hat{\beta}^{\textrm{NB}} \xrightarrow{P} \log\left(\frac{\mathbb{E}[w_i y_i]}{\mathbb{E}[w_i e^{o_i}]} \right).
		\end{equation}
		But the Poisson MLE $\hat{\beta}^{\textrm{Pois}}$ (\ref{pois_mle}) converges in probability to the same limit:
		$$ \hat{\beta}^{\textrm{pois}} =  \log \left(\frac{ (1/n) \sum_{i=1}^n w_i y_i}{(1/n)\sum_{i=1}^n w_i e^{o_i}} \right) \xrightarrow{P} \log \left(\frac{\mathbb{E}[w_i y_i]}{ \mathbb{E}[w_i e^{o_i}]} \right).$$ Therefore, for large $n$, we can approximate $\hat{\beta}^{\textrm{NB}}$ by $\hat{\beta}^{\textrm{pois}}$.	
		\iffalse
		Suppose the log likelihood is unweighted (i.e., $w_i = 1$ for all $i$). We present an alternate (and simpler, more general) derivation of the approximate formula for the MLE. Let $y_i$ be a random variable with finite first moment and conditional mean $\mathbb{E}[y_i | o_i] = \exp(\beta + o_i).$ By the law of total expectation,
		$$ \mathbb{E}[y_i] = \mathbb{E}[\mathbb{E}[ y_i | o_i]] = \mathbb{E}[\exp(\beta + o_i)] = \exp(\beta) \mathbb{E}(e^{o_i}),$$ or
		\begin{equation}\label{nb_mo_4}
		\beta = \log\left(\frac{\mathbb{E}[y_i]}{E[e^{o_i}]} \right).
		\end{equation}
		Assuming the MLE is well-behaved, we have that $\hat{\beta} \xrightarrow{P} \beta.$ On the other hand, by WLLN and (\ref{nb_mo_4}),
		\begin{equation}\label{nb_mo_5}
		\log\left( \frac{(1/n) \sum_{i=1}^n y_i }{(1/n) \sum_{i=1}^n e^{o_i}} \right) \xrightarrow{P}  \log\left(\frac{ \mathbb{E}[y_i] }{ \mathbb{E}[e^{o_i}] } \right) = \beta.
		\end{equation}
		Therefore, the LHS of (\ref{nb_mo_5}) approximates the MLE $\hat{\beta}$ for large $n$.
		\fi
		\paragraph{Application to GLM-EIV}
		The GLM that we seek to estimate (\ref{stat_acc_1}) is an approximate intercept-plus-offset model: $T_1(1), \dots, T_n(1)$ are the weights $w_1,\dots, w_n$, and $\hat{f}^m_1, \dots, \hat{f}^m_n$ are the offsets $o_1, \dots, o_m$. Of course, $T_1(1), \dots, T_1(n)$ are in general dependent random variables, as are $\hat{f}^m_1, \dots, \hat{f}^m_n.$ $T_i(1)$ depends on $m_i$ and $g_i$, as well as the final parameter estimate $(\hat{\pi}, \hat{\beta}_m, \hat{\beta}_g),$ which itself is a function of $m$ and $g$; the situation is similar for the $\hat{f}^m_i$s. In practice, we find that the intercept-plus-offset model is very good approximation to the GLM (\ref{stat_acc_1}), especially when the number of cells $n$ is large. Additionally, we note that the GLM (\ref{stat_acc_1}) is fitted as a subroutine of the algorithm for producing pilot parameter estimates (Algorithm \ref{algo:pilot_estimates_2}). The quality of the pilot parameter estimates does not affect the validity of the estimation and inference procedures (Algorithm \ref{algo:em_full}), barring issues related to convergence to local optima.
		
\subsection{Computing}\label{sec:computing}

We describe in detail the at-scale GLM-EIV pipeline. First, we run a round of ``precomputations'' on all $d_g$ genes and $d_p$ perturbations. The precomputations involve regressing the gene expressions (or gRNA counts) onto the technical factors, thereby ``factoring out'' Algorithm \ref{algo:pilot_estimates_1}. Next, we run differential expression analyses on the full set of gene-perturbation pairs; for a given pair, this amounts to obtaining the complete set of pilot parameters (by running a reduced GLM-EIV), fitting the GLM-EIV model (Algorithm \ref{algo:em_full}), and performing inference. The three loops in Algorithm \ref{algo:at_scale} are embarrassingly parallel and therefore can be massively parallelized. 

\begin{algorithm}
	\caption{Applying GLM-EIV at scale.}\label{algo:at_scale}
	\begin{algorithmic}
		\State $G \gets \{\textrm{gene}_1, \dots, \textrm{gene}_{d_g}\}; P \gets \{\textrm{perturbation}_1, \dots, \textrm{perturbation}_{d_p}\}$
		
		\For{gene $\in G$}
		\State Run precomputation (Algorithm \ref{algo:pilot_estimates_1}) on gene; save $\hat{f}^m$, $[\beta^m_0]^\textrm{pilot}$ and $[\gamma^T_m]^\textrm{pilot}$.
		\EndFor
		\For{perturbation $\in P$}
		\State Run precomputation  (Algorithm \ref{algo:pilot_estimates_1}) on perturbation; save $\hat{f}^g$, $[\beta^g_0]^\textrm{pilot}$ and $[\gamma^T_g]^\textrm{pilot}$.
		\EndFor
		\For{(gene, perturbation) $\in G \times P$}
		\State Load $\hat{f}^m, \hat{f}^g,$ $[\beta^m_0]^\textrm{pilot}$ $[\gamma^T_m]^\textrm{pilot}$, $[\beta^g_0]^\textrm{pilot}$ and $[\gamma^T_g]^\textrm{pilot}$.
		\State Compute $[\beta^m_1]^\textrm{pilot}, [\beta^g_1]^\textrm{pilot}, \pi^\textrm{pilot}$ by fitting a reduced GLM-EIV (Algorithm \ref{algo:pilot_estimates_2}).
		\State Run GLM-EIV using the pilot parameters (Algorithm \ref{algo:em_full}).
		\EndFor
	\end{algorithmic}
\end{algorithm}

\section{The Nat.\ Biotech.\ 2020 method}\label{sec:Replogle_method}

As described in the main text, the Nat.\ Biotech.\ 2020 method (of \citet{Replogle2020}) fits a Poisson-Gaussian mixture model to the log-2 transformed gRNA counts and then assigns gRNAs to cells based on the posterior perturbation probabilities. If a given cell has a posterior perturbation probability greater than 1/2, then the gRNA is assigned to that cell; otherwise, the gRNA is not assigned to that cell. Covariates (including gRNA library size, gene library size, batch, etc.) are not included in the model.

As mentioned in the main text, the Nat.\ Biotech.\ 2020 method poses several conceptual and practical challenges. First, the log-2 transformed gRNA counts are not integer-valued. Thus, it is unclear how the Poisson component of the mixture distribution is fitted to the data. Second, the authors of the Nat.\ Biotech.\ 2020 method used the Python package \texttt{Pomegranate} (\url{github.com/jmschrei/pomegranate}; version $<=$ 0.14.8) to implement their method. Unfortunately, due to recent updates to the \texttt{Pomegranate} package, we and others have been unable to install version $<=$ 0.14.8 (relevant Github issues: \url{github.com/jmschrei/pomegranate/issues/1052}, \url{github.com/jmschrei/pomegranate/issues/1057}). 

Thus, we attempted to implement the Nat.\ Biotech.\ 2020 method ourselves in R using the \texttt{flexmix} package, a popular package for mixture modeling. We found that \texttt{flexmix} throws an error when one attempts to fit a Poisson distribution to non-integer data. We therefore considered a modification to the Nat.\ Biotech.\ 2020 method in which we fitted a two-component Gaussian mixture to the log-transformed gRNA counts, adding a pseudocount of one to avoid taking the log of zero. Unfortunately, our modified version of the Nat.\ Biotech.\ 2020 method did not work well in practice, as it categorized all cells as unperturbed on both the simulated gRNA data (Figure \ref{main_text_sim}) and the low-MOI gRNA data (Figure \ref{fig:grna_mixture_model}). The default CellRanger method for gRNA assignment --- which is based on the Nat.\ Biotech.\ 2020 method --- uses a two-component Gaussian mixture model (\url{www.10xgenomics.com/support/software/cell-ranger/latest/algorithms-overview/cr-crispr-algorithm}). The CellRanger method became open-source shortly before the publication of this paper.

\newpage
\section{Data analysis details}\label{sec:data_analysis_details}
		
First, we performed quality control and basic pre-processing on both datasets. As is standard in single-cell analysis, we removed cells with a high fraction ($>8\%$) of mitochondrial reads \citep{choudhary2022}. We additionally excluded genes that were expressed in fewer than $10\%$ of cells or that had a mean expression level of less than $1$. We excluded cells in the Gasperini dataset with gene transcript UMI or gRNA counts below the 5th percentile or above the 95th percentile to reduce the effect of outliers. We did not repeat this latter quality control step on the Xie data because the Xie data appeared to be less noisy. The quality-controlled Gasperini and Xie datasets contained $n = 170,645$ (resp. $n = 101,508$) cells, $2,079$ (resp. $1,030$) genes, and $6,598$ (resp. $516$) distinct perturbations.
		
The Gasperini dataset came with $17,028$ candidate \textit{cis} pairs, $97,818$ negative control pairs, and $322$ positive control pairs. The \textit{cis} pairs consisted of genes paired to nearby enhancers with unknown regulatory effects. The negative control pairs consisted of non-targeting gRNAs paired to genes. The positive control pairs are described in the main text. The Xie data did not come with either \textit{cis}, negative control, or positive control pairs. Therefore, we constructed a set of $681$ candidate \textit{cis} pairs by pairing perturbations to nearby genes, and we constructed a set of $50,000$ \textit{in silico} negative control by pairing perturbations to genes on different chromosomes. See the \textit{Methods} section of \cite{Barry2021} for details on the construction of \textit{cis} and \textit{in silico} negative control pairs on the Xie data. Because the negative control pairs are not expected to exhibit a regulatory relationship, the ground truth fold change in gene expression for these pairs is taken to be unity.

We modeled the gene expression counts using a negative binomial distribution with unknown size parameter $s$; we estimated $s$ using the {glm.nb} package. \cite{choudhary2022} report that Poisson models accurately capture highly sparse single-cell data. Although Choudhary and Satija did not investigate the application of Poisson models gRNA data specifically, we modeled the gRNA counts using Poisson distributions, as the gRNA modality exhibited greater sparsity than the gene modality.
		
We applied GLM-EIV and the thresholding method to analyze the entire set of pairs in both datasets. We did not report results on the candidate \textit{cis} pairs in the text because we do not know the ground truth for these pairs, making them less useful for method assessment. We focused our attention instead on the negative control pairs in both datasets and the positive control pairs in the Gasperini dataset.
		
We describe in more detail how we conducted the ``excess background contamination'' analysis. For each positive control pair, we varied excess background contamination over the grid $[0.0, 0.05, 0.1, \dots, 0.4].$ For a given level of excess background contamination, we generated $B = 50$ synthetic gRNA datasets, holding fixed the raw gene expressions, covariates, library sizes, and fitted perturbation probabilities. We fitted GLM-EIV and the thresholding method to the data, yielding estimates $[\hat{\beta}^m_1]^{(1)}, \dots, [\hat{\beta}^m_1]^{(B)}$. Next, we averaged over the $[\hat{\beta}^m_1]^{(i)}$s to obtain the mean estimate for a given pair and level of background contamination, and we calculated the REC using these mean estimates.

\section{Additional related work}\label{sec:additional_related_work}
Several authors working on statistical methods for single-cell data recently have extended models that (implicitly or explicitly) assume Gaussianity and homoscedasticity to a broader class of exponential family distributions. For example, \cite{Lin2021} and \cite{Townes2019} (separately) developed eSVD and GLM-PCA, generalizations of SVD and PCA, respectively, to exponential family response distributions. Unlike their vanilla counterparts, eSVD and GLM-PCA can model gene expression counts directly, improving performance on dimension reduction tasks. We see our work (in part) as a continuation of this broad effort to ``port'' common statistical methods and models to single-cell count data. Our focus, however, is on regression rather than dimension reduction: we extend the classical errors-in-variables model in several key directions (see above), enabling its direct and natural application to multimodal single-cell data.

\section{Simulation study details and additional simulation studies}\label{sec:extra_sims}

\subsection{Main text simulation study parameter values}

We constructed a table (Table \ref{tab:param_mapping}) that maps each model parameter to its (i) main text simulation study value and (ii) estimated value on real data. We obtained the real-data parameter estimates by applying GLM-EIV to analyze a representative gRNA-gene pair from the \cite{Gasperini2019} data (namely, gene ``ENSG00000213931'' paired to positive control gRNA ``pos\_control\_Klannchr1\_HS4''). The main difference between the simulation parameter values and real-data parameter values is that the perturbation effect size on gRNA expression (i.e., $\exp(\beta^g_1)$) is smaller in the simulation study than on the real data. This difference has the effect of placing the simulation study into a more challenging region of the problem space.

\begin{center}
\begin{table}[h]
\begin{tabular}{ |c|c|c| c| } 
 \hline
 Parameter & Simulation value & Estimated real data value & Meaning \\ 
 \hline
 $\exp(\beta^m_0)$ & 0.01 & 0.02 & Gene model intercept \\ 
 $\exp(\beta^m_1)$ & 0.25 & 0.68 & Gene perturbation effect \\ 
 $\exp(\gamma^m_1)$ & 0.9 & 1.0 & Gene batch effect \\
 $\exp(\beta^g_0)$ & $5.0 \cdot 10^{-3}$ & $3.4 \cdot 10^{-6}$ & gRNA model intercept \\
 $\exp(\beta^g_1)$ & $[1.0, 1.5, \dots, 4.0]$ & $6,200$ & gRNA perturbation effect \\
 $\exp(\gamma^g_1)$ & $1.1$ & $1.05$ & gRNA batch effect \\
 $\pi$ & $0.02$ & $0.004$ & Perturbation probability \\
 \hline
\end{tabular}
\caption{A mapping of each model parameter to its (i) main text simulation study value and (ii) estimated value on real data.}\label{tab:param_mapping}
\end{table}
\end{center}

\subsection{Additional simulation studies}

We report the results of five additional simulation studies. Study 2 considers Gaussian (as opposed to negative binomial or Poisson) data; study 3 varies the negative binomial size parameter $s$; study 4 varies the effect size of the perturbation on gene expression $\beta^m_1$; and study five (resp., six) considers gRNA (resp., gene) expression data that are contaminated by doublets and an unmeasured covariate. In all simulation studies we deployed the accelerated version of GLM-EIV.

\begin{figure}[H]
\centering
\includegraphics[width=0.75\linewidth]{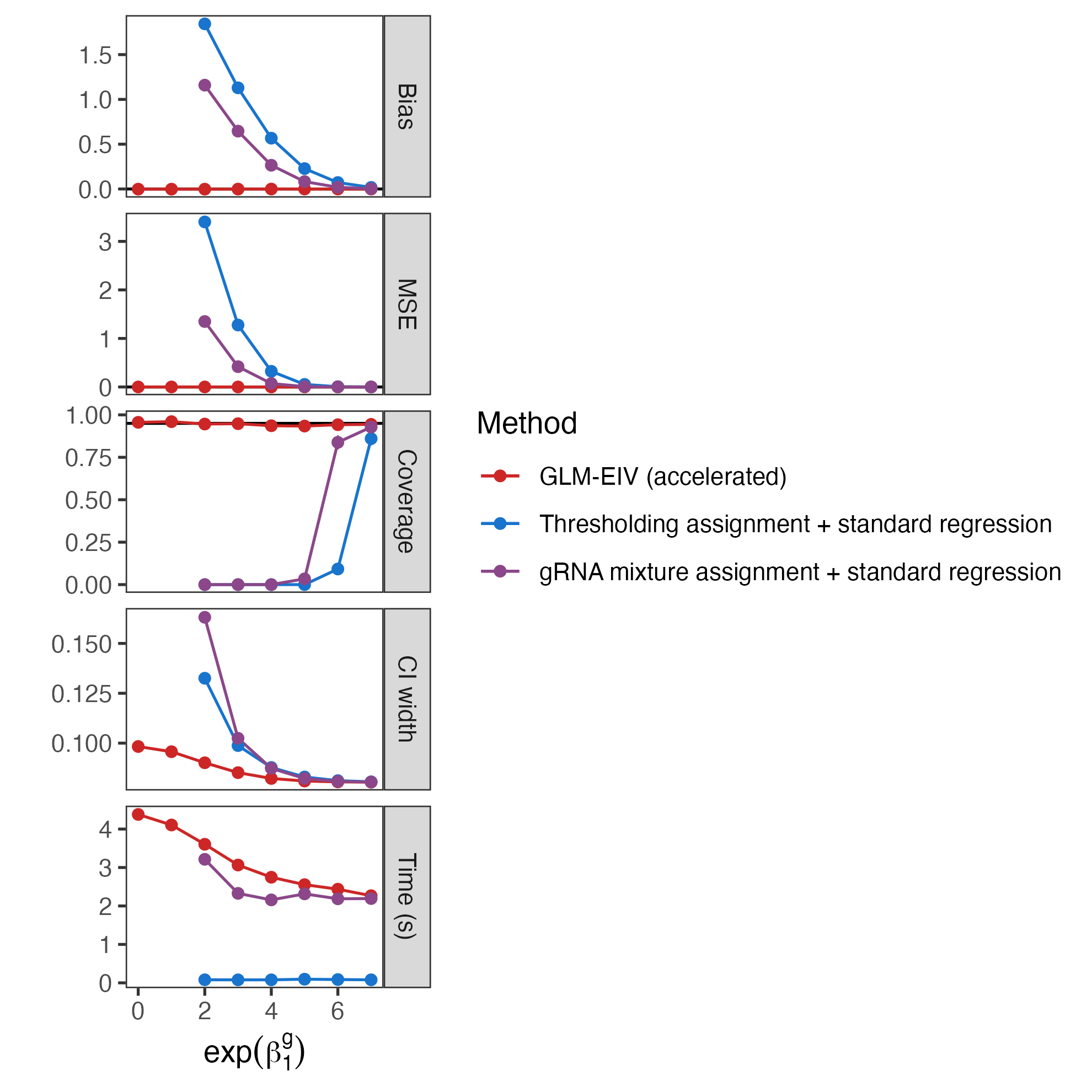}
\caption{\textbf{Simulation study 2}. Analyzing data generated from a linear Gaussian model. Rejection probability (not plotted) was strictly 1 across methods and parameter settings.}\label{fig:sim_study_2}
\end{figure}

\textbf{Simulation study 2.} In simulation study 2 we modeled the gene and gRNA expressions using a Gaussian distribution with an identity link. We generated data on $n = 50,000$ cells, fixing the target of inference $\beta^m_1$ to $-4$ and the probability of perturbation $\pi$ to $0.05$. We included ``sequencing batch'' (modeled as a Bernoulli-distributed variable) and ``sequencing depth'' (modeled as a Poisson-distributed variable) as covariates in the model. We did not include sequencing depth as an offset because use of the identity link renders offsets meaningless. We varied $\beta^g_1$ over a grid on the interval $[0,7].$ We applied GLM-EIV, thresholded regression, and the gRNA mixture assignment method (coupled to linear regression) to analyze the simulated data. The ranking of the methods was as follows: GLM-EIV (best), gRNA mixture assignment method (intermediate), and thresholding method (worst) (Figure \ref{fig:sim_study_2}).

\textbf{Simulation study 3.} Simulation study 3 was similar to the main text simulation study. The difference is that in simulation study 3, we held fixed $\beta^g_1 = \log(2.5)$ while varying the negative binomial size parameter $s$ over the grid $1 = 10^{0/9}, 10^{2/9}, 10^{4/9}, \dots, 10^{16/9}, 10^{18/9} = 100.$ We applied the three methods twice: once assuming known $s$ and once under unknown $s$. All methods demonstrated roughly uniform bias over the grid of $s$ values: the bias of GLM-EIV was near zero, while that of the thresholding method and the gRNA mixture method was about 0.02. As $s$ increased, the CI width of all methods decreased (as the gene expression data became more Poisson-like, causing standard errors to shrink). The confidence interval coverage of the thresholding method and the gRNA mixture method degraded, while that of GLM-EIV remained at the roughly nominal level. The former two methods likely lost coverage because their biased estimates caused the increasingly-narrow confidence intervals to be centered at the wrong location. The results were broadly similar across known $s$ and unknown $s$ (though slightly better under known $s$).

\textbf{Simulation study 4.} Simulation study 4 also was similar to the main text simulation study. The difference is that in simulation study 4, we held fixed the perturbation effect size on gRNA expression ($\exp(\beta^g_1) = 2.5$) and varied the perturbation effect size on gene expression $\exp(\beta^m_1)$ over the grid $0.2, 0.3, \dots, 0.9, 1.0$. We applied the three methods to analyze data generated from Poisson, negative binomial (with known $s$), and negative binomial (with unknown $s$) gene expression distributions. We observed that as the magnitude of the effect size increased (i.e., as $\exp(\beta^m_1)$ decreased from 1.0 to 0.2), GLM-EIV remained roughly unbiased, while the thresholding method and the gRNA mixture assignment method exhibited increasingly severe attenuation bias. Furthermore, GLM-EIV maintained coverage at the nominal level, while the coverage of the thresholding method and the gRNA mixture assignment method degraded due to the aforementioned attenuation bias. Results were broadly similar (albeit slightly worse) under estimated $s$ than known $s$.

We additionally plotted the rejection probability, i.e. the probability of rejecting the null hypothesis of $H_0: \exp(\beta^m_1) = 1$ at level 0.05. When $\exp(\beta^m_1) = 1$ (i.e., when we are under the null hypothesis), the rejection probability (which corresponds to type-I error) should be 0.05, the nominal level. When $\exp(\beta^m_1) < 1$ (i.e., when we are under the alternative hypothesis), the rejection probability (which corresponds to power) should be as large as possible (with a value of 1.0 being optimal). We observed that all methods exhibited a rejection probability of roughly 0.05 under the null hypothesis of $\exp(\beta^m_1) = 1$ and a rejection probability of 1.0 under the alternative hypotheses of $\exp(\beta^m_1) = 0.9, 0.8, \dots, 0.2, 0.1$. In other words, over the grid of values that we examined, each method performed optimally with respect to testing the hypothesis $\exp(\beta^m_1) = 1$. (We note that our goal in the simulation studies was to explore discrepancies in estimation accuracy and confidence interval coverage across methods, but we present type-I error control and power results for completeness.)

\textbf{Simulation study 5.} In simulation study 5 we applied the methods to analyze data drawn from a distribution that lay outside the GLM-EIV family of distributions. First, we simulated gRNA count data from a poisson GLM with two covariates: batch (modeled as a Bernoulli random variable with probability 1/2) and cell cycle (modeled as a uniform random variable on the interval [0,1]). We treated cell cycle as an unmeasured covariate, i.e. we did not give any of the methods access to cell cycle. Next, we randomly selected 1$\%$ of cells and doubled the gRNA count in these cells, thereby simulating the presence of doublets (i.e., droplets that contain two cells) in the data. We simulated the gene expression data from the same negative binomial model that we used in the main text simulation (and so the gene expression model was correctly specified.) For simplicity we assumed that the size parameter $s=20$ was known. We varied the perturbation effect size on gRNA expression $\exp(\beta_1^g)$ over the grid $1, 2, \dots, 7$ and the perturbation effect size on gene expression $\exp(\beta^m_1)$ over the grid $0.25, 0.5, 0.75, 1.0$.

We applied GLM-EIV, thresholded regression, and the gRNA mixture assignment method to analyze the data. GLM-EIV exhibited generally lower bias, lower mean squared error, and better confidence interval coverage than the other methods. The rightmost panel (i.e, $\exp(\beta_1^m) = 1$) corresponds to the null hypothesis of no perturbation effect on gene expression; the left panels (i.e., $\exp(\beta_1^m) = 0.75, 0.5, 0.25$), by contrast, correspond to alternative hypotheses of varying strength. All methods controlled type-I error at the nominal level of 0.05. GLM-EIV demonstrated equal or greater power than the competing methods.

\textbf{Simulation study 6}. Simulation study 6 was similar to simulation study 5, the difference being that simulation study 6 considered a misspecified gene expression model (while simulation study 5 considered a misspecified gRNA count model). We generated the gene expression data from a negative binomial GLM containing the unmeasured covariate of cell cycle, and we doubled the gene expression count in 1$\%$ of randomly selected cells to simulate doublets. We generated gRNA counts from the same gRNA model that we used in the main text simulation (and so the gRNA count model was correctly specified.) Again, we varied $\exp(\beta_1^g)$ over the grid $1, 2, \dots, 7$ and $\exp(\beta^m_1)$ over the grid $0.25, 0.5, 0.75, 1.0$. We found that GLM-EIV generally performed best: GLM-EIV exhibited lower bias, lower mean squared error, and better confidence interval coverage than the other methods. There was one setting for $\beta^g_1$ (namely, $\exp(\beta^g_1) = 1.5$) for which GLM-EIV did not control type-I error under the null hypothesis of $\exp(\beta^m_1) = 1.$ However, this was an extreme value for $\beta^g_1$, and GLM-EIV controlled type-I error under all other values of $\beta^g_1$.

\begin{figure}[H]
\centering
\includegraphics[width=0.55\linewidth]{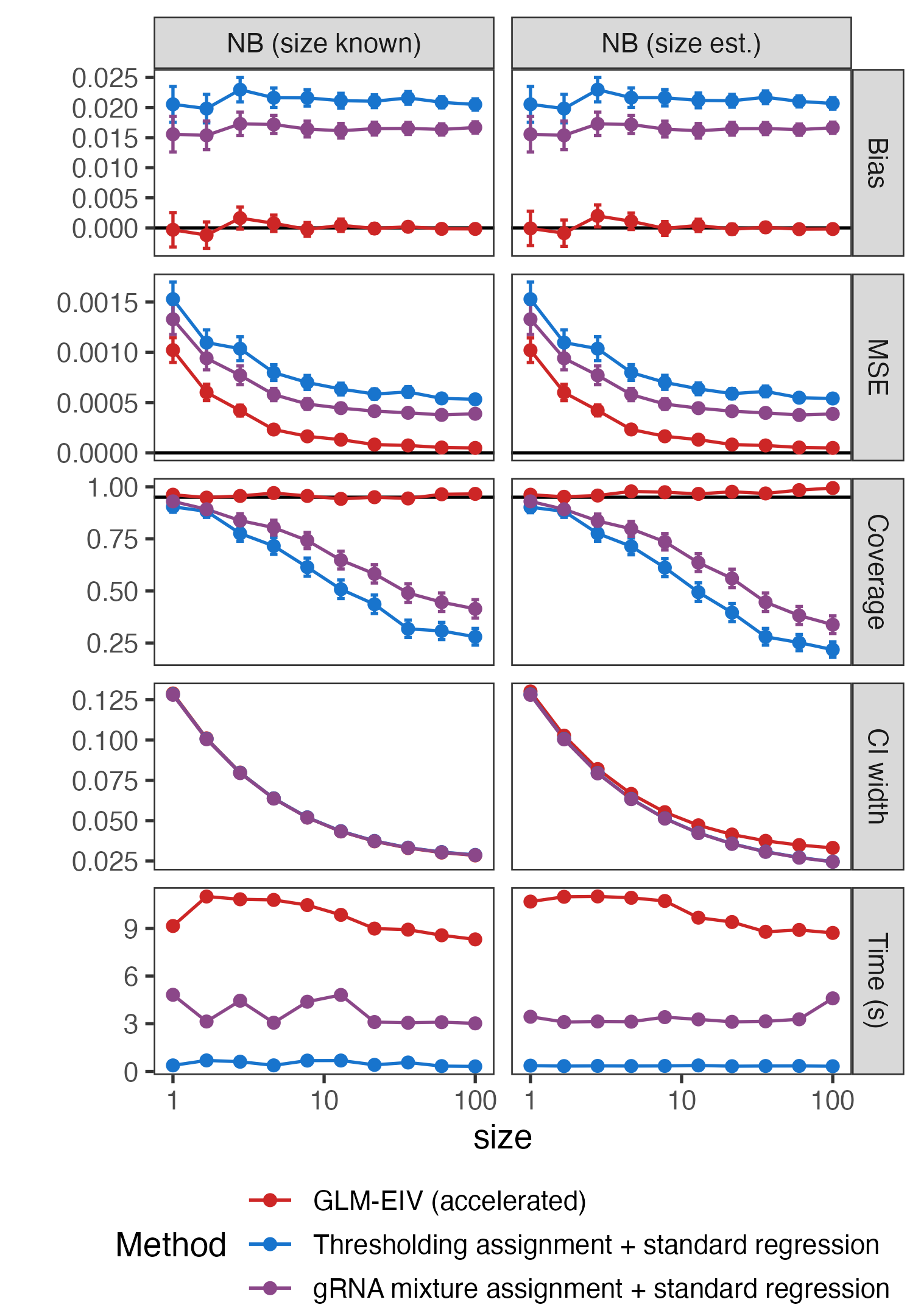}
\caption{\textbf{Simulation study 3}. Varying the negative binomial size parameter $s$. Rejection probability (not plotted) was strictly 1 across methods and parameter settings.}\label{fig:sim_study_3}
\end{figure}

\begin{figure}[H]
\centering
\includegraphics[width=0.7\linewidth]{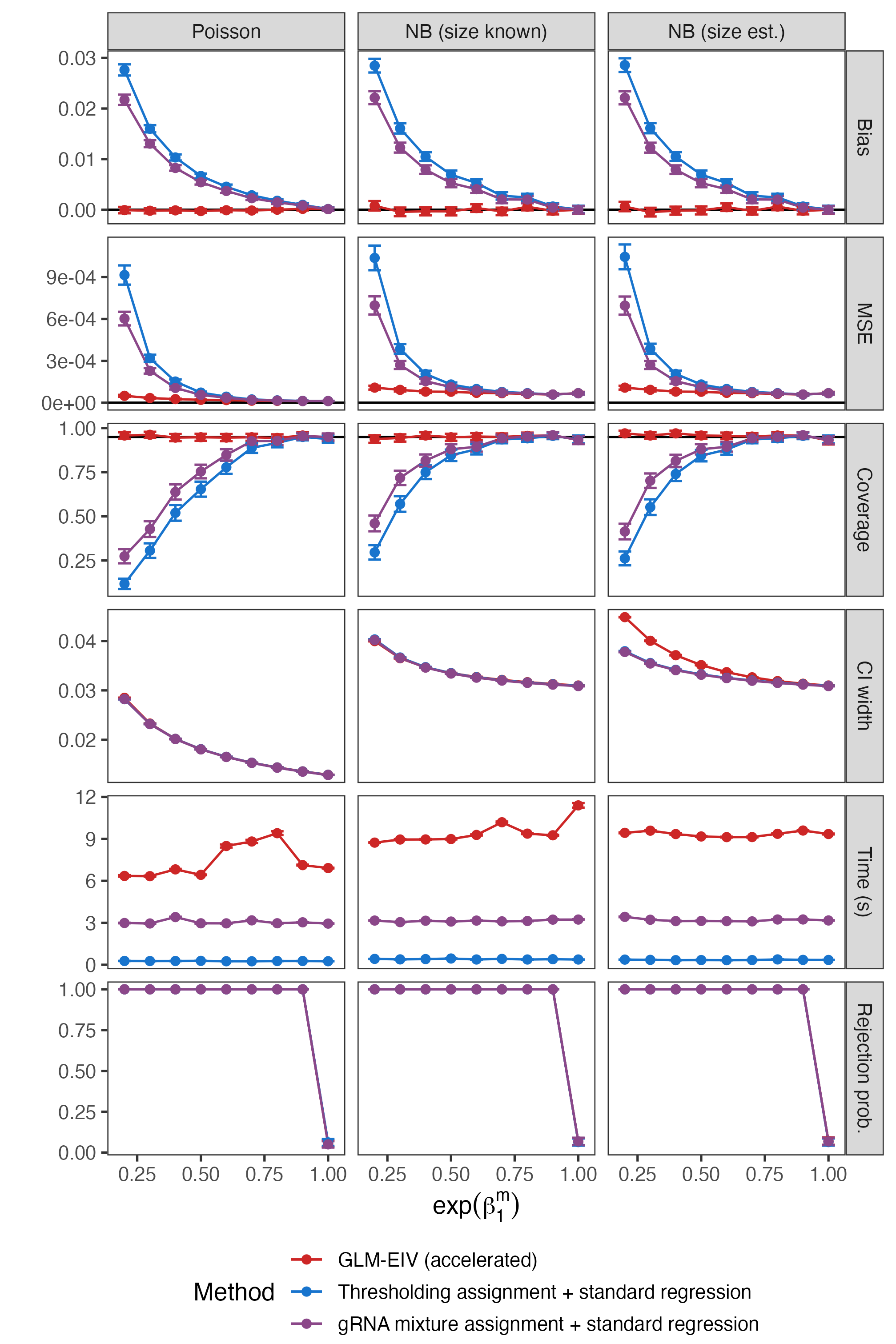}
\caption{\textbf{Simulation study 4}. Varying the perturbation effect size on gene expression, $\beta^m_1$.}\label{fig:sim_study_4}
\end{figure}

\begin{figure}[H]
\centering
\includegraphics[width=0.8\linewidth]{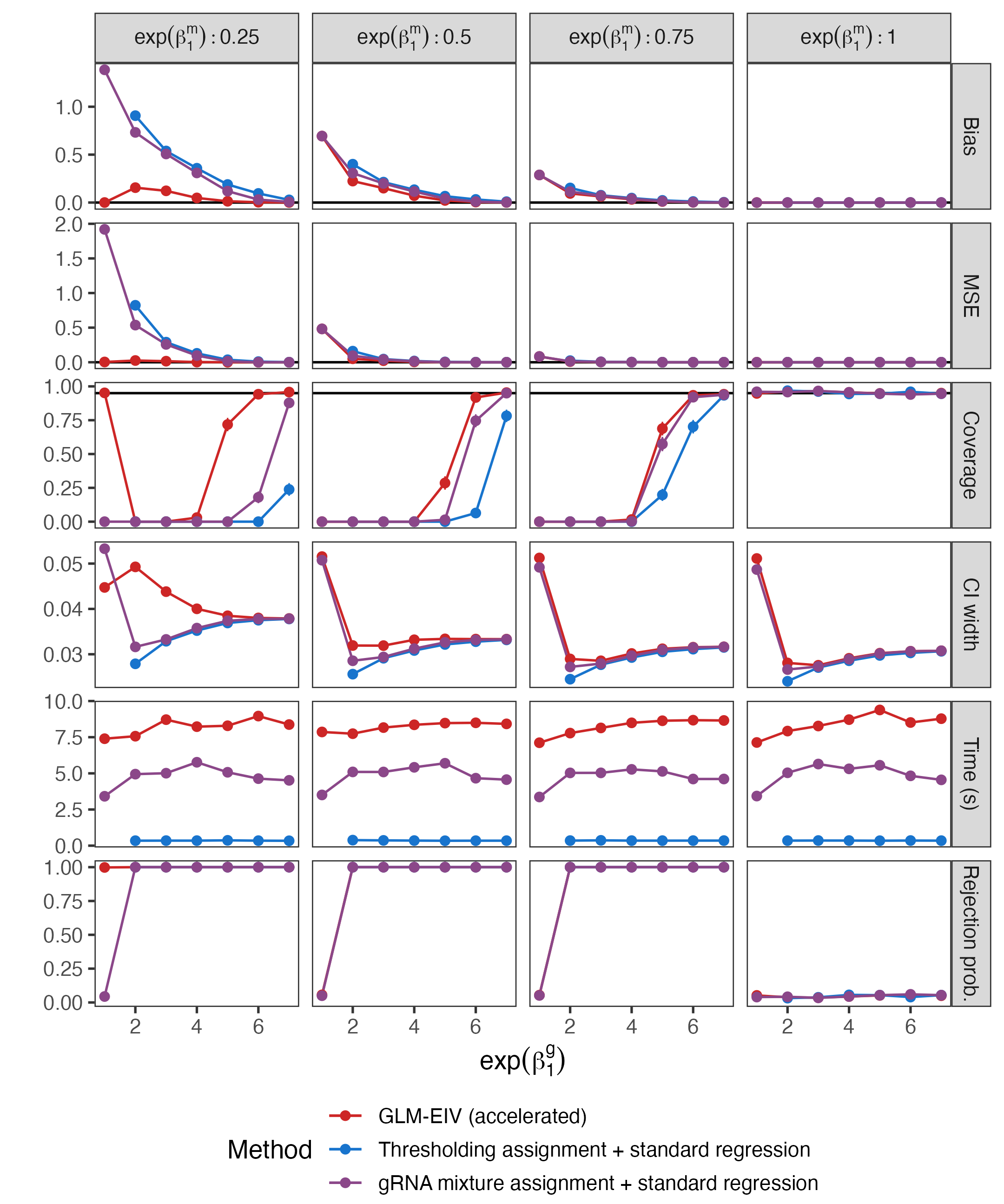}
\caption{\textbf{Simulation study 5}. Analyzing data using a misspecified gRNA count model.}\label{fig:sim_study_5}
\end{figure}

\begin{figure}
\centering
\includegraphics[width=0.8\linewidth]{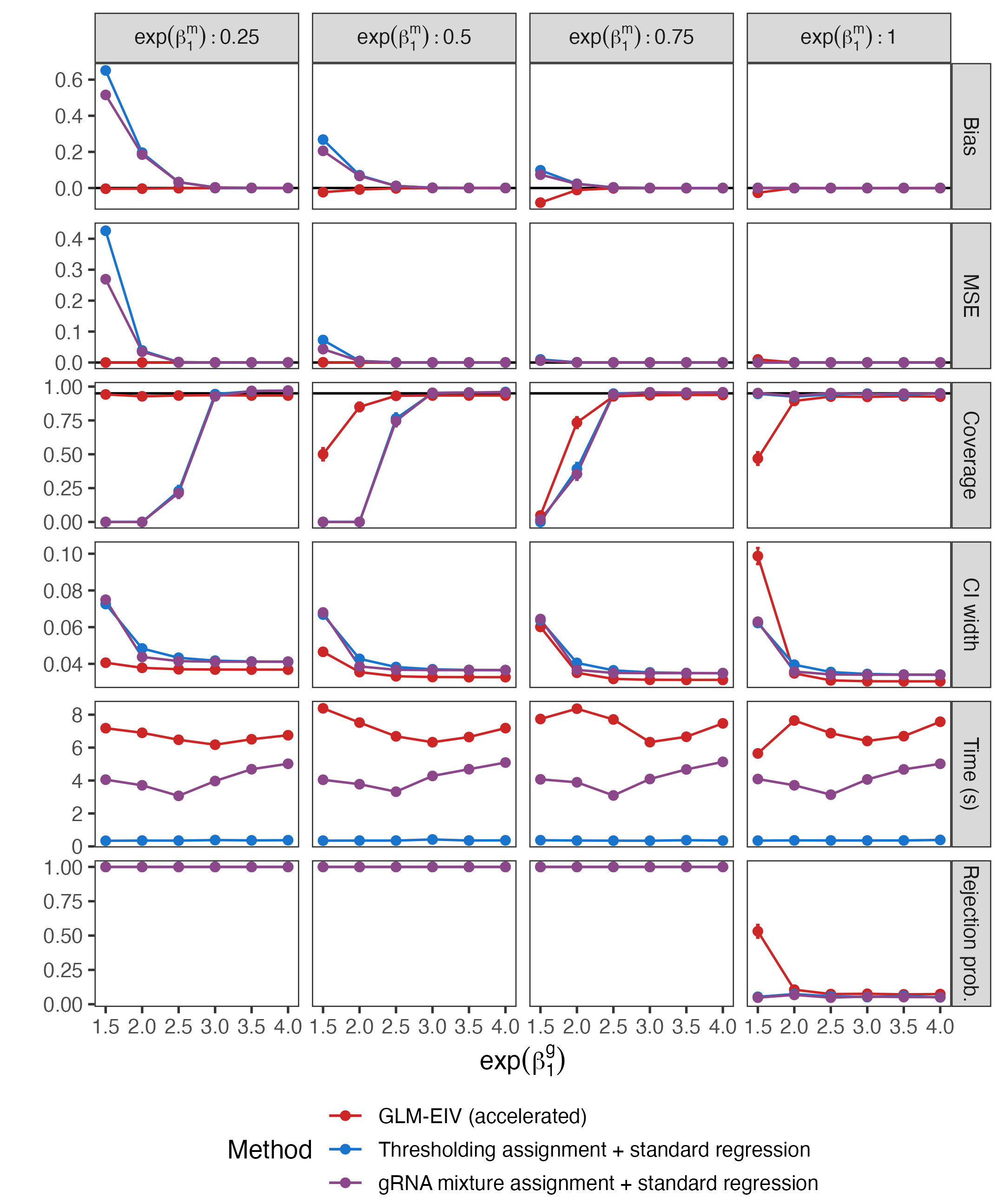}
\caption{\textbf{Simulation study 6}. Analyzing data using a misspecified gene expression model.}\label{fig:sim_study_6}
\end{figure}

% \end{appendices}
\end{document}